%% file: FreqPap.tex
\documentclass[prb,preprint,superscriptaddress,showpacs]{revtex4}

\usepackage{amssymb}
\usepackage{amsmath}
\usepackage{epsfig}
\usepackage{bbm}
\usepackage{graphicx}
\usepackage[usenames]{color}

\newcommand{\mtin}[1]{\mbox{\tiny {#1}}}
\newcommand{\ca}[1]{{\cal #1}}
\newcommand{\ovl}[1]{\overline{#1}}
\newcommand{\sfrac}[2]{{\textstyle\frac{#1}{#2}}}

\newcommand{\bok}{\mathbf{k}}         
\newcommand{\bol}{\mathbf{l}}         
\newcommand{\bop}{\mathbf{p}}

\newcommand{\bzero}{\mathbf{0}}
\newcommand{\bpi}{\hat{\pi}}
\newcommand{\intd}{\textrm{d}}                  
\newcommand{\undemi}{\frac{1}{2}}  		        % 1/2
  		        
\newcommand{\ie}{i.e.\ }

\newcommand{\rIm}{\mathop{\mathrm{Im}}}
\newcommand{\rRe}{\mathop{\mathrm{Re}}}      
\newcommand{\dero}[1]{\frac{\textrm{d}}{\textrm{d}#1}}	% d/d...
\newcommand{\ZZ}{\mathbb{Z}}      
\newcommand{\RR}{\mathbb{R}}      
\newcommand{\om}{\omega} 
\newcommand{\Om}{\Omega}
\newcommand{\eps}{{\varepsilon}}        
\newcommand{\cO}{\mathcal{O}}         
         
\newcommand{\abs}[1]{\left| #1 \right|}
\newcommand{\sgn}{\mathop{\mathrm{sgn}}}
\newcommand{\wbox}[1]{\fcolorbox{white}{white}{#1}}

\begin{document}

\bibliographystyle{apsrev}
\title{Frequency Dependent Vertex Functions of the $(t,t')$--Hubbard Model
 at Weak Coupling}
\author{C.~Husemann}
\affiliation{Max Planck Institute for Solid State Research,
 D-70569 Stuttgart, Germany}
\author{K.-U.~Giering}
\author{M.~Salmhofer}
\affiliation{Institute for Theoretical Physics, Universit\"at Heidelberg, D-69120 Heidelberg, Germany}
\date{\today}

\begin{abstract}
We present a functional renormalization group calculation for the two-dimensional $(t,t')$-Hubbard model at Van Hove filling. Using a channel decomposition we describe the momentum and frequency dependence of the vertex function in the normal phase. Compared to previous studies that neglect frequency dependences we find higher pseudo-critical scales and a smaller region of $d$-wave superconductivity. A large contribution to the effective interaction is given by a forward scattering process with finite frequency exchange. We test different frequency parameterizations and in a second calculation include the frequency dependence of the imaginary self-energy. We also generalize the channel decomposition to frequency-dependent fermion-boson vertex functions.
\end{abstract}

\maketitle
%\tableofcontents
\section{Introduction}
The phase diagrams of currently investigated
quasi-two-dimensional correlated electron systems,
such as the cuprates \cite{LeeNagaosaWen2006}, ruthenates\cite{MackenzieMaeno}, and
 pnictides \cite{pnictides,PaglioneGreene2010},
exhibit a variety of dynamically generated
scales and phenomena caused by competing correlation effects.
The fermionic functional renormalization group (RG) method
has become a useful tool for the theoretical
investigation of  models for these materials, and more generally, low-dimensional
correlated fermion models, both in and out of equilibrium  (for a recent review, see Ref. \onlinecite{RGreview2011}).

The RG is a natural approach for understanding scale-dependent phenomena.
Specific advantages of the method in its application to correlated fermion
systems are that it does not require a priori
assumptions about which correlations will eventually  dominate, that it applies to
a wide range of models, including multiband systems and models with general two-
or even higher-body interactions, and that it allows to calculate the vertex
functions of the associated quantum field theory, which are directly
related to measurable quantities, as a function of momenta and
(Matsubara) frequencies.
Moreover, its application to weakly coupled Fermi systems is backed up by
mathematical results implying full theoretical control,
in the sense that for systems with weak, short-range interactions, convergent expansions
of the RG map exist and can be used in the model class relevant for the
above materials \cite{FT1990, FMRT,FKTperturbationbound2004,SalmhoferWieczerkowski2000,DR2000,BenfattoGiulianiMastropietro2006,PedraSalmhofer2008}.
 Practically, this means that the method is robust with respect to small
changes and has good stability properties.

In its most widely used form, the fermionic RG takes the form of an
infinite hierarchy of integro-differential equations for the vertex functions
\cite{RGreview2011}.
Even in an approximation by truncation to a finite system of equations,
it is difficult to take the physically important dependence of the
vertex functions on momenta and frequencies accurately into account.
In this paper, we report the results of a study for the two-dimensional
Hubbard model in which we have taken into account the frequency
and momentum dependence of the vertex functions, using the standard
level-2 and Katanin truncation schemes described in Ref.~\onlinecite{RGreview2011},
as well as the vertex parametrization of Ref.~\onlinecite{decomposition}.
In the following we briefly review
some important general arguments, issues, and results, first in the
translation-invariant two-dimensional case and then in a more general context.

Weak-coupling power counting indicates that in the RG flow at low energy
scales,
the largest contributions to the flow of the vertex functions come from
the close vicinity of the Fermi surface and small Matsubara frequencies.
This suggests to start with a vertex function that depends only on the
projection of momentum to the Fermi surface and of frequency to
zero frequency as a first approximation. The vertex functions still
depend on where the fermion momenta are on the Fermi surface.
It is then natural to discretize this dependence, which converts the
integro-differential equation into a system of many ODE.
In the rigorous treatment of Ref.~\onlinecite{FMRT}, the number of such points
on Fermi surfaces with curvature is chosen to scale as $\Lambda^{-1/2}$
as a function of the RG energy scale $\Lambda$.
In practical applications to more general Fermi surface shapes,
the number of patches $N$ is taken fixed, but as large as possible. 
This $N$-patch method avoids any simplifying assumptions
on the functional form of the projected vertex functions,
and it has been very useful in understanding instabilities
and drawing conclusions for phase diagrams  \cite{ZanchiSchulz1,ZanchiSchulz2000,HalbothMetzner,Umklapp,HonerkampSalmhofer_QuasiparticleWeight,HonerkampSalmhofer_Tflow}.
Its implementation at sufficiently large $N$ is, however,
numerically expensive for
two-dimensional systems, since the number of components of a fermionic
two-body interaction grows with the third power of $N$.
Using parallelization and present-day computing power, values of $N$ up to 144 have been attained, 
and the radial dependence has partly been taken into account \cite{HonerkampInteractionFlow2004}.
The parametrization of Ref.~\onlinecite{decomposition} approximates the four-point vertex by
terms of the form of a boson exchange, which reduces the growth in $N$
in the resulting equations from cubic to linear. The price to pay is an additional
error term, since representing the vertex function exactly in this way would
require an infinite number of such terms. In the models we study here,
however, it turns out that a finite, not very large number of them provides
a good approximation in a large parameter range \cite{remainder}.

The careful weak-coupling power counting argument is subtle,
as it involves Taylor expansions of the vertex functions and a detailed
weighing of gains and losses in scale behavior in showing that projections
can be taken.
The argument is rigorous in the case of a single dominant instability
and when the effective interaction is weak.
However, the symmetry breaking at low temperatures
manifests itself via a growth of the effective interactions
for certain combinations of momenta as the RG scale parameter
is decreased, so that the effective vertex develops singularities
corresponding to a slow decay of the effective interaction in position
space. Although power counting improvements allow to continue the flow
in the above-mentioned truncation \cite{salmfrgtt}
beyond the region of small couplings,
the truncation becomes unreliable already before a true singularity is reached.
If one insisted on continuing the flow to arbitrarily low scales one would
lose control over the errors. This so-called flow to strong coupling
is not a problem of the RG method as a whole, but
can be mended when composite fields are treated appropriately.
Then, the flow to strong coupling is found to indicate
that correlations of composite fields become long-range, leading to
phase transitions. Since the dynamics of long-range correlations
strongly depends on space-time dimension, the frequency dependence
of the vertex functions becomes an essential issue in this regime.
Besides allowing to check previous instability analyses, the results
given here also provide vertex functions in a useful form for stopping
the flow before the truncation fails, and for introducing bosonic order parameter
fields via a Hubbard-Stratonovich transformation.
A good description of the frequency dependence is also important
in the technical implementation of RG flows for irreducible vertex functions,
because their functional behavior at low scales is in general not simple,
which raises the question how good simple parameterizations
based on Taylor expansion around zero frequency really are.
Secondly, the 1PI equations never completely decouple the
high-energy and high-frequency modes from the low-energy ones,
so a quantitative description requires more than just the asymptotic
low-scale information, but a good approximation in the entire frequency
range.

For the 2D lattice fermion systems, the frequency
dependence was studied in various truncations.
Besides being important for the analysis at low energy scales,
tracking the frequency dependence of the vertex also becomes an issue 
when the frequency dependence of the fermionic self-energy is investigated.
This is important in the study of quasi-particle weights and lifetimes.
Inserting a frequency-independent four-point vertex in the standard 
equation for the self-energy flow in the 1PI hierarchy gives a frequency-independent
self-energy. 
In previous studies, the frequency-dependence of the self-energy was obtained in an
approximation where bare propagators are used and the vertex flow is kept frequency
independent,
but the integrated flow equation for the vertex 
(which changes the one-loop frequency dependence) is inserted into the flow equation for 
the self-energy.
In this way, the self-energy gets a frequency-dependence as in sunset diagrams,
in which the momentum dependence is that of the flowing vertex.
This approximation was used to calculate the flow of 
quasiparticle scattering rates\cite{Honerkamp_ScatRate}, $Z$ factors\cite{HonerkampSalmhofer_QuasiparticleWeight}
and to continue to real frequencies\cite{RoheMetzner,KataninKampf2004}
(Ref. \onlinecite{RoheMetzner} uses an alternative formulation in the Wick ordered scheme).
In a partially bosonized formulation, exchange propagators with a quadratic frequency dependence 
have been used\cite{BaierBickWetterich,Friederich2011}.

In various further situations RG flows with frequency-dependent vertices have been studied in some detail.
In the case of the single-impurity Anderson model coupled to noninteracting
leads, integration over the degrees of freedom in the leads produces an
effective local model in which all dependence is in the frequency variables.
The dependence on the frequencies has been studied in great detail 
\cite{KarraschHeddenMeden2008} and also at real frequency \cite{SeverinPletyukovSchoeller2010}.
A representation with boson fields was studied in Ref.~\onlinecite{BartoschKopietz2009}.
Very recently, the polaron problem in ultracold gases was addressed
\cite{SchmidtEnss}.

This paper begins with a short recap of fermionic RG equations in Sec.~\ref{sec:FermionicRG}. In Sec.~\ref{sec:ChannelDecomp} we describe our parameterization of the effective two-fermion vertex. The corresponding flow equations are derived in Sec.~\ref{sec:flowequations}. In Sec.~\ref{sec:FreqParam} we discuss different frequency parameterizations of effective boson exchange interactions. After numerical details are given in Sec.~\ref{sec:Numerical} we present results at Van Hove filling in Sec.~\ref{sec:results}.

We consider the extended $(t,t')$-Hubbard model on a 2D square lattice with unit spacing
\begin{align}
 H[a^\dagger,a] = \sum_{\mathbf{k},\sigma} \epsilon_{\mathbf{k}} a^{\dagger}_\sigma(\mathbf{k}) a_\sigma(\mathbf{k}) 
 + U \sum_x n_{x,+} n_{x,-}\ ,
\end{align}
where the kinetic term $\epsilon_{\mathbf{k}} = -2t (\cos k_x +\cos k_y) +4t' \cos k_x \cos k_y$ originates from a tight binding approximation with  nearest neighbor and next to nearest neighbor hopping. While $t>0$ sets the energy scale, $t'\in (0,0.5)t $ measures the curvature of the Fermi surface $\{\mathbf{k}\in (-\pi,\pi]^2:e({\mathbf{k}})=\epsilon_{\mathbf{k}} -\mu =0\}$. 
We adjust the system to Van Hove filling, $\mu = -4t'$, such that the free density of states is logarithmically divergent at the Fermi niveau.
The interaction with $U>0$ is an onsite Coulomb repulsion of local densities 
$n_{x, \sigma} =  a^{\dagger}_{x,\sigma } a_{x,\sigma }$.
%%%%%%%%%%%%%%%%%%%%%%%%%%%%%%%%%%%%%%%%%%%%%%%%%%%%%%%%%%%%%%%%%%%%%%%%%%

\section{Fermionic RG\label{sec:FermionicRG}}
We apply the fermionic Wilsonian RG based on integrating out fluctuations at scale $\Lambda$, see Ref. \onlinecite{RGreview2011} for a review. A finite $\Lambda>0$ regularizes all graphs in the infrared, for example, cuts off momentum shells of thickness $\Lambda$ around the Fermi surface. Our specific choice of introducing the scale $\Lambda$ is different and explained below. We analyze the change of correlation functions when $\Lambda$ is lowered in the RG flow within the level-2 truncation of the infinite hierarchy of 
flow equations \cite{salmfrgtt,RGreview2011}.

The $SU(2)$ and $U(1)$ invariant two-particle vertex is parameterized as
\begin{align}\label{eq:vertexfunction}
\ca{V}_{\Lambda}[\Psi]&= \frac 12 \int \mathrm{d}p_1\ldots \mathrm{d}p_4
\;\delta(p_1+p_2-p_3-p_4)\; V_{\Lambda}(p_1,p_2,p_3) \\ \nonumber
&\hspace{5cm} \times \sum_{\substack{\sigma,\sigma' \\  \in\{+,-\}}}
\ovl{\psi}_{\sigma}(p_1) \ovl{\psi}_{\sigma'} (p_2)
\psi_{\sigma'}(p_3) \psi_{\sigma}(p_4) \, ,
\end{align}
where $\ovl{\psi}$ and $\psi$ are anti-commuting Grassmann variables that replace the fermion operators of the Hamiltonian when writing the grand canonical partition function as a functional integral \cite{ZinnJustin}. The electron spin is denoted by $\sigma$ and $\sigma'\in\{\pm\}$ and the $p_i=(p_0^i,\mathbf{p}_i)$ include both Matsubara frequency $p_0^i$ and momentum $\mathbf{p}_i$. So the integrals $\int\mathrm{d}p_i$ are shorthand notation for $\int\frac{\mathrm{d}^2\mathbf{p}_i}{(2\pi)^2} \frac{1}{\beta} \sum_{p_0^i}$ in the thermodynamic limit with inverse temperature $\beta$.  One frequency and momentum is fixed because of energy and momentum conservation. The effective vertex function $V_{\Lambda}(p_1,p_2,p_3)$ thus depends on three independent momenta and frequencies and the RG scale $\Lambda$.

In our truncation the effective vertex function is determined by the initial condition $V_{\Lambda}(p_1,p_2,p_3)=U$ for $\Lambda\to \infty$ and its flow equation \cite{salmfrgtt}
\begin{align}\label{eq:OneLoopVertexFlow}
\dot{V}_{\Lambda}(p_1,\ldots p_3) &= \ca{T}_{\mtin{pp}}(p_1,\ldots p_3)+
\ca{T}_{\mtin{ph}}^{\mtin{d}}(p_1,\ldots p_3) +
\ca{T}_{\mtin{ph}}^{\mtin{cr}}(p_1,\ldots p_3) \, ,
\end{align}
where the dot denotes a derivative with respect to the scale
$\Lambda$. The particle--particle contribution and the crossed and
direct particle--hole contributions are respectively given by
\begin{align}
\ca{T}_{\mtin{pp}}(p_1,\ldots p_3) &= -\int \mathrm{d}p
\left[{\textstyle \frac{d}{d\Lambda}} G(p) G(p_1+p_2-p)\right] V_{\Lambda}(p_1,p_2,p)\,
V_{\Lambda}(p_1+p_2-p,p,p_3)\nonumber\\
\ca{T}_{\mtin{ph}}^{\mtin{cr}} (p_1,\ldots p_3) &= -\int \mathrm{d}p
\left[{\textstyle\frac{d}{d\Lambda}} G(p) G(p+p_3-p_1)\right]
V_{\Lambda}(p_1,p+p_3-p_1,p_3)\nonumber\\ &\hspace{7cm}\times \, V_{\Lambda}(p,p_2,p+p_3-p_1)\nonumber\\
\ca{T}_{\mtin{ph}}^{\mtin{d}} (p_1,\ldots p_3) &= \int \mathrm{d}p
\left[{\textstyle\frac{d}{d\Lambda}} G(p) G(p+p_2-p_3)\right] \Big[
2V_{\Lambda}(p_1,p+p_2-p_3,p) \,V_{\Lambda}(p,p_2,p_3) \nonumber\\
&\hspace{3cm} -V_{\Lambda}(p_1,p+p_2-p_3,p_1+p_2-p_3)\,V_{\Lambda}(p,p_2,p_3)
\nonumber\\& \hspace{3cm}-V_{\Lambda}(p_1,p+p_2-p_3,p)\,
V_{\Lambda}(p,p_2,p+p_2-p_3)\Big] \, ,
\end{align}
where the scale derivative acts on both propagators but not on the vertex functions.
Here $G(p)$ are fully dressed but regularized fermion propagators
\begin{align}
G(p) &= \frac{\chi_{\Lambda}(p)}{ip_0 -e(\mathbf{p}) -\chi_{\Lambda}(p)\Sigma_{\Lambda}(p)},
\end{align}
where $\chi_{\Lambda}(p)$ is a multiplicative regulator function, which sets the scale $\Lambda$ by suppressing
frequency modes $|p_0|\ll \Lambda$ or momentum shells with energy $|e(\bop)|\ll \Lambda$. For notational brevity, we have not indicated the $\Lambda$-dependence of $G$ in the notation, but emphasize that both $G$ and $S$ (defined below) depend on $\Lambda$.
For $\Lambda\to 0$ the regulator function becomes unity so that $G$ becomes the full propagator. The self-energy $\Sigma_{\Lambda}(p)$ is independent of spin because of $SU(2)$ invariance and can be calculated via
\begin{align}
\dot{\Sigma}_{\Lambda}(p) = \int \mathrm{d} k \; S(k) \Big[2 V_{\Lambda}(k,p,p) -V_{\Lambda}(p,k,p)\Big],
\end{align}
where $S(k)=\dot{G}(k)-G(k)\dot{\Sigma}_{\Lambda}G(k)$ is the single scale propagator, which is non-zero only where  $\dot{\chi}_{\Lambda}(k)\neq 0$ because %formally
\begin{align}
 S(k)=
\dot{\chi}_{\Lambda}(k) \frac{ik_0 - e(\bok)}{(ik_0-e(\bok)-\chi_\Lambda(k)\Sigma_\Lambda(k))^2}\, .
\end{align}
 Notice that we replaced single scale propagators by $\dot{G}$ in the flow equation for the effective vertex function. This is consistent in the level-2 truncation since $\dot{\Sigma}$ contains higher order terms \cite{KataninWard,symmetrybrokenBCS}.

Although on the right hand side of Eq.~(\ref{eq:OneLoopVertexFlow}) only one-loop graphs appear, 
higher loop graphs are generated by solving the differential equation on iteration. 
The initial condition $V_{\Lambda\to \infty}(p_1,p_2,p_3)=U$ is a constant but the vertex function develops a singular momentum and frequency structure as $\Lambda$ is decreased. 
A singularity in the flow signals the breakdown of the level-2 truncation and 
forces us to stop the RG flow at a stopping scale $\Lambda_*$. 
By analyzing the momentum and frequency structure of the effective vertex at the scale $\Lambda_*$  we determine possible instabilities of the Landau Fermi liquid. 
Note that $\Lambda_*$ and the pseudo-critical scale, where a singularity appears in the vertex function, are only upper bounds of the critical scale for a possible phase transition.

Already the bare one-loop bubbles $\Phi_{\pm}(l)=\int \mathrm{d}p \;G(p)G(l\pm p)$ diverge logarithmically 
for certain values of $l$ in the limit $\Lambda\to 0$. At Van Hove filling the density of states diverges logarithmically, which leads to an enhanced divergence of the particle-particle bubble $\Phi_-(0)\sim (\ln \Lambda)^2$. Also the particle-hole bubble $\Phi_+(l)$ diverges logarithmically 
at $l=0$ and $l=(0,\hat{\pi})$, 
with $\bpi = (\pi, \pi)$,
for  $t'/t \in (0,\sfrac{1}{2})$ at Van Hove filling. For finite but low enough $\Lambda$ and $t'\neq 0$, the regularized bubble $\Phi_+(l)$ exhibits a maximum close to, but not exactly at, $\bol = \hat\pi$.

\section{Channel Decomposition\label{sec:ChannelDecomp}}
At not too low scales $\Lambda\gg\Lambda_*$ the vertex function is still regular and the potential singular structure of Eq.~(\ref{eq:OneLoopVertexFlow}) is given by transfer momenta $p_1+p_2$, $p_3-p_1$, and $p_2-p_3$ that propagate through the fermion bubbles. In order to parameterize this frequency and momentum dependence we introduce three distinct channels \cite{decomposition}
\begin{align}\nonumber
\dot{\Phi}_{\mtin{SC}}^{\Lambda}(p_1,p_3,p_1+p_2)
&=-\ca{T}_{\mtin{pp}}(p_1,p_2,p_3)\\ \label{eq:channelevolution}
\dot{\Phi}_{\mtin{M}}^{\Lambda}(p_1,p_4,p_3-p_1)
&=\ca{T}_{\mtin{ph}}^{\mtin{cr}}(p_1,p_2,p_3)\\\nonumber
\dot{\Phi}_{\mtin{K}}^{\Lambda}(p_1,p_3,p_2-p_3)
&=-2\ca{T}_{\mtin{ph}}^{\mtin{d}}(p_1,p_2,p_3)
+\ca{T}_{\mtin{ph}}^{\mtin{cr}}(p_1,p_2,p_1+p_2-p_3) ,
\end{align}
which we loosely call the superconducting, magnetic, and density channel, respectively. Each channel absorbs the graphs with one particular transfer momentum. A linear combination of these channels gives the flow of the vertex function, see  Eqs.~(\ref{eq:OneLoopVertexFlow}) and (\ref{eq:vertexdecomposition}) below. 
The vertex function satisfies the symmetries
\begin{align}\nonumber
 V_{\Lambda}(p_1,p_2,p_3) &= V_{\Lambda}(p_3,p_1+p_2-p_3,p_1) \\ \label{eq:VertexSymmetries}
			 &= V_{\Lambda}(p_2,p_1,p_1+p_2-p_3)\\ \nonumber
			&=V_{\Lambda}(\tilde{p}_1,\tilde{p}_2,\tilde{p}_3) ,
\end{align}
that is, particle-hole symmetry, an anti-symmetrization property and time reversal symmetry, respectively,
and $\tilde{p}=(-p_0,\mathbf{p})$. As a consequence of the first two symmetries
the functions $\dot{\Phi}_{\mtin{SC}}^{\Lambda}$,  $\dot{\Phi}_{\mtin{M}}^{\Lambda}$, and $\dot{\Phi}_{\mtin{K}}^{\Lambda}$ are symmetric in their first two momentum and frequency indices. This frequency and momentum dependence is assumed to be regular in each channel. The regularity is known to hold as long as the main momentum dependence is generated by a single channel in the fermion bubble. In case of competing instabilities at lower RG scales $\Lambda$ this becomes a non-trivial assumption that, in the end, has to be checked.
We will use this regularity to expand each channel in a sum of boson exchange interactions.

In generalization of Ref. \onlinecite{decomposition} we allow the fermion-boson vertex to depend on momentum, frequency, and the scale. For example, a general expansion of the superconducting channel is given by
\begin{align}\label{eq:SCExpansion}
 {\Phi}_{\mtin{SC}}^{\Lambda}(q,q',l) &= \sum_{m,n} D_{mn}^{\Lambda}(l) \tilde{\Gamma}_m^{\Lambda}(q,l) \tilde{\Gamma}_n^{\Lambda}(q',l) + \ca{R}^{\Lambda}_{\mtin{SC}}(q,q',l).
\end{align}
The interpretation of the $(m,n)$-term in the sum is that two Cooper pairs corresponding to fermionic bilinears
$
\int \mathrm{d}q \; \ovl{\psi}_+(q) \tilde{\Gamma}_m^{\Lambda}(q,l) \ovl{\psi}_-(l-q)
$
couple in the form of a boson exchange, where the boson has propagator $D_{mn}^{\Lambda}(l)$. Here, $m$ labels different gap symmetries (and possibly other details) of the pair. Note that in order to be a true boson propagator, $D^{\Lambda}(l)$ must satisfy positivity requirements, so that the Gaussian integral in a Hubbard-Stratonovich transformation in convergent. Because we stay within a fermionic formulation, we do not need to assume any positivity, so the ansatz is more general. In fact, not all terms come out positive in certain parameter regions (see below). For brevity, we shall still refer to them as boson propagators throughout.

In this paper, we parameterize the fermion-boson vertex function
\begin{align}
 \tilde{\Gamma}_m^{\Lambda}(q,l)=f_m(\sfrac{\mathbf{l}}{2}-\mathbf{q})\Gamma_m^{\Lambda}(q_0,l_0,\mathbf{l})
\end{align}
as a product of momentum-dependent form factors $f_n$ that are orthogonal on $[-\pi,\pi]^2$ and boson-fermion vertex functions $\Gamma_n^{\Lambda}(q_0,l_0,\mathbf{l})$ that depend on frequency and momentum. We choose a fixed orthonormal set $(f_n)_n$ independent of scale. This would be no loss of generality if one retained infinitely many terms in the sum over $m$ and $n$, but we shall keep only a small number of terms in this sum. This amounts to the assumption that the vertex function $\tilde{\Gamma}^{\Lambda}_n$ is regular.

Apart from scale dependent coefficients $D_{mn}^{\Lambda}(l)$ we also allow each $\Gamma_n^{\Lambda}$ to be scale dependent in order to capture part of the frequency dependence on $q_0$ and $q_0'$.
Due to particle-hole symmetry, i.e. the first line in Eq.~\eqref{eq:VertexSymmetries}, $\Gamma_n(q_0,l_0,\mathbf{l})=\Gamma_n(l_0-q_0,q_0,\mathbf{l})$ for singlet superconductivity, where $f_n(\mathbf{q})=f_n(-\mathbf{q})$. For simplicity, we expand in symmetric form factors only. A generalization to triplet superconductivity is straightforward since $D^{\Lambda}(l)$ consists of block matrices for singlet and triplet superconductivity.

The scale-dependent remainder function $\ca{R}_{\mtin{SC}}^{\Lambda}$ accounts for the fact that ${\Phi}_{\mtin{SC}}^{\Lambda}$ is not really a sum of finitely many boson exchange interactions. In the following we neglect this remainder. Ideally, the form factors $f_n$ can be chosen such that $\ca{R}_{\mtin{SC}}^{\Lambda}$ contains only subleading terms. In practice, we want to deal with only a few expansion terms. How well the effective vertex function of the Hubbard model can be approximated by a few boson exchange interactions is analyzed in detail elsewhere \cite{remainder}.

Differentiating Eq.~(\ref{eq:SCExpansion}) with respect to the scale and integrating over two form factors by using their orthogonality gives
\begin{align}\label{eq:freqvertexderivation}
 \int \frac{\mathrm{d}^2\mathbf{q}}{(2\pi)^2} \frac{\mathrm{d}^2\mathbf{q}'}{(2\pi)^2} &\; f_m(\sfrac{\mathbf{l}}{2}-\mathbf{q})  f_n(\sfrac{\mathbf{l}}{2}-\mathbf{q}')\dot{\Phi}^{\Lambda}_{\mtin{SC}}(q,q',l) = \dot{D}^{\Lambda}_{mn}(l) \Gamma^{\Lambda}_m(q_0,l_0,\mathbf{l}) \Gamma^{\Lambda}_n(q_0',l_0,\mathbf{l})  \\
\nonumber &\hspace{3cm} +D^{\Lambda}_{mn}(l) \Big[ \dot{\Gamma}^{\Lambda}_m (q_0,l_0,\mathbf{l})\Gamma^{\Lambda}_n(q_0',l_0,\mathbf{l}) +\Gamma^{\Lambda}_m(q_0,l_0,\mathbf{l})\dot{\Gamma}^{\Lambda}_n(q_0',l_0,\mathbf{l}) \Big].
\end{align}

We use the normalization $\Gamma_m^{\Lambda}(\sfrac{l_0}{2},l_0,\mathbf{l})=1$ such that
the boson propagators $D_{mn}^{\Lambda}$ are defined as in Ref. \onlinecite{decomposition}. For $T=0$ we set $q_0=q_0'=\sfrac{l_0}{2}$ in Eq.~(\ref{eq:freqvertexderivation}) and obtain the flow equation for the boson propagators in the superconducting channel
\begin{align}
 \dot{D}^{\Lambda}_{mn}(l)&=  \int\frac{\mathrm{d}^2\mathbf{q}}{(2\pi)^2} \frac{\mathrm{d}^2\mathbf{q}'}{(2\pi)^2} \; f_m(\sfrac{\mathbf{l}}{2}-\mathbf{q})  f_n(\sfrac{\mathbf{l}}{2}-\mathbf{q}')\dot{\Phi}_{\mtin{SC}}^{\Lambda}(q,q',l)_{\big|q_0=q_0'=\sfrac{l_0}{2}}.
\end{align}

By setting only $q_0'=\sfrac{l_0}{2}$ but $m=n$ in Eq.~(\ref{eq:freqvertexderivation}) and using the flow equation for $\dot{D}_{mn}^{\Lambda}$ we find the flow equation for the boson-fermion vertex function
\begin{align}\label{eq:FlowEqGamma}
 \dot{\Gamma}^{\Lambda}_{n}(q_0,l_0,\mathbf{l}) &= \frac{1}{D^{\Lambda}_{nn}(l)} \int \frac{\mathrm{d}^2\mathbf{q}}{(2\pi)^2} \frac{\mathrm{d}^2\mathbf{q}'}{(2\pi)^2} \; f_n(\sfrac{\mathbf{l}}{2}-\mathbf{q})  f_n(\sfrac{\mathbf{l}}{2}-\mathbf{q}')  \Big[\dot{\Phi}^{\Lambda}_{\mtin{SC}}(q,q',l)_{\big|q_0'=\sfrac{l_0}{2}} \\ \nonumber &\hspace{6cm}- \Gamma_{n}^{\Lambda}(q_0,l_0,\mathbf{l}) \dot{\Phi}^{\Lambda}_{\mtin{SC}}(q,q',l)_{\big|q_0=q_0'=\sfrac{l_0}{2}}\Big].
\end{align}
Inserting $q_0=\sfrac{l_0}{2}$ gives  $\dot{\Gamma}^{\Lambda}_n(\sfrac{l_0}{2},l_0,\mathbf{l})=0$, which ensures that the normalization remains unchanged in the flow.
For $T>0$ the frequency $\frac{l_0}{2}$ is not a fermion frequency in general. In this case we project to $q_0=\pm \pi T$ and symmetrize over the sign.
This corresponds, at $T = 0$, to a projection $q_0 = 0$ independent of $l_0$,
which yields results very similar to the projection $q_0 = l_0/2$.

Setting $m=n$ is necessary since $\Gamma_n$ does not depend on $m$ and Eq.~(\ref{eq:freqvertexderivation}) could otherwise not be satisfied.
Later on we restrict to only two form factors describing $s$- and $d$-wave superconductivity. In this and other cases the boson propagators are nearly diagonal \cite{decomposition}. We further use that the boson propagators 
have maxima in a neighborhood $\ca{U}$ around $0$ and $\hat{\pi}$ in momentum space to choose only two values $\mathbf{l}=0$ and $\mathbf{l}=\hat{\pi}$ for the dependence of $\Gamma_n$ on the transfer momentum $\mathbf{l}$. We further restrict the frequency dependence to the relative frequency inside a Cooper pair. That is, we define $\gamma^{\mtin{SC},\mathbf{a}}_n(q_0):= \Gamma_n^{\Lambda}(q_0,0,\mathbf{a})$ and write
\begin{align}
 \Gamma_n^{\Lambda}(q_0,l_0,\mathbf{l}) = \sum_{\mathbf{a}=0,\hat{\pi}} \gamma^{\mtin{SC},\mathbf{a}}_n(\sfrac{l_0}{2}-q_0)  \mathbbm{1}(l\in \ca{U}(\mathbf{a})),
\end{align}
where $\ca{U}(\mathbf{a})=\{\mathbf{p}\in (-\pi,\pi]^2: |p_x+p_y-a_x-a_y|\le\pi\}$
(periodically continued) are neighborhoods of $\mathbf{a}=0,\hat{\pi}$. The functions $\gamma^{\mtin{SC},\mathbf{a}}_n$ are symmetric due to time reversal symmetry (third line in Eq.~(\ref{eq:VertexSymmetries})) and normalized to $\gamma^{\mtin{SC},\mathbf{a}}_n(0)=1$. Although we suppress this in the notation, the $\gamma_n$ depend on the scale. Inserting into (\ref{eq:FlowEqGamma}) and evaluating at $l_0=0$ gives
\begin{align}\nonumber
 \dot{\gamma}^{\mtin{SC},\mathbf{a}}_n(q_0)&=\frac{1}{D_{nn}^{\Lambda}(a)} \int \frac{\mathrm{d}^2\mathbf{q}}{(2\pi)^2}  \frac{\mathrm{d}^2\mathbf{q}'}{(2\pi)} \; f_n(\sfrac{\mathbf{a}}{2}-\mathbf{q})  f_n(\sfrac{\mathbf{a}}{2}-\mathbf{q}') \\ & \hspace{3cm} \Big[\dot{\Phi}_{\mtin{SC}}^{\Lambda}(q,q',a)_{\big|q_0'=0} - \gamma_n^{\mtin{SC},\mathbf{a}}(q_0) \dot{\Phi}_{\mtin{SC}}^{\Lambda}(q,q',a)_{\big|q_0=q_0'=0}\Big]
\end{align}
with $a=(0,\mathbf{a})$.

In complete analogy we expand the magnetic and density channel in spin and density operators as well
\begin{align}\nonumber
 {\Phi}_{\mtin{M}}^{\Lambda}(q,q',l) &= \sum_{\mathbf{a}=0,\hat{\pi}} \mathbbm{1}\big(\mathbf{l}\in\ca{U}(\mathbf{a})\big)\sum_{m,n} M_{mn}^{\Lambda}(l+a) f_m(\sfrac{\mathbf{l}+\mathbf{a}}{2}+\mathbf{q}) f_n(\sfrac{\mathbf{l}+\mathbf{a}}{2}+\mathbf{q}') \\&\hspace{2cm}\times \gamma_m^{\mtin{M},\mathbf{a}}(q_0+\sfrac{l_0}{2}) \gamma_n^{\mtin{M},\mathbf{a}}(q_0'+\sfrac{l_0}{2}) + \ca{R}^{\Lambda}_{\mtin{M}}(q,q',l)\\ \nonumber
 {\Phi}_{\mtin{K}}^{\Lambda}(q,q',l) &= \sum_{\mathbf{a}=0,\hat{\pi}} \mathbbm{1}\big(\mathbf{l}\in\ca{U}(\mathbf{a})\big)\sum_{m,n} K_{mn}^{\Lambda}(l+a) f_m(\sfrac{\mathbf{l}+\mathbf{a}}{2}+\mathbf{q}) f_n(\sfrac{\mathbf{l}+\mathbf{a}}{2}+\mathbf{q}') \\&\hspace{2cm}\times \gamma_m^{\mtin{K},\mathbf{a}}(q_0+\sfrac{l_0}{2}) \gamma_n^{\mtin{M},\mathbf{a}}(q_0'+\sfrac{l_0}{2}) + \ca{R}^{\Lambda}_{\mtin{K}}(q,q',l),
\end{align}
where one may choose different form factors than in the superconducting channel. As $\ca{R}^{\Lambda}_{\mtin{SC}}$ we also neglect $\ca{R}^{\Lambda}_{\mtin{M}}$ and $\ca{R}^{\Lambda}_{\mtin{K}}$ in the following. For the boson propagators we obtain
\begin{align}
 \dot{M}^{\Lambda}_{mn}(l)&=  \int \frac{\mathrm{d}^2\mathbf{q}}{(2\pi)^2} \frac{\mathrm{d}^2\mathbf{q}'}{(2\pi)^2} \; f_m(\sfrac{\mathbf{l}}{2}+\mathbf{q})  f_n(\sfrac{\mathbf{l}}{2}+\mathbf{q}')\dot{\Phi}_{\mtin{M}}^{\Lambda}(q,q',l)_{\big|q_0=q_0'=-\sfrac{l_0}{2}}\\ \nonumber
\dot{K}^{\Lambda}_{mn}(l)&=  \int \frac{\mathrm{d}^2\mathbf{q}}{(2\pi)^2} \frac{\mathrm{d}^2\mathbf{q}'}{(2\pi)^2} \; f_m(\sfrac{\mathbf{l}}{2}+\mathbf{q})  f_n(\sfrac{\mathbf{l}}{2}+\mathbf{q}')\dot{\Phi}_{\mtin{K}}^{\Lambda}(q,q',l)_{\big|q_0=q_0'=-\sfrac{l_0}{2}}
\end{align}
and likewise for the frequency dependent part of the fermion-boson vertex
\begin{align}
 \dot{\gamma}^{\mtin{M},\mathbf{a}}_n(q_0) &= \frac{1}{M^{\Lambda}_{nn}(a)} \int \frac{\mathrm{d}^2\mathbf{q}}{(2\pi)^2} \frac{\mathrm{d}^2\mathbf{q}'}{(2\pi)} \; f_n(\mathbf{q})  f_n(\mathbf{q}')  \Big[\dot{\Phi}^{\Lambda}_{\mtin{M}}(q,q',a)_{\big|q_0'=0} - \gamma^{\mtin{M},\mathbf{a}}_n(q_0) \dot{\Phi}^{\Lambda}_{\mtin{M}}(q,q',a)_{\big|q_0=q_0'=0}\Big]\\ \nonumber
 \dot{\gamma}^{\mtin{K},\mathbf{a}}_n(q_0) &= \frac{1}{K^{\Lambda}_{nn}(a)} \int \frac{\mathrm{d}^2\mathbf{q}}{(2\pi)^2} \frac{\mathrm{d}^2\mathbf{q}'}{(2\pi)^2} \; f_n(\mathbf{q})  f_n(\mathbf{q}')  \Big[\dot{\Phi}^{\Lambda}_{\mtin{K}}(q,q',a)_{\big|q_0'=0} - \gamma^{\mtin{K},\mathbf{a}}_n(q_0) \dot{\Phi}^{\Lambda}_{\mtin{K}}(q,q',a)_{\big|q_0=q_0'=0}\Big]
\end{align}

\section{Detailed Set-up and Flow Equations \label{sec:flowequations}}

We choose a minimal set of form factors that describe the leading instabilities of the $(t,t')$-Hubbard model at Van Hove filling, and at not too large $t'$ for all fillings. In the superconducting channel $f_1(\mathbf{p})=1$ describes $s$-wave superconductivity with a peak at $\mathbf{l}=0$ and $s$-wave alternating pairing with a peak near $\bol=\hat{\pi}$. Both couplings are subleading but are essential at higher scales, and ultimately suppress the pseudo-critical scale to lower values. 
A second form factor $f_2(\mathbf{p})=\cos p_x-\cos p_y$ in the superconducting channel is used to describe $d$-wave superconductivity. In the two particle-hole channels we only use $f_1(\mathbf{p})=1$. Then a singularity of $M^{\Lambda}_{11}(l)$ at $l=0$ signals ferromagnetism and a singularity near $(0,\hat{\pi})$ (incommensurate) antiferromagnetism.

In principle $D_{12}^{\Lambda}(l)$ couples $s$- and $d$-wave superconductivity. The coupling, however, is small and exactly zero for $|l_x|=|l_y|$ at $l_0=0$, which includes the peaks at $0$ and $\pi$. We therefore neglect this coupling and consider only diagonal boson propagators.
In order to parameterize the momentum dependence of boson propagators $B=D,M$ or $K$ we distinguish their values close to $0$ and $\hat{\pi}$
\begin{align}
B_{nn}^{\Lambda}(l) = \sum_{\mathbf{a}=0,\hat{\pi}} B_{n}^{\mathbf{a}}(l-a) \mathbbm{1}\big(\bol\in \ca{U}(\mathbf{a}) \big),
\end{align}
where from now on we drop the explicit notation of the scale dependence.

The flow equations for the boson propagators, fermion-boson vertices, and the fermionic self-energy can now be derived by inserting the channel decomposition of the effective vertex function
\begin{align}\label{eq:vertexdecomposition}
 V_{\Lambda}(p_1,p_2,p_3) &= U- \Phi_{\mtin{SC}}^{\Lambda}(p_1,p_3,p_1+p_3)+\Phi_{\mtin{M}}^{\Lambda}(p_1,p_4,p_3-p_1)\\ &\qquad \nonumber +\frac 12 \Phi_{\mtin{M}}^{\Lambda}(p_1,p_3,p_2-p_3) -\frac 12 \Phi_{\mtin{K}}^{\Lambda}(p_1,p_3,p_2-p_3) 
\end{align}
in the equations of the last section. We use that 
the boson propagators are symmetric in their dependence on each momentum $l_x$ and $l_y$ separately. In the superconducting channel we arrive at
\begin{align}
\dot{D}_n^{\mathbf{a}}(l) &=  \int \mathrm{d} p \left[\frac{\mathrm{d}}{\mathrm{d}\Lambda} G(\mathbf{p},\sfrac{l_0}{2}-p_0) G(\mathbf{l}+\mathbf{a}-\mathbf{p},\sfrac{l_0}{2}+p_0)\right] f_n(\sfrac{\mathbf{l}+\mathbf{a}}{2}-\mathbf{p})^2 \\ \nonumber &\hspace{6cm}\Big[ D_n^{\mathbf{a}}(l) \gamma_n^{\mtin{SC},\mathbf{a}}(p_0)  - U\delta_{n=1} -\tilde{\alpha}_n^{\mtin{SC}}(p_0,l_0)\Big]^2
\end{align}
with the following contributions to the frequency dependence from box and vertex diagrams
\begin{align}
 \tilde{\alpha}_n^{\mtin{SC}}(p_0,l_0)&= \frac 12 \int_{\ca{U}(0)} \frac{\mathrm{d}^2 \mathbf{k}}{(2\pi)^2}  A_n(\mathbf{k}) \Big[ 3M^0_1(\mathbf{k},p_0) \gamma_{\mtin{M}}^{0}(\sfrac{l_0-p_0}{2}) \gamma_{\mtin{M}}^{0}(\sfrac{l_0+p_0}{2}) \\ \nonumber &\hspace{5cm} + (-1)^{n-1} 3M^{\hat{\pi}}_1(\mathbf{k},p_0) \gamma_{\mtin{M}}^{\hat{\pi}}(\sfrac{l_0-p_0}{2}) \gamma_{\mtin{M}}^{\hat{\pi}}(\sfrac{l_0+p_0}{2})
\\ \nonumber &\hspace{5cm} - K^{0}_1(\mathbf{k},p_0) \gamma_{\mtin{K}}^{0}(\sfrac{l_0-p_0}{2}) \gamma_{\mtin{K}}^{0}(\sfrac{l_0+p_0}{2})
\\ \nonumber &\hspace{5cm} - (-1)^{n-1} 3K^{\hat{\pi}}_1(\mathbf{k},p_0) \gamma_{\mtin{K}}^{\hat{\pi}}(\sfrac{l_0-p_0}{2}) \gamma_{\mtin{K}}^{\hat{\pi}}(\sfrac{l_0+p_0}{2}) \Big]
\end{align}
with $A_1(\mathbf{k})=1$ and $A_2(\mathbf{k})=\sfrac{\cos k_x +\cos k_y}{2}$ for $s$- and $d$-wave superconductivity. Notice that the projected momentum dependence is the same for direct, box, and vertex diagrams and is just given by a form factor. The function $A_2$ is obtained by using simple trigonometric identities and the symmetric momentum dependence of the boson propagators mentioned above.

The flow of the frequency dependent part of the boson-fermion vertex function is given by
\begin{align}\nonumber
\dot{\gamma}^{\mtin{SC},\mathbf{a}}_n(q_0) &= \frac{1}{D_n^{\mathbf{a}}(0)}\int \mathrm{d} p \left[\frac{\mathrm{d}}{\mathrm{d}\Lambda} G(p) G(a-p)\right] f_n(\sfrac{\mathbf{a}}{2}-\mathbf{p})^2 \Big[ D_n^{\mathbf{a}}(0)\gamma_n^{\mtin{SC},\mathbf{a}}(p_0) - U\delta_{n=1} -\tilde{\alpha}_n^{\mtin{SC}}(p_0,0)\Big] \\
&\hspace{3cm} \times \Big[ U\delta_{n=1} \big(\gamma_n^{\mtin{SC},\mathbf{a}}(q_0) -1) +\gamma_n^{\mtin{SC},\mathbf{a}}(q_0) \tilde{\alpha}_n^{\mtin{SC}}(p_0,0) - \tilde{A}_n^{\mtin{SC}}(p_0,q_0)\Big]  \label{eq:flowgammaSC}
\end{align}
with frequency dependent (unprojected) box and vertex contributions
\begin{align}
 \tilde{A}_n^{\mtin{SC}}(p_0,q_0)&= \frac 12 \int_{\ca{U}(0)} \frac{\mathrm{d}^2 \mathbf{k}}{(2\pi)^2}  A_n(\mathbf{k}) \Big[ 3M^0_1(\mathbf{k},p_0-q_0) \gamma_{\mtin{M}}^{0}(\sfrac{p_0+q_0}{2})^2  \\ \nonumber &\hspace{5cm} + (-1)^{n-1} 3M^{\hat{\pi}}_1(\mathbf{k},p_0-q_0)  \gamma_{\mtin{M}}^{\hat{\pi}}(\sfrac{p_0-q_0}{2})^2
\\ \nonumber &\hspace{1cm} - K^{0}_1(\mathbf{k},p_0-q_0) \gamma_{\mtin{K}}^{0}(\sfrac{p_0+q_0}{2})^2
 - (-1)^{n-1} 3K^{\hat{\pi}}_1(\mathbf{k},p_0-q_0) \gamma^{\hat{\pi}}_{\mtin{K}}(\sfrac{p_0+q_0}{2})  \Big].
\end{align}
Notice that $\tilde{A}_n^{\mtin{SC}}(p_0,0)=\tilde{\alpha}_n^{\mtin{SC}}(p_0,0)$ represents the frequency projection.

Similar flow equations are obtained for the magnetic propagator
\begin{align}
 \dot{M}^{\mathbf{a}}_1(l)&=-\int \mathrm{d} p \left[\frac{\mathrm{d}}{\mathrm{d}\Lambda} G(\mathbf{p},p_0-\sfrac{l_0}{2}) G(\mathbf{p}+\mathbf{l}+\mathbf{a},p_0+\sfrac{l_0}{2})\right] \\ & \hspace{3cm} \times\Big[ U+M^{\mathbf{a}}_1(l)\gamma^{\mathbf{a}}_{\mtin{M}}(p_0) +\alpha^{\mtin{M}}_1(p_0,l_0) +\alpha^{\mtin{M}}_2(p,l+a)\Big]^2 \nonumber
\end{align}
with
\begin{align}\nonumber
 \alpha^{\mtin{M}}_1(p_0,l_0) &= \frac 12 \sum_{\mathbf{a}=0,\hat{\pi}} \int \frac {\mathrm{d}^2\mathbf{k}}{(2\pi)^2} \left[ \gamma^{\mathbf{a}}_{\mtin{M}}(\sfrac{p_0+l_0}{2}) \gamma^{\mathbf{a}}_{\mtin{M}}(\sfrac{p_0-l_0}{2}) M^{\mathbf{a}}_1(\mathbf{k},p_0) \right. - \gamma^{\mathbf{a}}_{\mtin{K}}(\sfrac{p_0+l_0}{2}) \gamma^{\mathbf{a}}_{\mtin{K}}(\sfrac{p_0-l_0}{2}) K^{\mathbf{a}}_1(\mathbf{k},p_0)  \\ &\hspace{3cm} -2 \left.\gamma^{\mathbf{a}}_{\mtin{SC},1}(\sfrac{p_0+l_0}{2}) \gamma^{\mathbf{a}}_{\mtin{SC,1}}(\sfrac{p_0-l_0}{2}) D_1^{\mathbf{a}}(\mathbf{k},p_0) \right] \\ \nonumber
 \alpha^{\mtin{M}}_2(p,l) &= - \sum_{\mathbf{a}=0,\hat{\pi}} \int \frac {\mathrm{d}^2\mathbf{k}}{(2\pi)^2} \; \gamma^{\mathbf{a}}_{\mtin{SC,2}}(\sfrac{p_0+l_0}{2}) \gamma^{\mathbf{a}}_{\mtin{SC,2}}(\sfrac{p_0-l_0}{2}) D^{\mathbf{a}}_2(\mathbf{k},p_0) f_2(\mathbf{p}-\sfrac{\mathbf{k}}{2}) f_2(\mathbf{p}+\mathbf{l}-\sfrac{\mathbf{k}}{2}).
\end{align}
 Using trigonometric identities the momentum dependence of $\alpha^{\mtin{M}}_2(p,l)$ can be made explicit.
The flow equation of fermion-boson vertex function reads
\begin{align}
 \dot{\gamma}^{\mathbf{a}}_{\mtin{M}}(q_0) &= \frac{1}{M^{\mathbf{a}}_1(0)} \int \mathrm{d} p \left[\frac{\mathrm{d}}{\mathrm{d}\Lambda} G(p) G(p+a)\right] \Big[ U+M^{\mathbf{a}}_1(0)\gamma^{\mathbf{a}}_{\mtin{M}}(p_0) +\alpha^{\mtin{M}}_1(p_0,0) +\alpha^{\mtin{M}}_2(p,a)\Big]\\ \nonumber
&\hspace{0.7cm} \times\left[ U\Big(1-\gamma_{\mtin{M}}^{\mathbf{a}}(q_0)\Big) +A^{\mtin{M}}_1(p_0,q_0) +A^{\mtin{M}}_2(p,a,q_0) -\gamma_M^{\mathbf{a}}(q_0) \Big(\alpha^{\mtin{M}}_1(p_0,0)+\alpha^{\mtin{M}}_2(p,a)\Big) \right]
\end{align}
with
\begin{align}\nonumber
 A^{\mtin{M}}_1(p_0,q_0) &= \frac 12 \sum_{\mathbf{a}=0,\hat{\pi}} \int \frac {\mathrm{d}^2\mathbf{k}}{(2\pi)^2} \left[ \gamma^{\mathbf{a}}_{\mtin{M}}(\sfrac{p_0+q_0}{2})^2 M^{\mathbf{a}}_1(\mathbf{k},p_0-q_0) \right. - \gamma^{\mathbf{a}}_{\mtin{K}}(\sfrac{p_0+l_0}{2})^2  K^{\mathbf{a}}_1(\mathbf{k},p_0-q_0)  \\ &\hspace{3cm} -2 \left.\gamma^{\mathbf{a}}_{\mtin{SC},1}(\sfrac{p_0-l_0}{2})^2  D_1^{\mathbf{a}}(\mathbf{k},p_0+q_0) \right] \\ \nonumber
 A^{\mtin{M}}_2(p,a,q_0) &= - \sum_{a=0,\hat{\pi}} \int \frac {\mathrm{d}^2\mathbf{k}}{(2\pi)^2} \; \gamma^{\mathbf{a}}_{\mtin{SC,2}}(\sfrac{p_0-q_0}{2})^2  D^{\mathbf{a}}_2(\mathbf{k},p_0+q_0) f_2(\mathbf{p}-\sfrac{\mathbf{k}}{2}) f_2(\mathbf{p}+\mathbf{a}-\sfrac{\mathbf{k}}{2}).
\end{align}
Similar flow equations are obtained in the density channel
\begin{align}\label{eq:KflowEq}
 \dot{K}_1^{\mathbf{a}}(l)&=-\int \mathrm{d} p \left[\frac{\mathrm{d}}{\mathrm{d}\Lambda} G(\mathbf{p},p_0-\sfrac{l_0}{2}) G(\mathbf{p}+\mathbf{l}+\mathbf{a},p_0+\sfrac{l_0}{2})\right] \\ \nonumber & \hspace{3cm} \times\Big[ K^{\mathbf{a}}_1(l)\gamma^{\mathbf{a}}_{\mtin{M}}(p_0) -U  -\alpha^{\mtin{K}}_1(p_0,l_0) -\alpha^{\mtin{M}}_2(p,l+a)\Big]^2
\end{align}
with
\begin{align}\nonumber
 \alpha^{\mtin{K}}_1(p_0,l_0) &= \frac 12 \sum_{\mathbf{a}=0,\hat{\pi}} \int \frac {\mathrm{d}^2\mathbf{k}}{(2\pi)^2} \left[ 3\gamma^{\mathbf{a}}_{\mtin{M}}(\sfrac{p_0+l_0}{2}) \gamma^{\mathbf{a}}_{\mtin{M}}(\sfrac{p_0-l_0}{2}) M^{\mathbf{a}}_1(\mathbf{k},p_0) \right. + \gamma^{\mathbf{a}}_{\mtin{K}}(\sfrac{p_0+l_0}{2}) \gamma^{\mathbf{a}}_{\mtin{K}}(\sfrac{p_0-l_0}{2}) K^{\mathbf{a}}(\mathbf{k},p_0)  \\ &\hspace{3cm} -2 \left.\gamma^{\mathbf{a}}_{\mtin{SC},1}(\sfrac{p_0+l_0}{2}) \gamma^{\mathbf{a}}_{\mtin{SC,1}}(\sfrac{p_0-l_0}{2}) D_1^{\mathbf{a}}(\mathbf{k},p_0) \right].
\end{align}
The boson-fermion flow is given by
\begin{align}
 \dot{\gamma}^{\mathbf{a}}_{\mtin{K}}(q_0) &= -\frac{1}{K^{\mathbf{a}}_1(0)} \int \mathrm{d} p \left[\frac{\mathrm{d}}{\mathrm{d}\Lambda} G(p) G(p+a)\right] \Big[ K^{\mathbf{a}}_1(0)\gamma^{\mathbf{a}}_{\mtin{K}}(p_0)-U -\alpha^{\mtin{K}}_1(p_0,0) -\alpha^{\mtin{M}}_2(p,a)\Big]\\ \nonumber
&\hspace{0.7cm} \times\left[ U\Big(1-\gamma_{\mtin{K}}^{\mathbf{a}}(q_0)\Big) +A_1^{\mtin{K}}(p_0,q_0) +A_2^{\mtin{M}}(p,a,q_0) -\gamma_K^{\mathbf{a}}(q_0) \Big(\alpha^K_1(p_0,0)+\alpha^M_2(p,a)\Big) \right]
\end{align}
with
\begin{align}\nonumber
 A^{\mtin{K}}_1(p_0,q_0) &= \frac 12 \sum_{\mathbf{a}=0,\hat{\pi}} \int \frac {\mathrm{d}^2\mathbf{k}}{(2\pi)^2} \left[ 3\gamma^{\mathbf{a}}_{\mtin{M}}(\sfrac{p_0+q_0}{2})^2 M^{\mathbf{a}}_1(\mathbf{k},p_0-q_0) \right. + \gamma^{\mathbf{a}}_{\mtin{K}}(\sfrac{p_0+l_0}{2})^2  K^{\mathbf{a}}_1(\mathbf{k},p_0-q_0)  \\ &\hspace{2cm} -2 \left.\gamma^{\mathbf{a}}_{\mtin{SC},1}(\sfrac{p_0-l_0}{2})^2  D_1^{\mathbf{a}}(\mathbf{k},p_0+q_0) \right].
\end{align}

Suppose that $\chi_{\Lambda}\neq 0$ for all finite $\Lambda$, that is, arbitrarily large scales have to be taken into account. In principle, the initial condition could be stated for $\Lambda\to \infty$ as the bare action coming from the Hubbard Hamiltonian. We find it more convenient to choose a relatively large scale $\Lambda_0\gg U$ and treat all scales $\Lambda>\Lambda_0$ by perturbation theory. This is possible because $\Lambda_0$ is an infrared cutoff, such that the small parameter is effectively given by $\sfrac{U}{\Lambda_0}$. The effective vertex in second order perturbation theory reads
\begin{align}
 V_{\Lambda_0}(p_1,p_2,p_3) = U - U^2 \Phi^{\Lambda_0}_-(p_1+p_2) -U^2 \Phi^{\Lambda_0}_+(p_3-p_1) + \ca{O}(\sfrac{U}{\Lambda_0})^3
\end{align}
with contributions from the particle-particle and the particle-hole channel. We read off the initial condition for $s$-wave superconducting boson propagator
\begin{align}
 D_{1,0}^{\mathbf{a}}(l)=\Phi^{\Lambda_0}_-(l+a)
\end{align}
and choose for the magnetic and density boson propagators
\begin{align}
 M_{1,0}^{\mathbf{a}}(l) = K_{1,0}^{\mathbf{a}}(l)=\Phi^{\Lambda_0}_+(l+a) \, .
\end{align}
Since we treat both particle-hole channels in the same approximation, a different assignment of the initial condition among the magnetic and density channel would not influence the outcome of the flow. In all these $s$-wave channels the initial boson-fermion vertex function is equal to one,
\begin{align}
\gamma^{\mtin{SC},\mathbf{a}}_1(q_0)=\gamma^{\mathbf{a}}_{\mtin{M}}(q_0)=\gamma^{\mathbf{a}}_{\mtin{K}}(q_0)=1
\end{align}
for $\mathbf{a}=0$ or $\hat{\pi}$.

Initially, there is no coupling for $d$-wave superconductivity, so $D_{2,0}^{\mathbf{a}}(l)=0$, where for the remainder of this section we only consider $\mathbf{a}=0$ in the $d$-wave channel and suppress the explicit notation. Therefore the fermion-boson vertex function is undefined in this channel. In order to circumvent $\sfrac{1}{0}$ in Eq.~(\ref{eq:flowgammaSC}) we propose two solutions
\begin{enumerate}
 \item We arbitrarily set $\gamma_2^{\mtin{SC}}(q_0)=1$ and compute the flow with constant $\gamma_2^{\mtin{SC}}$ from $\Lambda_0$ down until $D_2(0)$ reaches a minimal value $D_{\mtin{min}}$. This defines a scale $\Lambda_1$, where we switch on the flow of $\gamma_2^{\mtin{SC}}$.
The flow rapidly adjusts from a constant $\gamma_2^{\mtin{SC}}(q_0)=1$ to a decaying function. We checked that the result does not depend on reasonably chosen minimal values $D_{\mtin{min}}\in [10^{-6},10^{-3}]$.

\item In order for $\lim_{D_2(0)\to 0} \dot{\gamma}^{\mtin{SC}}_2(q_0)$ to exist it is necessary that
\begin{align}
 \lim_{D_2(0)\to 0}  D_2(0) \dot{\gamma}^{\mtin{SC}}_2(q_0)  \overset{!}{=}0 .
\end{align}
This condition is satisfied if
\begin{align}
 \gamma^{\mtin{SC}}_2(q_0) = \frac{\int \mathrm{d} p \left[\frac{\mathrm{d}}{\mathrm{d}\Lambda} G(p) G(-p)\right] f_2(p)^2\tilde{\alpha}_2^{\mtin{SC}}(p_0,0) \tilde{A}_2^{\mtin{SC}}(p_0,q_0)}{\int \mathrm{d} p \left[\frac{\mathrm{d}}{\mathrm{d}\Lambda} G(p) G(-p)\right] f_2(p)^2\tilde{\alpha}_2^{\mtin{SC}}(p_0,0)^2} .
\end{align}
Taking this as the initial condition
ensures that $\gamma^{\mtin{SC}}_2(q_0)$ has the same frequency decay as the vertex and box diagrams (but normalized) already at the beginning of the flow. This in principle allows to start the full flow at $\Lambda_0$ with $D_2(l)=0$ and $\gamma_2^{\mtin{SC}}(q_0)$ as above.
\end{enumerate}

For the time being we have applied the first variant for simplicity. Since the effect of non-constant fermion-boson vertex functions turns out to be not very large we resign from implementing the second variant, which is fundamentally more sound.

The flow equations and initial conditions can be evaluated for different regulator functions $\chi_{\Lambda}(p)$. Physically, the result should not depend on the choice of regularization. However, because of the level-2 truncation and the requirement to stop the flow once singularities appear there is a cutoff dependence of quantitative results. In order to check our results qualitatively we use two very different regularization schemes. At temperature zero we use a soft frequency regularization. Here the RG scale $\Lambda$ is denoted by $\Omega$, and
\begin{align}
 \chi_{\Omega}(p)=\frac{p_0^2}{p_0^2+\Omega^2} \, .
\end{align}
In a second scheme we use temperature $\Lambda=T>0$ as a regularization as introduced in Ref. \onlinecite{HonerkampSalmhofer_Tflow}.
Here the Grassmann fields $\Psi$ are rescaled such that the quartic part of the microscopic action does not explicitly depend on $T$.
This results in a regularized free propagator that depends on $T$ through the discrete frequency variables
and an overall factor $\sqrt{T}$.

To complete the set-up of our study, we discuss our determination of the 'transition' scale $\Omega_*$ in the $\Omega$-scheme, and the 'transition' temperature $T_*$ in the $T$-scheme. As in previous studies, we never run a flow up to, or even very close to the point $\Omega_{\mtin{C}}$ (resp. $T_{\mtin{C}}$) where any part of the vertex function is divergent, because the truncation we use breaks down before that point. Instead we define a stopping condition, namely that the largest value of one of the exchange propagators reaches $V_{\max}=20t$, 
which corresponds to about 2.5 times the free bandwidth. 
While the choice of $V_{\max}$ is arbitrary to a certain extent, we have verified that varying $V_{\max}>16t$ does not change our results qualitatively.

%%%%%%%%%%%%%%%%%%%%%%%%%%%%%%%%%%%%%%%%%%%%%%%%%%%%%%%%%%%%%
\section{Transfer Frequency Dependence\label{sec:FreqParam}}

\subsection{Lorentzians as Vertex Functions\label{sec:Lorentz}}
In this section we motivate an explicit frequency parameterization of the boson propagators. The goal is to capture the leading frequency behavior for small scales near the critical scale.

Consider the case where one channel is clearly dominant, for example, a curved and regular Fermi surface, for which Umklapp scattering can be neglected. There, the onset of superconductivity determines the stopping scale. Neglecting the marginal particle-hole graphs corresponds to neglecting vertex and box diagrams in the flow equation for the superconducting boson propagator, which can then be solved to give the RPA result
\begin{align}\label{eq:SCRPA}
D_{nn}(l) = \frac{D_{nn}^0(l)}{1-D_{nn}^0(l) \Phi_{-}^{nn}(l)} \, ,
\end{align}
where $\Phi_{-}^{nn}(l)=\int \mathrm{d}p\; G(p) G(l-p) f_n(\sfrac{\mathbf{l}}{2}-\mathbf{p})^2$ is the particle-particle bubble
and $D^0_{nn}(l)$ is the  initial interaction. For notational simplicity we restrict to only one form factor. The following analysis generalizes to more form factors since the boson propagators are approximately diagonal.

Suppose that $D^0_{nn}(l)=u$ is a positive constant, that is, an attractive
initial interaction.
For momentum $\mathbf{l}=0$, small scales $\Omega$, and small frequencies $l_0$ the particle-particle bubble
at Van Hove filling is roughly given by
\begin{align*}
\rRe \Phi_{-}^{nn}(l_0, \bzero) \sim  + C \ln^2 \big( \Omega^2 + \tilde Z l_0^2 \big) \, ,
\end{align*}
where $C$ and $\tilde Z$ are positive constants and we have neglected imaginary parts of the bubble. 
 That is, the bubble diverges for $\Omega=0$ as $\ln^2 |l_0|$.
The leading singular structure of $D(l)$, however, does not
 contain a logarithmic frequency term.
Instead a singularity at a scale $\Omega_{\mtin{C}}$ builds up.
It is characterized by a mass term $m_\Om^2=1 - uC \ln^2\Om^2$ that vanishes
at this critical scale $\Omega_{\mtin{C}}$.
Just above the critical scale the denominator of Eq.~(\ref{eq:SCRPA}) at $\mathbf{l}=0$ can be written as
\begin{align*}
   1 - uC \ln^2 (\Om^2 + \tilde Z l_0^2)
&=
   m_\Om^2 - uC\ln \left( 1 + \tilde Z\  \frac{l_0^2}{\Om^2}  \right)
      \left( \ln (\Om^2 + \tilde Z l_0^2) + \ln \Om^2 \right)
\nonumber\\
&=
   m_\Om^2 - 4uC\tilde Z \ln\Om\cdot \left( \frac{l_0}{\Om} \right)^2
   + \cO\big( ( l_0 / \Om_c )^4 \big),
\qquad
   \Om \ge \Om_c > 0
.
\end{align*}
A similar argument holds for the particle-hole channels, where the particle-hole bubble diverges with $\ln \Omega$ for $l=0$ at van Hove filling. Due to the $\Omega$-regularization there is no damping term proportional to $\frac{l_0}{|\mathbf{l}|}$ for small $l_0$, $|\mathbf{l}|$, and $\frac{l_0}{|\mathbf{l}|}$, at $\Omega>\Omega_{\mtin{C}}$.

The imaginary part of the particle-particle bubble vanishes at zero frequency.
However, it exhibits singular behavior in its first frequency derivative if the density of states is not symmetric. 
At Van Hove filling and for curved Fermi surfaces, $0 < t' < t/2$,
\begin{equation}\label{eq:ppBubble_om-diff}
  (\partial_{l_0}\rIm \Phi_-^{11})(l_0 = 0, \bzero) = -C_1 \ln^2\Om + \cO(\ln\Om),
  \quad C_1 > 0
.
\end{equation}

This kind of small frequency behavior can be captured by parameterization of
the frequency dependence of the boson propagators in each channel with ``Lorentzians''
\begin{align}
   B_{mm}(l_0, \bol) = \big[ m^2_\Om(\bol) + ia_\Om(\bol) l_0 + b^2_\Om(\bol) l_0^2\big]^{-1}
\end{align}
with momentum dependent parameters $m_\Om, a_\Om, b_\Om \in \RR$. 
A flow equation for these parameters is obtained by considering
the RG equation for $B_{mm}(l_0, \bol)$ and its first two frequency derivatives
at zero frequency $l_0 = 0$.
This corresponds to a second order derivative expansion with multiple bosonic fields.
Parameters are thus determined from the small frequency behavior only.
In addition to small frequencies also the $l_0^{-2}$ decay of $s$-wave channels for large frequencies $l_0$ is described correctly but with a wrong prefactor. The $d$-wave superconducting boson propagator is solely generated by other channels that decay in frequency themselves. It therefore has an  $l_0^{-4}$ decay at finite $\Omega>0$.

In RPA, the flow equations are decoupled such that the leading instability is determined by
the evaluation of direct graphs in the channel decomposition only. %
Here only the singular value of boson propagators at frequency zero determines the instability.
In contrast, the fermionic RG couples different channels via box and vertex diagrams. The value of these diagrams is  computed by a convolution in loop and vertex frequencies. Therefore, the coupling of different channels may be estimated wrongly when intermediate to large frequencies are described inadequately. In the next section we compare Lorentz curves to the full frequency dependence of boson propagators calculated in the RG flow.

\subsection{Comparison to Actual Dependence}

The Lorentz parameterization of the transfer frequency has been motivated above
for the small frequency regime. There is no obvious reason why it should extend to an accurate description of exchange propagators for arbitrary frequency values.
This is why, in a second calculation, we discretize the transfer frequency dependence and then trace it fully during the flow. This allows us to check the Lorentz ansatz.

 \begin{figure}
  \input{FreqPap_fitsBfreq_withfreq_prec4_moderate_t20_2}
  \caption{(Color online) Frequency dependence of the magnetic propagator $l_0 \mapsto M_{11}(l_0, \bpi)$
for parameter values $U = 3t, t' = 0.2t$ at the stopping scale $\Omega_* = 0.089\ t$.
Plus marks indicate discrete frequency data from the RG calculation.
A simple Lorentz parameterization from the small frequency expansion (dashed line) accurately describes
the small frequency behavior. 
The least squares fit to the sum of \emph{two} Lorentz distributions (solid line)
also captures well the frequency dependence away from the small frequency region, it is composed of the two dotted curves.}
\label{fig:fits_Bfreq}
\end{figure}
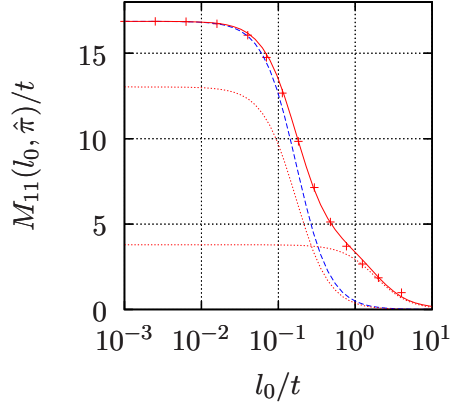

In anticipation of the full discussion of the RG flow with frequency dependent vertices
we show in Fig.~\ref{fig:fits_Bfreq} several approximations to the discretized frequency data
for a typical example, the antiferromagnetic exchange propagator at the stopping scale $\Omega_*$.
By construction, a single Lorentz curve determined from 
numerical extraction of Taylor coefficients
reproduces the discrete reference data well for small frequencies (dashed line in Fig.~\ref{fig:fits_Bfreq}).
However, this is not true for frequencies $l_0 \gtrsim \Omega_*$.
At $l_0 \approx 2\Omega_*$ the relative error is about $15\%$ and it becomes very large at higher frequencies.
We find a similar situation in the temperature flow setup, this scheme is shortly discussed
in Sec.~\ref{sec:Tflow}.
Fitting a single Lorentz curve to the discrete data with a least square condition
brings no improvement. 

Compared to other RG schemes,
in the 1PI scheme contributions to the flow at scale $\Lambda$ are not restricted to
low frequency or low energy processes with $\abs{e}, \abs{\om} \le \Lambda$.
Moreover, the $\Om$-scheme provides just a mild regularization of the free propagator 
such that its scale derivative as well as the corresponding single scale propagator 
do not have compact support in frequency space.
Hence, proper frequency parameterization away from the small frequency region possibly is important.

Such a parameterization can be accomplished by a sum of two Lorentz curves (solid line in Fig.~\ref{fig:fits_Bfreq}).
One of these curves can be interpreted as a small frequency process and the other one as a large frequency process.
Parameters in this ansatz
have been determined by a weighted least-squares fit of the discrete data,
with a ten times larger weighting factor for frequencies with $\abs{\om} < \Om$.

In Sec. \ref{sec:results} we compute the flow with a single Lorentz curve and compare to the flow with discretized frequencies.
Surprisingly, the simple Lorentz curve ansatz captures well the flow
of the leading couplings of the interaction vertex in most parameter regions.
However, it does not detect a new type of scattering singularity, detailed below,
that we encounter in the setup with frequency discretization.
Also it overestimates the flow of the $Z$ factor at momentum $(0, \pi)$ for all scales $\Om$ and produces
scale derivatives $\dot Z$ with a relative error of up to 40\%, see section \ref{sec:ImSigma}.

On the other hand,
we find that quantitatively accurate results can be obtained with the frequency parameterization
by a sum of two Lorentz curves.
Although its effectiveness for the extraction of information about arbitrary observables is not tested,
this parameterization reproduces very well 
the flow of the interaction vertex
as well as the flow of the $Z$ factor. 

This functional form may turn out to be a promising generalization of bosonic field theories
for future applications in condensed matter theory.
A simple strategy for determing its parameters,
independent of a prior frequency discretization procedure,
needs yet to be established.

\section{Numerical implementation\label{sec:Numerical}}

\subsection{Transfer Momentum Parameterization}

\begin{figure}
  \centering
  \scalebox{0.75}{
       \includegraphics{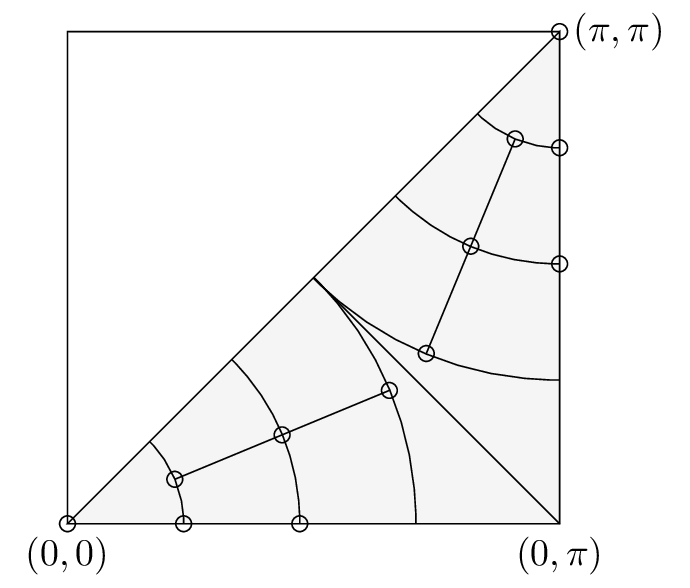}
  }
  \caption{Illustration of momentum discretization for exchange propagators in 12 segments
about $\bok = (0, 0)$ and $\bok = (\pi, \pi)$. Small circles mark the representative momentum
associated with each segment. In calculations we use an analogous discretization in 32 segments.
}
\label{fig:sector-discretisation}
\end{figure}

The momentum dependence of boson propagators is discretized using step functions.
By reflection symmetries about the coordinate axes and the Brillouin zone diagonal one can restrict
to discretization of one eighth of the Brillouin zone, see Fig.~\ref{fig:sector-discretisation}.
The discretization in radial segments about $\bok = (0, 0)$ and $(\pi, \pi)$
permits a detailed momentum resolution in the vicinity of these points.
In case of incommensurate antiferromagnetism the maximum of the magnetic boson propagator can move far away from $(\pi,\pi)$ in the $(1,0)$-direction. Thus we place several representative momenta
along the coordinate axes.
In the calculations below we use a discretization in 32 segments per one eighth of the Brillouin zone
for each exchange propagator.

\subsection{Frequency Discretization}

Within each momentum segment, the frequency dependence of exchange propagators is discretized.
By symmetry, consideration of transfer frequencies $l_0 \ge 0$ is sufficient.
We use a logarithmic grid in the frequency range $10^{-3} \le l_0/t \le 3 \cdot 10^2$
and include the frequency value $l_0 = 0$.

We have checked that our numerical results are not affected by the concrete implementation
of frequency dependences. For this purpose we have
(a) used a frequency grid of 20 points, which still allows application of adaptive numerical integrators to the
loop frequency integral on the rhs of flow equations,
and
(b) used a frequency grid of higher resolution, with evaluation of frequency integrals
as a discrete sum.

The fermion-boson vertex functions $\gamma(q_0)$ are symmetric around $q_0=0$ as well.  
We choose a frequency grid of 34 base points for positive frequencies. 
For the evaluation of the flow equations the boson-fermion vertex functions have to be evaluated at $q_0$, $\frac{q_0-p_0}{2}$, and $\frac{q_0+p_0}{2}$, where $p_0$ is the loop frequency and $q_0$ the external frequency. The latter two values are obtained by linear interpolation.

\subsection{Loop Integrals}\label{sec:loop_integrals}

The evaluation of loop frequency integrals in 1PI flow equations by contour techniques is still possible
in the presence of frequency dependent vertex functions
but becomes more involved.
In calculations where we use a momentum independent cutoff and neglect the self-energy the evaluation of loop integrals can be simplified by first integrating over the loop momentum and then over the loop frequency. The crucial point is here that the momentum integrals are independent of the RG scale and have to be evaluated only once.
More precisely, the free propagator with $\Om$ cutoff $C_\Om(p) = \chi_\Om(p_0)\ ( ip_0 - \eps_\bop)^{-1}$
allows the factorization
\begin{equation}
  \dero{\Om} \Big( C_\Om(p)C_\Om(k) \Big)
=
   C_0(p)C_0(k)\ \dero{\Om} \Big( \chi_\Om(p_0)\chi_\Om(k_0) \Big)
\end{equation}
in a momentum independent and a scale independent term.

The flow equations for exchange propagators have the general form
\begin{align}\label{eq:floweq-loopcalc}
  \dot B_{mm}(l)
&=
  \int_\RR\frac{\intd p_0}{2\pi}\  \dero{\Om} \Big( \chi_\Om(p_0-\sfrac{l_0}{2})\chi_\Om(p_0+\sfrac{l_0}{2}) \Big)
   \sum_{j_1, j_2} \alpha_{j_1}^\Om(l_0, p_0) \alpha_{j_2}^\Om(l_0, p_0)\
   I^\pm_{\psi_{j_1}, \psi_{j_2}}(l, p_0),
\nonumber\\
  I^\pm_{\psi_{j_1}, \psi_{j_2}}(l, p_0)
&=
  \mp\int\frac{\intd\bop}{(2\pi)^2}\ C_0( \pm(p-\sfrac{l}{2})) C_0(p+\sfrac{l}{2})\ \psi_{j_1}(\bol, \bop) \psi_{j_2}(\bol, \bop)
.
\end{align}
Here the dependence of the projected interaction vertex on loop and external momenta is extracted analytically using trigonometric identities \cite{decomposition}.
This produces two sums of frequency dependent functions $\alpha_j^\Om$
and momentum dependent functions $\psi_j$ in \eqref{eq:floweq-loopcalc},
where $\psi_j$ is scale independent.
Therefore, the momentum integrals $I^\pm_{\psi_{j_1}, \psi_{j_2}}(l, p_0)$ are scale independent
and can be calculated before starting the flow.
This needs to be done for all discrete frequency-momenta $l = (l_0, \bol)$ where exchange propagators
are calculated and in addition for all combinations $(\pm, \psi_{j_1}, \psi_{j_2})$ that occur.

We discretize the momentum integral expression in the frequency variable $p_0$
as follows:
By symmetry, only $p_0, l_0 \ge 0$ is needed.
$I^\pm_{\psi_{j_1}, \psi_{j_2}}(l, p_0)$ is regular in $p_0$
except for possible singular behavior at $p_0 = l_0/2$,
which drives the flow.
The asymptotics of the momentum integral
for zero external momentum $\bol = \bzero$ and
loop frequencies $p_0$ close to $l_0/2$
can be explicitly calculated.
At Van Hove filling and with 
$\psi_{j_1}(\bol, \bop) = \psi_{j_2}(\bol, \bop) = 1$ we find
\begin{equation}
  I_{11}^+\big( (l_0, \bzero), p_0 \big)
=
  \begin{cases}
     C_1\ p_0^{-1} + \cO( \ln p_0 )
     &:\ l_0 = 0,\\
     -C_2 \sgn(p_0 - l_0/2)\ln\abs{p_0 - l_0/2} + \cO\big( (p_0 - l_0/2)^0 \big)
     &:\ l_0 > 0,
  \end{cases}
\end{equation}
with positive constants $C_1, C_2$.
Typically, $I^\pm_{\psi_{j_1}, \psi_{j_2}}(l, p_0)$ is
a monotonic function in $p_0$ at both sides of the singularity point $p_0 = l_0/2$.

To capture the structure of such momentum integrals we discretize
the frequency variable $p_0$ on a grid that depends on $l_0$.
We use logarithmic spacing and about 100 data points at both sides of the point $p_0 = l_0/2$.
The minimal distance to this point in the grid is adjusted depending on the expected scale
$\Om_*$, usually $\Om_*/10$ is sufficient.
This ensures that during the flow the full structure of integrands
in the RG equations is properly taken into account at all scales.

%%%%%%%%%%%%%%%%%%%%%%%%%%%%%%%%%%%%%%%%%%%%%%%%%%%%%%%%%%%%%%%%%%%%%%%%%%%%%%%%%%%%%%%%%%%%%%%%%%%%%%%%%%%%%%%%%
\section{Results at Van Hove Filling\label{sec:results}}

\subsection{Results in the $\Om$-Scheme}

In this section we discuss several frequency parameterizations of the interaction vertex,
going from simple setups to more detailed parameterizations, and compare the resulting flows
for systems at Van Hove filling and temperature zero.
Here, the self-energy and imaginary parts of the vertex function are neglected at first,
both will be discussed in the next sections.

\begin{figure}
  \hspace{-2cm}
  \input{FreqPap_Omcrit_compare}
  \input{FreqPap_withfreqNoSigma_Omcrit_compare}
  \caption{(Color online) Stopping scales for systems at Van Hove filling.
Left: Comparison of different frequency parameterizations:
(i) Frequency independent vertex,
(ii) Vertex with simple Lorentz parameterization of the transfer frequency,
(iii) Vertex with discretized transfer frequency,
(iv, symbols) Vertex with discretized transfer and non-transfer frequencies.
Right: Setup (iii) for interaction parameters $U/t = 3$, $2.5$ and $2$ (top-down).
Most dominant ordering tendencies: commensurate AFM (dashes, filled square),
incommensurate AFM (dotted, open square), $d$-SC (solid, circle),
scattering instability (dash-dot, triangle), FM (short dashes, diamond).
}
\label{fig:Omcrit_FreqParam}
\end{figure}
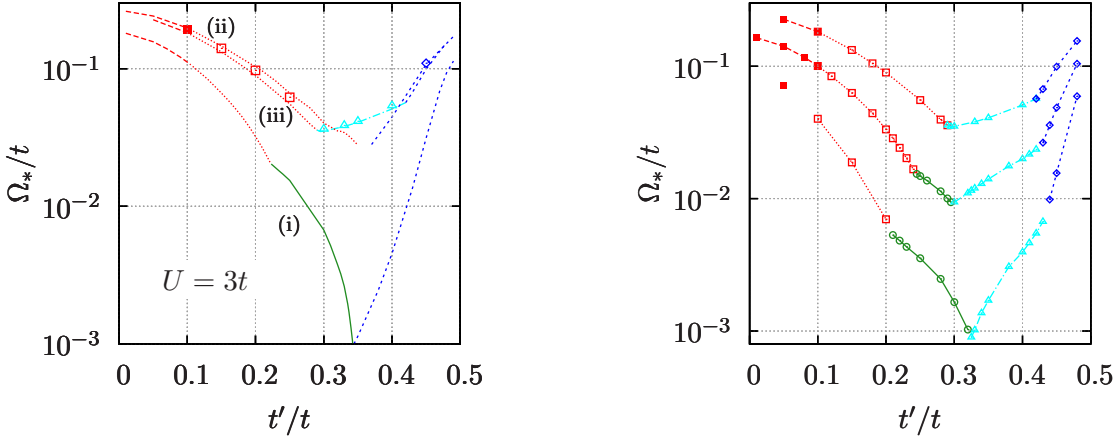

 In the left graph of Fig.~\ref{fig:Omcrit_FreqParam} the stopping scale $\Omega_*$ is plotted over next to nearest neighbor hopping $t'$ at $U=3t$. The lowest line corresponds to frequency independent vertex functions. This approximation is motivated by weak coupling power counting and has been used in early RG applications to the 2D Hubbard model \cite{ZanchiSchulz1,ZanchiSchulz2000,HalbothMetzner,Umklapp,HonerkampSalmhofer_QuasiparticleWeight,HonerkampSalmhofer_Tflow}. In agreement with Ref. \onlinecite{HonerkampSalmhofer_Tflow} we find regions of antiferromagnetism (AFM), $d$-wave superconductivity ($d$-SC), and ferromagnetism (FM) at Van Hove filling. That is, for low $t'$ the dominant coupling is the boson propagator in the magnetic channel with form factor $f_1(\bop)=1$ at (commensurate AFM) or near (incommensurate AFM) momentum transfer $(\pi,\pi)$.
For intermediate hopping ratios $t'/t$
the fermionic interaction has a sharp peak in the superconducting channel with $d$-wave form factor $f_2(\bop)=\cos p_x -\cos p_y$ at zero transfer momentum.
For $\frac 13<t'/t < \frac{1}{2}$
the ferromagnetic instability is signaled by a divergence of the magnetic boson propagator with form factor $f_1(\bop)=1$ at zero transfer momentum.  
In the crossover region between superconductivity and ferromagnetism the stopping scale is strongly suppressed compared to $t' = 0$.

Our simplest frequency parameterization considers frequency independent
fermion-boson vertex functions, $\Gamma_m^\Om(q_0, l_0, \bol) \equiv 1$, and describes the frequency dependence of the boson propagators with one Lorentz curve. The stopping scale obtained in this approximation is the upper most curve (ii) plotted in Fig.~\ref{fig:Omcrit_FreqParam}. There are two major differences to a frequency independent approximation, line (i). First, the stopping scale is much higher, especially for intermediate next to nearest neighbor hopping. Secondly, there is no region of dominant $d$-wave superconductivity anymore. Instead, incommensurate antiferromagnetic couplings with transfer momenta quite far from $(\pi,\pi)$ are dominant for intermediate $t'$.

\begin{figure}
\begin{center}
\includegraphics[width=0.8\textwidth,angle=0]{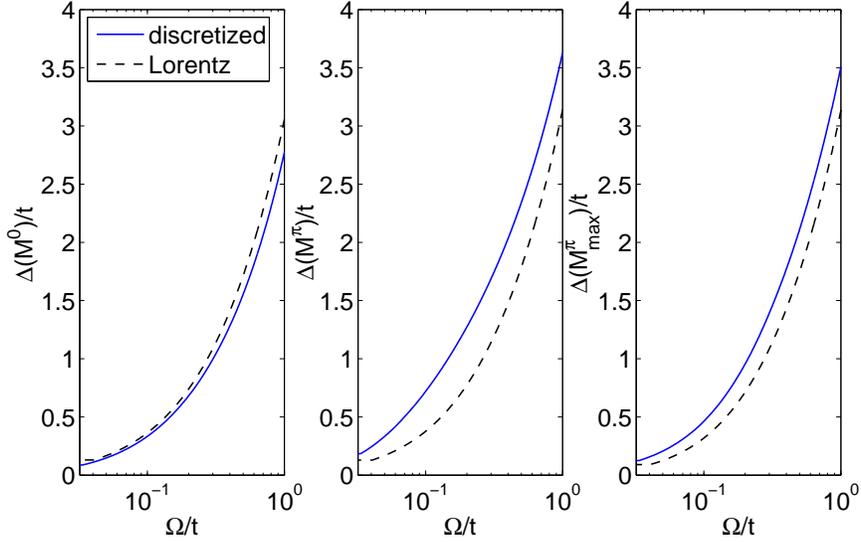}
\caption{(Color online) Comparison of effective widths $\Delta(B)$ of boson propagators $M_{11}(l_0,0)$, $M_{11}(l_0,\hat{\pi})$, and $\max_{\bop\approx \hat{\pi}}  M_{11}(l_0,\bop)$ (from left to right) obtained from Lorentz (black dashed line) and discrete (blue line) parameterizations at $t'=0.3t$ and $U=3t$. Here $\Delta(B)$ is defined
by  $B(\Delta,\bop)=\sfrac{1}{2} B(0,\bop)$.
Generally, the effective width decreases with $\Omega$ leading to narrow peaks at low RG scales\label{fig:effectiveWidth}.
}
\end{center}
\end{figure}

In our understanding both effects are closely related.
In general, the static approximation overestimates the vertex function.
Consideration of a frequency dependent vertex function leads to an additional decay of the
frequency integrand in box and vertex diagrams,
unless they are generated directly by $U$.
 The latter is the case for all $s$-wave channels with form factor $f_1(\bop)=1$ but not for the $d$-wave superconducting channel. Therefore, box and vertex diagrams that generate an attractive  $d$-wave interaction contribute less after the loop frequency integration. More generally, the coupling of different channels is reduced due to the frequency dependence of box and vertex diagrams since they are not evaluated at their maximal value only.
The effective width of boson propagators approaching a singularity tends to zero, but is also reduced for all other boson propagators at low scales, see Fig.~\ref{fig:effectiveWidth}.
This leads to a reduction of screening and mutual coupling between the channels and consequently to a higher pseudo-critical scale.

We stress that $d$-wave superconductivity is still generated by the RG flow, but to a much lesser extent. Due to the higher stopping scale this generating process has not enough RG time to become a leading instability. If we change our definition of $\Omega_*$ to allow lower scales then $d$-wave superconductivity becomes dominant in a small parameter region of $t'$ eventually. For example, this can be achieved by not taking the maximal value of the dominant boson propagator at frequency zero as the stop condition but rather a
frequency mean value over a not too small neighborhood around zero. (This would not change $\Omega_*$ in case of a frequency independent calculation.) However, the so defined stopping scale is still much larger than the stopping scale obtained with frequency independent vertices and the region of $d$-wave superconductivity is substantially smaller.

For lower initial interaction values $U$ the effects of frequency dependence becomes less drastic. In particular, for $U=2.5t$ and $U=2t$ we find a $d$-wave superconducting region, which is, however, smaller than in the frequency-independent calculation.

In a next step we improve the frequency parameterization of boson propagators by discretizing the transfer frequency dependence, rather than
assuming a Lorentz form.
 The corresponding stopping scale is plotted in Fig.~\ref{fig:Omcrit_FreqParam} (line iii).
For $t'<0.3t$ and $t'>0.42t$
the flow of the most singular couplings obtained in the simple Lorentz parameterization scheme
is essentially reproduced.
This is remarkable because, as has already been pointed out above,
simple Lorentz distributions describe the dependence on transfer frequency well only
in the small frequency regime.
The full discretized frequency parameterization gives a slower frequency decay for intermediate to large frequencies for the antiferromagnetic boson propagator $M_{11}(l_0,\bpi)$. Furthermore, $M_{11}(l_0,0)$ decays a little faster, see Fig \ref{fig:effectiveWidth}. This, however, does not entail a significantly larger $d$-wave superconducting coupling. In the antiferromagnetic and in the ferromagnetic regime the stopping scale is reduced only by a negligible amount.

While this seems to encourage a simpler Lorentz curve setup over a more involved frequency parameterization, the former does not allow sign changes of boson propagators in their frequency dependence.
This restriction is not present in the discretized parameterization. In fact, in the density channel at zero transfer momentum we find a sign change of $K_{11}(l_0, 0)$ at a non-zero transfer frequency.
The channel decomposition is set up such that the boson propagators are positive for transfer frequency zero. 
Positive values of $K_{11}$ in the $s$-wave channel correspond to an attractive interaction of density pairs, 
the flow of which is dominated by the local Coulomb repulsion $U$ and can therefore not diverge. 
Once $K_{11}(l_0, 0)$ becomes negative, \ie repulsive, it will be enhanced by $U$. 
The sign change of the right hand side in Eq.~\eqref{eq:KflowEq} is due to the frequency dependence of boson propagators that enter vertex and box diagrams via $\alpha_1^{\mtin{K}}$, which decay in the loop frequency. The biggest contribution to this effect comes from the magnetic channel. The bare particle-hole bubble vanishes at momentum $\bol=\bzero$ for frequencies $l_0\neq 0$. Although this is changed by the $\Omega$-regularization we expect only minor contributions from the direct graph. The direct graph vanishes for the $T$-scheme, see below.

\begin{figure}
\hspace{-2cm}
\input{FreqPap_K11strange_freqdep1}
\hspace{-1cm}
\input{FreqPap_K11strange_freqdep2}
  \caption{Scale dependence of the function $l_0 \mapsto K_{11}(l_0, \bol = \bzero)$ for
    $U = 2.5 t$, $t'/t = 0.33$. The stopping scale is about $\Om_* \approx 0.012t$,
the initial scale is $\Om_0= 40t$.
    The function is shown in the course of the flow for
    $\Om / \Om_* \approx 1600, 400, 150, 40, 20$ (left) and
    $\Om / \Om_* \approx 20, 4, 2.5, 1.5, 1$ (right)
    with progressive forming of the minimum at non-zero frequency.
}\label{fig:K11strange_freqdep}
\end{figure}
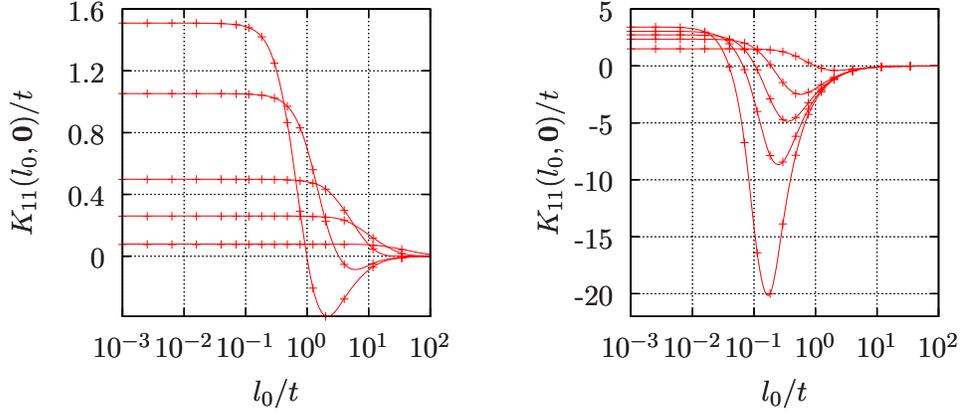

\begin{figure}
\begin{center}
\includegraphics[width=0.65\textwidth,angle=0]{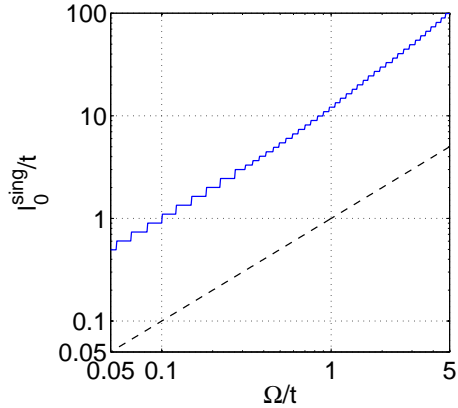}
\caption{(Color online) The frequency $l_0^{\mtin{sing}}$ of minimal $K_{11}(l_0, \bzero)$
 in dependence of the RG scale $\Omega$
 for $t'=0.3t$ and $U=3t$ (blue line). Due to the finite frequency grid the dependence is a step function. For comparison the black dashed line denotes $l_0^{\mtin{sing}}=\Omega$.
\label{fig:l0singdependence}}
\end{center}
\end{figure}

We find that a negative value of $K_{11}(l_0,\bol = \bzero)$
becomes the leading coupling in the parameter range
$0.3 \lesssim t'/t \lesssim 0.42$ for $U = 3t$.
Here, the scattering propagator $K_{11}(l_0, \bol = \bzero)$
slowly grows to positive values at $l_0 = 0$ during the flow,
whereas in a certain frequency range away from $l_0 = 0$
it quickly runs to negative values, compare Fig.~\ref{fig:K11strange_freqdep}.
The frequency $l_0^{\mtin{sing}}(\Om)$ of minimal
$K_{11}(l_0, \bol = \bzero)$ is roughly proportional to $\Om$, see Fig.~\ref{fig:l0singdependence}.
We have verified that this scattering divergence is robust
against refinement of the discretization of exchange propagators
in frequency and momentum space.

By decreasing $U$ we again find a growing region where $d$-wave superconductivity is the leading instability, see the right plot in Fig.~\ref{fig:Omcrit_FreqParam}. The $K$-scattering remains leading in a region between $d$-wave superconductivity and ferromagnetism. 	

\begin{figure}
\begin{center}
\includegraphics[width=0.55\textwidth,angle=0]{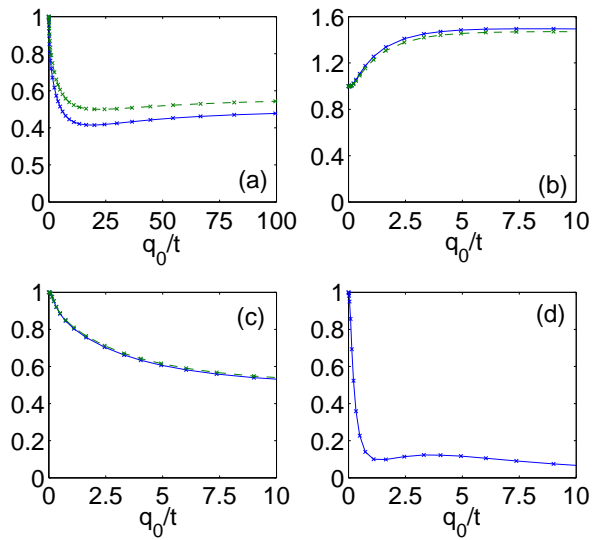}
\caption{(Color online) Frequency dependence of boson-fermion vertex functions at $\Omega_* \approx 0.037t$ for $U=3t$ and $t'=0.3$ at Van Hove filling: (a)  $\gamma^{\mtin{SC},\mathbf{a}}_{1}$ , (b) $\gamma_{\mtin{M}}^{\mathbf{a}}$ , (c) $\gamma_{\mtin{K}}^{\mathbf{a}}$ , and (d) $\gamma^{\mtin{SC},\mathbf{a}}_{2}$.
The blue curves correspond to $\mathbf{a}=0$ and the green dashed lines to $\mathbf{a}=\hat{\pi}$.\label{fig:bosefermi_freqdep}}
\end{center}
\end{figure}

Up to now we have assumed frequency independent boson-fermion vertex functions, which were simply given by scale independent form factors. In a next step we evaluate the full flow equations given in Sec. \ref{sec:flowequations}, where in addition to boson propagators we also discretize the frequency dependence of $\gamma^{\mtin{SC},\mathbf{a}}_{1,2}$, $\gamma_{\mtin{M}}^{\mathbf{a}}$, and $\gamma_{\mtin{K}}^{\mathbf{a}}$. Their frequency dependence is plotted in Fig.~\ref{fig:bosefermi_freqdep}. The $s$-wave channels deviate only by a small amount from unity (which is the normalization at frequency zero). For large frequencies, attractive channels saturate at a value below 1 and repulsive channels above 1. In the $d$-wave superconducting channel the boson-fermion vertex function decays to zero. 
Despite the frequency dependence of the boson-fermion vertex functions we find only minor changes in the stopping scale, see symbols in Fig.~\ref{fig:Omcrit_FreqParam}. Most importantly, the $K$-scattering process remains leading in the same parameter region as before.

We cannot exclude the possibility that the $K$-scattering singularity with sign change in the density channel is an artifact of the channel decomposition that is not present in the full level-2 truncation of the RG. However, our result is stable against the inclusion of frequency dependent fermion-boson vertex functions. For determining singularities in the RG flow the dependence of the vertex function on non-transfer frequencies seems negligible. This dependence may become important in a more quantitative analysis of observables.

Even if $K(l_0^{\mtin{sing}}, \bzero)$ is not the leading coupling at the stopping scale, its sizeable contribution to the effective interaction cannot be disregarded. In the conventions of Ref.~\onlinecite{decomposition} it reads
\begin{align}
 \ca{V}_{\mtin{K}}[\Psi] =-\frac 14 \int\mathrm{d}l\; K(l) Y(l) Y(-l)
\end{align}
with density operators $Y(l)=\int\mathrm{d}p \sum_{\sigma}\ovl{\Psi}_{\sigma}(p) \Psi_{\sigma}(p+l)$. 
Suppose that $K(l_0, \bzero)$ has a strong negative peak at $l_0^{\mtin{sing}}$. In the following we neglect contributions from other channels and assume a mean-field approximation
\begin{align}
 K(l)=-4\lambda\ \delta(\bol)\ \delta(\abs{l_0}-l_0^{\mtin{sing}})
\end{align}
at the critical scale with $\lambda>0$.
Then the $K$-scattering singularity corresponds to a repulsion of time modulated densities
\begin{align}
 \ca{V}_{\mtin{K}}^{\mtin{MF}}[\Psi]= \lambda \sum_{x,x',\tau,\tau'} n(x,\tau) n(x',\tau') e^{il_0^{\mtin{sing}}(\tau-\tau')}
\end{align}
where $x$ denotes lattice sites and $\tau$ is the imaginary time.
For $l_0^{sing} = 0$, mean field theory produces only a shift of the chemical potential.
In contrast, $l_0^{sing} > 0$, which is the situation that we encounter,
yields off-diagonal frequency terms in the two-point function and hence changes its structure in a
non-trivial way.
A detailed analysis of this scattering singularity, in particular regarding its influence
on an RG flow below the stopping scale is left for future work.

\subsection{Influence of Imaginary Exchange Propagators}

Unlike in RG setups where frequency dependences are neglected,
the frequency dependent interaction vertex acquires an imaginary part during the flow.
By symmetry, in our parameterization
particle-hole exchange propagators $M_{11}$, $K_{11}$ are
real-valued whereas $D_{11}$, $D_{22}$ in the particle-particle channel are not.
Still, $\rIm D_{mm}(l_0 = 0, \bol) = 0$ is imposed by symmetry.

Perturbation theory and RPA resummation  yield a singular behavior in the first frequency
derivative of $\rIm D_{11}$, originating from an asymmetry in the density of states
about zero energy,
 see Eq.~\eqref{eq:ppBubble_om-diff}.
We have examined the influence of this singularity in the RG calculation
at the data point $t'/t = 0.3$, $U/t = 3$.
Here the stopping scale is relatively low such that an effect, if present,
should be visible.
Furthermore, a strong scattering singularity has been encountered in this
parameter region and we study how it is influenced.

The fact that $\rIm D_{mm}(l)$ exactly vanishes at $l_0 = 0$
already suggests that $\rIm D_{mm}$ possibly has limited impact.
This is confirmed by carrying out the RG calculation:
Although $(\partial_{l_0}\rIm D_{mm})(l_0 = 0, \bol)$ shows the expected singular behavior,
$\rIm D_{mm}$ remains small during flow.
At the stopping scale,
$\abs{\rIm D_{11}(l)} < 5t$ with the maximum at $\bol = \bpi$ and
$l_0 \approx 10\Om_*$,
as well $\abs{\rIm D_{22}(l)} < 0.05t$.

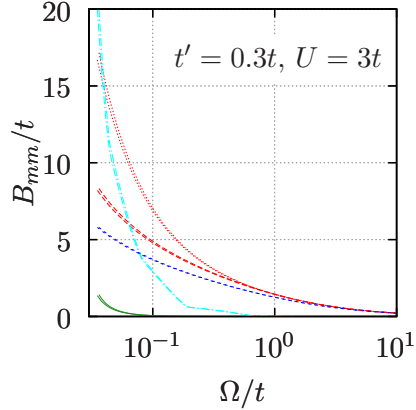
\begin{figure}
  \begin{center}
      \input{FreqPap_LeadingFlow_with_ImDmm}
   \end{center}
   \caption{(Color online) Influence of $\rIm D_{mm}$ on the flow of the most singular couplings
     for frequency dependent interaction vertex.
     The flow with slightly lower stopping scale takes into account $\rIm D_{11}$ and $\rIm D_{22}$,
     the second one neglects them. Line conventions as in Fig.~\ref{fig:Omcrit_FreqParam}.}
   \label{fig:flow_with_ImDmm}
\end{figure}

As a consequence, these quantities have little influence on the flow of the leading couplings,
see Fig.~\ref{fig:flow_with_ImDmm}
where the flows including and neglecting the imaginary parts of the SC channel
are compared.
In particular, the $K$-scattering singularity remains unchanged.
Although we have not performed a  detailed scan of the whole parameter space
we expect $\rIm D_{mm}$ to be of minor influence.

\subsection{Comparison to $T$-Scheme}\label{sec:Tflow}

We need to check whether the $K$-scattering singularity is an artifact of the $\Omega$-regularization. To this end, we repeat the calculation
in the temperature flow scheme\cite{HonerkampSalmhofer_Tflow},
which uses temperature $T$ as the scale parameter.
This scheme is used frequently in RG studies of the Hubbard model\cite{HonerkampSalmhofer_Tflow, Katanin_SigmaFM},
so far mostly in the static approximation.
For our purpose the frequency dependence of the interaction vertex needs
to be taken into account as well.
In terms of rescaled fields, the free propagator in the $T$ scheme reads
$C_T(p) = T^{1/2}[ip_0 - e(\bop) ]^{-1}$ and fermionic Matsubara frequencies
$p_0 \in \pi T (2\ZZ-1)$ are discrete.

The flow equation \eqref{eq:OneLoopVertexFlow} can be rewritten
by setting up a channel decomposition as before.
Exchange propagators now carry discrete bosonic frequencies  $l_0 \in \pi T\ 2\ZZ$.
We project non-transfer frequencies to the lowest possible frequency value,
\ie $p_0 = \pm \pi T$, and symmetrize over the sign.
Self-energy effects and imaginary parts of the interaction vertex are neglected. 
We calculate the flow for $0.3 \le t'/t \le 0.35$,
\ie in the parameter region where a strong scattering process has been found in the
$\Om$-scheme,
and for several Hubbard parameters $U$.
The initial condition of the flow is determined from second order perturbation theory
at a high temperature $T_0 = 100t$.
Technical details of the calculation are given in Ref. \onlinecite{giering_phd}.

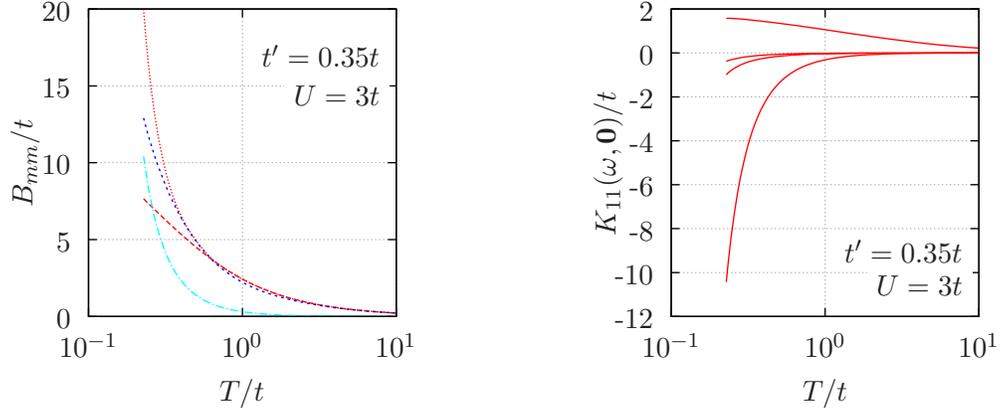
\begin{figure}
  \input{FreqPap_precompT_2htol_light_t20_35_leading_flow}
  \input{FreqPap_precompT_2htol_light_t20_35_K11_flow}
  \caption{(Color online) Temperature flow at Van Hove filling.
    Left: Flow of most dominant instabilities. The stopping temperature
    here is $T_* \approx 0.22t$. Line conventions as in Fig.~\ref{fig:Omcrit_FreqParam}.
    Right: Flow of  $K_{11}(l_0, \bzero)$ for the lowest possible
  frequency values $l_0/ \pi T = 0, 6, 4, 2$ (from top to bottom).
  The scattering singularity shows at the lowest non-zero
  transfer frequency.
}
\label{fig:K11flow_Tscheme}
\end{figure}

Firstly, we observe that the $T$-flow diverges faster than the $\Om$-flow in the sense that,
when comparing scales as energy variables, $T_*$ is much larger than $\Om_*$ (whereas in the static approximation $T_*\approx \Omega_*$). 
We interpret this as a consequence of the specific form of transfer frequency dependence,
which is quite different in both schemes.
Whereas we observe that exchange propagators are monotonic functions of the scale parameter
in the $T$-scheme, we find non-monotonic behavior for the $\Om$-scheme.
Here, exchange propagators at a specific transfer frequency $l_0$ during the flow
usually first grow for $\Om \gtrsim l_0$ and then shrink for $\Om \lesssim l_0$. 
Consequently, as a function of frequency, exchange propagators in the $\Om$-scheme are typically broader
than in the $T$-scheme
and hence closer to the approximation of
frequency independent exchange propagators.

Furthermore, the zero-momentum scattering exchange propagator $K_{11}(l_0, \bzero)$ again
shows two different behaviors: At zero frequency it slowly flows towards
positive values while for the smallest non-zero exchange frequency,
\ie $l_0 = \pm 2\pi T$, it strongly grows to negative values,
compare Fig.~\ref{fig:K11flow_Tscheme}. 
This corresponds to $l_0^{\mtin{sing}} = 2\pi T$ in the T-scheme 
and is consistent with 
the approximately linear decrease of $l_0^{\mtin{sing}}$ as a function of scale in the $\Om$-scheme. 

We have not found parameter values where this scattering process
 becomes the dominant coupling
in the $T$-flow. However, it is present and gets strong in this scheme as well.
For that reason we do not consider it an artefact of a particular regularisation.

\subsection{Consideration of Imaginary Self-energy}\label{sec:ImSigma}

A non-trivial flow equation for the frequency dependent self-energy
can be constructed already with a static approximation
of the interaction vertex\cite{HonerkampSalmhofer_QuasiparticleWeight}.
Knowledge about the full frequency dependent interaction vertex
now allows a detailed study of the frequency dependent self-energy.
Analysis of $\Sigma(p)$ in second order perturbation theory \cite{FeldmannSalmhofer2008b}
for a system at van Hove filling and with square Fermi surface shows
a logarithmic singularity in the first frequency derivative at momentum $\bop = (0, \pi)$.
It turns out that this singularity gets enhanced in the RG setup and 
results in a suppression of the interaction vertex during the flow. 
This suppression can become rather strong.

We calculate the flow of the imaginary part of the self-energy, for which most singular behavior is expected,
 \begin{equation}
   \rIm \dot\Sigma(p)
    =
    - \int\intd l\
       \rIm S(l+p)
      \left( (\rRe D + \undemi K + \frac{3}{2} M)_{11}(l)
             + \rRe D_{22}(l) f_2^2( \frac{\bol}{2} + \bop ) \right)
.
\end{equation}
Imaginary parts of exchange propagators in the pairing channel are neglected here
since they remain small\cite{GieringSalmhofer2011}.
Exchange propagators are discretized in frequency and momentum space as before.
We then trace the function $\om \mapsto \rIm\Sigma(\om, \bop)$ by frequency discretization
for several momenta $\bop$.
Note that a simple parametrization of $\rIm\Sigma$ using only a linear (momentum-dependent)
frequency term is insufficient and leads to an artificial suppression of the flow.

It turns out that the presence of the full propagator (\ie including $\rIm\Sigma$)
instead of the bare one makes the evaluation of the flow equations relatively
time-consuming since all loop frequency-momentum integrations then need to be
performed numerically at each RG step.
At low scales the involved triple integration in the full flow equations
was not done because of long computing times. 
As an approximation, for parameter values with flow to low scales, we restrict to a
momentum independent feed-back of $\rIm\Sigma(p)$ to the flow. This approximation
again allows factorization of the propagator and largely simplifies numerics by permitting
computation of loop momentum integrals beforehand,
similar to the strategy of Sec. \ref{sec:loop_integrals}.
Thus, we replace $\rIm\Sigma(\om, \bop) \to \rIm\Sigma(\om, (0, \pi))$ in the right hand side of the flow equations.
We have chosen the momentum  $(0, \pi)$ for the self-energy feed-in since the saddle point region,
for systems at Van Hove filling, mainly drives the flow and
should receive special attention.
Because in this region $\rIm\Sigma$ also shows its most singular behavior,
this approximation overestimates the suppression of the flow.

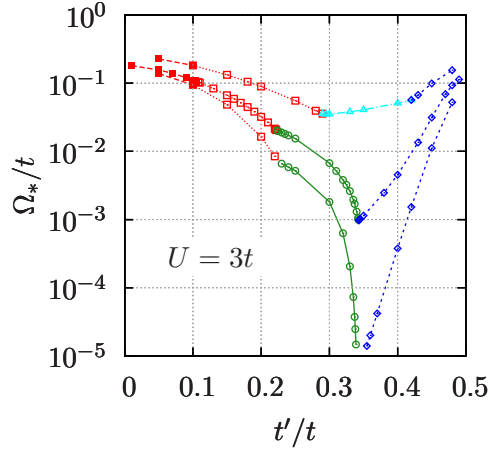
\begin{figure}
   \input{FreqPap_withfreqwithSigma_Omcrit_compare}
  \caption{(Color online) Comparison of stopping scales for different RG setups at Van Hove filling.
Top-down:
(i) frequency dependent vertex disregarding self-energy,
(ii) frequency independent vertex disregarding self-energy,
(iii) frequency dependent vertex including $\rIm\Sigma(\om, (0, \pi))$.
Line conventions as in Fig.~\ref{fig:Omcrit_FreqParam}.
}
\label{fig:Omcrit_ImSigma-flow}
\end{figure}

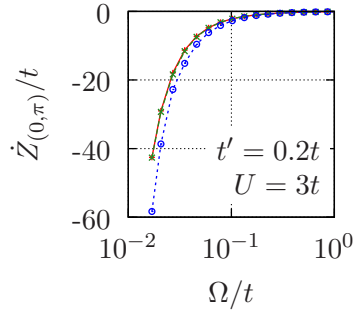
\begin{figure}
  \begin{center}
      \input{FreqPap_Zdot_LorentzTest}
   \end{center}
   \caption{(Color online) Dependence of the $Z$-factor scale derivative on different frequency parameterizations of exchange propagators.
   Compared to the correct scale derivative (computed from frequency discretized exchange propagators,
   + marks), the parameterization by a sum of two Lorentz curves (x marks) yields precise results.
   A single Lorentz curve determined from the small frequency behavior (circles) overestimates the flow.
   } \label{fig:Zdot_LorentzTest}
\end{figure}

The resulting stopping scale is shown in Fig.~\ref{fig:Omcrit_ImSigma-flow} for initial interaction $U = 3t$.
We observe three main differences as compared to the frequency dependent setup disregarding
self-energy effects.
First, the stopping scale $\Om_*$ is significantly lowered and
drops drastically in the parameter region of competing pairing and ferromagnetic
ordering tendencies.
Furthermore, a region of dominant $d$-wave pairing is recovered. On the contrary,
the $K$-scattering is weakened and no longer becomes dominant.

The flow of the self-energy is very sensitive to proper frequency parameterization
of exchange propagators. 
In the calculation of the flow as described above we additionally determine
the scale derivative of the $Z_{(0, \pi)}$ factor with alternative frequency parameterizations
of exchange propagators and compare to the correct result in Fig.~\ref{fig:Zdot_LorentzTest}.
This test does not take into account an error propagation because at each scale the true self-energy is used to
calculate scale derivatives.
We observe that the parameterization of exchange propagators by a sum of two Lorentz curves
produces quantitatively accurate results.

A more detailed study about self-energy flows as well as
a discussion about the neglect of the momentum dependence of $\rIm\Sigma$ at low scales
can be found in Ref. \onlinecite{GieringSalmhofer2011}.

\section{Conclusion}

We have presented a functional RG calculation for the two-dimensional Hubbard model
at Van Hove filling. All calculations use a channel decomposition for the interaction vertex,
which provides an efficient parameterization of momenta and frequency.
We studied the effective two-fermion interaction and compared quantitatively the effect of different approximations of the frequency dependence.

In general, including frequency dependences of vertices in an RG flow with bare propagators raises the (pseudo-)critical scale compared to frequency independent calculations. The higher scale is accompanied by a reduction of the $d$-wave superconductivity coupling that becomes leading in a smaller region of the $(U,t')$-parameter space. In particular, for $U=3t$ we did not find a $d$-wave instability
at Van Hove filling. Lower initial repulsive couplings $U$ re-establish a $d$-wave instability and the approximation of neglecting the frequency dependence leads to less drastic effects.

In a second calculation we have included the imaginary part of the fully frequency discretized 
self-energy, evaluated at the Van Hove points.
The divergence of the corresponding $Z$ factor then leads to a suppression of the flow of the
frequency dependent interaction vertex.
The stopping scale is reduced and can drop to rather small values,
notice that the evaluation of the self-energy at the Van Hove points overestimates this reduction.
Besides the change in the stopping scale, the flow qualitatively agrees well with previous
flows that neglect frequency dependences at all.
Again we find regions of dominating antiferromagnetism, $d$-wave superconductivity and ferromagnetism,
whereas couplings in the scattering channel remain subleading.
In this view, neglect of frequency dependences and self-energy contributions at the same time
seems to be a rather good approximation for the full RG flow. The reason for this is not fully understood yet. It is, however, clear from the lowest equation in the RG hierarchy that keeping a frequency-independent self-energy in a calculation with a frequency-dependent vertex cannot be exactly correct. In fact, such an approximation violates the Ward identity between the $\tilde{\Gamma}$-vertex and $\Sigma$. Including the frequency dependence in our case restores the results of the static approximation qualitatively.

For future applications of the functional RG it is important to develop feasible approximations of the frequency dependence. The simplest parameterization with a single Lorentz curve for each boson propagator already gives reasonable results for the flow of dominant couplings in the vertex function. Since this ansatz does not correctly describe intermediate to large frequencies, which contribute more to the flow equation for the self-energy than in the vertex function, the error for the $Z$-factor becomes large.
With two Lorentz curves each boson propagator can already be described well enough to compute
the imaginary self-energy and the leading instabilities of the interaction vertex reliably.

The benchmark for testing frequency parameterizations was set by a full discretization of frequency and momenta for each boson propagator. Imaginary contributions to the interaction vertex seem to have little impact on the RG flow. While single Lorentz curves do not allow sign changes in the frequency dependence, we find indeed such a sign change in the density forward scattering channel in the discretized flow. The corresponding coupling at transfer momentum zero and transfer frequency $l_0^{\mtin{sing}}$ can in principle diverge and becomes large in a region between superconductivity and ferromagnetism. The frequency $l_0^{\mtin{sing}}$ is roughly proportional to the RG scale.

While in the $\Omega$-regularization scheme the coupling of this this finite frequency $K$-scattering becomes leading in a flow with bare propagators, this is not the case in the temperature regularization scheme or when the imaginary self-energy is included. In any case we find a large contribution to the effective interaction at finite scales. Its influence on low scales and on observables is left for future work.

All calculations were performed using a channel decomposition of the level-2 truncation, which made the parameterization of momenta and frequencies more manageable. We generalized the
method to be able to describe non-transfer frequencies. Unlike non-transfer momenta they are not expanded in orthonormal functions. Instead we derived flow equations for fermion-boson vertex functions, that is, frequency-dependent Yukawa couplings. It turned out that their frequency dependence does not significantly change results for the RG flow of dominant couplings in the Hubbard model.

We thank A. Eberlein, J. Ortloff, and W. Metzner for fruitful discussions. 
This work was supported by the DFG research unit FOR 723.

\bibliography{frequency}
\end{document}

%% file: FreqPap_fitsBfreq_withfreq_prec4_moderate_t20_2.tex
% GNUPLOT: LaTeX picture with Postscript
\begingroup
  \makeatletter
  \providecommand\color[2][]{%
    \GenericError{(gnuplot) \space\space\space\@spaces}{%
      Package color not loaded in conjunction with
      terminal option `colourtext'%
    }{See the gnuplot documentation for explanation.%
    }{Either use 'blacktext' in gnuplot or load the package
      color.sty in LaTeX.}%
    \renewcommand\color[2][]{}%
  }%
  \providecommand\includegraphics[2][]{%
    \GenericError{(gnuplot) \space\space\space\@spaces}{%
      Package graphicx or graphics not loaded%
    }{See the gnuplot documentation for explanation.%
    }{The gnuplot epslatex terminal needs graphicx.sty or graphics.sty.}%
    \renewcommand\includegraphics[2][]{}%
  }%
  \providecommand\rotatebox[2]{#2}%
  \@ifundefined{ifGPcolor}{%
    \newif\ifGPcolor
    \GPcolortrue
  }{}%
  \@ifundefined{ifGPblacktext}{%
    \newif\ifGPblacktext
    \GPblacktexttrue
  }{}%
  % define a \g@addto@macro without @ in the name:
  \let\gplgaddtomacro\g@addto@macro
  % define empty templates for all commands taking text:
  \gdef\gplbacktext{}%
  \gdef\gplfronttext{}%
  \makeatother
  \ifGPblacktext
    % no textcolor at all
    \def\colorrgb#1{}%
    \def\colorgray#1{}%
  \else
    % gray or color?
    \ifGPcolor
      \def\colorrgb#1{\color[rgb]{#1}}%
      \def\colorgray#1{\color[gray]{#1}}%
      \expandafter\def\csname LTw\endcsname{\color{white}}%
      \expandafter\def\csname LTb\endcsname{\color{black}}%
      \expandafter\def\csname LTa\endcsname{\color{black}}%
      \expandafter\def\csname LT0\endcsname{\color[rgb]{1,0,0}}%
      \expandafter\def\csname LT1\endcsname{\color[rgb]{0,1,0}}%
      \expandafter\def\csname LT2\endcsname{\color[rgb]{0,0,1}}%
      \expandafter\def\csname LT3\endcsname{\color[rgb]{1,0,1}}%
      \expandafter\def\csname LT4\endcsname{\color[rgb]{0,1,1}}%
      \expandafter\def\csname LT5\endcsname{\color[rgb]{1,1,0}}%
      \expandafter\def\csname LT6\endcsname{\color[rgb]{0,0,0}}%
      \expandafter\def\csname LT7\endcsname{\color[rgb]{1,0.3,0}}%
      \expandafter\def\csname LT8\endcsname{\color[rgb]{0.5,0.5,0.5}}%
    \else
      % gray
      \def\colorrgb#1{\color{black}}%
      \def\colorgray#1{\color[gray]{#1}}%
      \expandafter\def\csname LTw\endcsname{\color{white}}%
      \expandafter\def\csname LTb\endcsname{\color{black}}%
      \expandafter\def\csname LTa\endcsname{\color{black}}%
      \expandafter\def\csname LT0\endcsname{\color{black}}%
      \expandafter\def\csname LT1\endcsname{\color{black}}%
      \expandafter\def\csname LT2\endcsname{\color{black}}%
      \expandafter\def\csname LT3\endcsname{\color{black}}%
      \expandafter\def\csname LT4\endcsname{\color{black}}%
      \expandafter\def\csname LT5\endcsname{\color{black}}%
      \expandafter\def\csname LT6\endcsname{\color{black}}%
      \expandafter\def\csname LT7\endcsname{\color{black}}%
      \expandafter\def\csname LT8\endcsname{\color{black}}%
    \fi
  \fi
  \setlength{\unitlength}{0.0500bp}%
  \begin{picture}(4320.00,3024.00)%
    \gplgaddtomacro\gplbacktext{%
      \csname LTb\endcsname%
      \put(1242,704){\makebox(0,0)[r]{\strut{} 0}}%
      \csname LTb\endcsname%
      \put(1242,1348){\makebox(0,0)[r]{\strut{} 5}}%
      \csname LTb\endcsname%
      \put(1242,1992){\makebox(0,0)[r]{\strut{} 10}}%
      \csname LTb\endcsname%
      \put(1242,2637){\makebox(0,0)[r]{\strut{} 15}}%
      \csname LTb\endcsname%
      \put(1374,484){\makebox(0,0){\strut{}$10^{-3}$}}%
      \csname LTb\endcsname%
      \put(1954,484){\makebox(0,0){\strut{}$10^{-2}$}}%
      \csname LTb\endcsname%
      \put(2534,484){\makebox(0,0){\strut{}$10^{-1}$}}%
      \csname LTb\endcsname%
      \put(3113,484){\makebox(0,0){\strut{}$10^{0}$}}%
      \csname LTb\endcsname%
      \put(3693,484){\makebox(0,0){\strut{}$10^{1}$}}%
      \put(604,1863){\rotatebox{-270}{\makebox(0,0){\strut{}$M_{11}(l_0, \bpi) /t$}}}%
      \put(2533,154){\makebox(0,0){\strut{}$l_0 / t$}}%
    }%
    \gplgaddtomacro\gplfronttext{%
    }%
    \gplgaddtomacro\gplbacktext{%
      \csname LTb\endcsname%
      \put(1242,704){\makebox(0,0)[r]{\strut{} 0}}%
      \csname LTb\endcsname%
      \put(1242,1348){\makebox(0,0)[r]{\strut{} 5}}%
      \csname LTb\endcsname%
      \put(1242,1992){\makebox(0,0)[r]{\strut{} 10}}%
      \csname LTb\endcsname%
      \put(1242,2637){\makebox(0,0)[r]{\strut{} 15}}%
      \csname LTb\endcsname%
      \put(1374,484){\makebox(0,0){\strut{}$10^{-3}$}}%
      \csname LTb\endcsname%
      \put(1954,484){\makebox(0,0){\strut{}$10^{-2}$}}%
      \csname LTb\endcsname%
      \put(2534,484){\makebox(0,0){\strut{}$10^{-1}$}}%
      \csname LTb\endcsname%
      \put(3113,484){\makebox(0,0){\strut{}$10^{0}$}}%
      \csname LTb\endcsname%
      \put(3693,484){\makebox(0,0){\strut{}$10^{1}$}}%
      \put(604,1863){\rotatebox{-270}{\makebox(0,0){\strut{}$M_{11}(l_0, \bpi) /t$}}}%
      \put(2533,154){\makebox(0,0){\strut{}$l_0 / t$}}%
    }%
    \gplgaddtomacro\gplfronttext{%
    }%
    \gplbacktext
    \put(0,0){\includegraphics{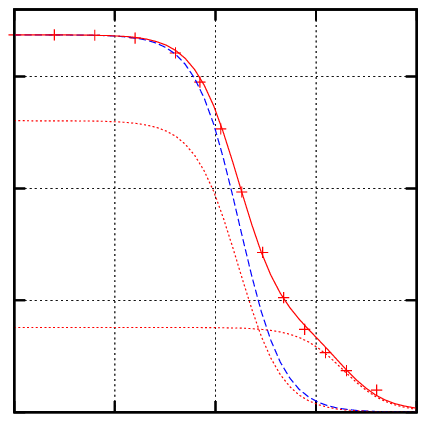}}%
    \gplfronttext
  \end{picture}%
\endgroup

%% file: FreqPap_Omcrit_compare.tex
% GNUPLOT: LaTeX picture with Postscript
\begingroup
  \makeatletter
  \providecommand\color[2][]{%
    \GenericError{(gnuplot) \space\space\space\@spaces}{%
      Package color not loaded in conjunction with
      terminal option `colourtext'%
    }{See the gnuplot documentation for explanation.%
    }{Either use 'blacktext' in gnuplot or load the package
      color.sty in LaTeX.}%
    \renewcommand\color[2][]{}%
  }%
  \providecommand\includegraphics[2][]{%
    \GenericError{(gnuplot) \space\space\space\@spaces}{%
      Package graphicx or graphics not loaded%
    }{See the gnuplot documentation for explanation.%
    }{The gnuplot epslatex terminal needs graphicx.sty or graphics.sty.}%
    \renewcommand\includegraphics[2][]{}%
  }%
  \providecommand\rotatebox[2]{#2}%
  \@ifundefined{ifGPcolor}{%
    \newif\ifGPcolor
    \GPcolortrue
  }{}%
  \@ifundefined{ifGPblacktext}{%
    \newif\ifGPblacktext
    \GPblacktexttrue
  }{}%
  % define a \g@addto@macro without @ in the name:
  \let\gplgaddtomacro\g@addto@macro
  % define empty templates for all commands taking text:
  \gdef\gplbacktext{}%
  \gdef\gplfronttext{}%
  \makeatother
  \ifGPblacktext
    % no textcolor at all
    \def\colorrgb#1{}%
    \def\colorgray#1{}%
  \else
    % gray or color?
    \ifGPcolor
      \def\colorrgb#1{\color[rgb]{#1}}%
      \def\colorgray#1{\color[gray]{#1}}%
      \expandafter\def\csname LTw\endcsname{\color{white}}%
      \expandafter\def\csname LTb\endcsname{\color{black}}%
      \expandafter\def\csname LTa\endcsname{\color{black}}%
      \expandafter\def\csname LT0\endcsname{\color[rgb]{1,0,0}}%
      \expandafter\def\csname LT1\endcsname{\color[rgb]{0,1,0}}%
      \expandafter\def\csname LT2\endcsname{\color[rgb]{0,0,1}}%
      \expandafter\def\csname LT3\endcsname{\color[rgb]{1,0,1}}%
      \expandafter\def\csname LT4\endcsname{\color[rgb]{0,1,1}}%
      \expandafter\def\csname LT5\endcsname{\color[rgb]{1,1,0}}%
      \expandafter\def\csname LT6\endcsname{\color[rgb]{0,0,0}}%
      \expandafter\def\csname LT7\endcsname{\color[rgb]{1,0.3,0}}%
      \expandafter\def\csname LT8\endcsname{\color[rgb]{0.5,0.5,0.5}}%
    \else
      % gray
      \def\colorrgb#1{\color{black}}%
      \def\colorgray#1{\color[gray]{#1}}%
      \expandafter\def\csname LTw\endcsname{\color{white}}%
      \expandafter\def\csname LTb\endcsname{\color{black}}%
      \expandafter\def\csname LTa\endcsname{\color{black}}%
      \expandafter\def\csname LT0\endcsname{\color{black}}%
      \expandafter\def\csname LT1\endcsname{\color{black}}%
      \expandafter\def\csname LT2\endcsname{\color{black}}%
      \expandafter\def\csname LT3\endcsname{\color{black}}%
      \expandafter\def\csname LT4\endcsname{\color{black}}%
      \expandafter\def\csname LT5\endcsname{\color{black}}%
      \expandafter\def\csname LT6\endcsname{\color{black}}%
      \expandafter\def\csname LT7\endcsname{\color{black}}%
      \expandafter\def\csname LT8\endcsname{\color{black}}%
    \fi
  \fi
  \setlength{\unitlength}{0.0500bp}%
  \begin{picture}(4680.00,3276.00)%
    \gplgaddtomacro\gplbacktext{%
      \csname LTb\endcsname%
      \put(1582,704){\makebox(0,0)[r]{\strut{}$10^{-3}$}}%
      \csname LTb\endcsname%
      \put(1582,1742){\makebox(0,0)[r]{\strut{}$10^{-2}$}}%
      \csname LTb\endcsname%
      \put(1582,2780){\makebox(0,0)[r]{\strut{}$10^{-1}$}}%
      \csname LTb\endcsname%
      \put(1714,484){\makebox(0,0){\strut{} 0}}%
      \csname LTb\endcsname%
      \put(2228,484){\makebox(0,0){\strut{} 0.1}}%
      \csname LTb\endcsname%
      \put(2742,484){\makebox(0,0){\strut{} 0.2}}%
      \csname LTb\endcsname%
      \put(3257,484){\makebox(0,0){\strut{} 0.3}}%
      \csname LTb\endcsname%
      \put(3771,484){\makebox(0,0){\strut{} 0.4}}%
      \csname LTb\endcsname%
      \put(4285,484){\makebox(0,0){\strut{} 0.5}}%
      \put(944,1989){\rotatebox{-270}{\makebox(0,0){\strut{}$\Omega_*/t$}}}%
      \put(2999,154){\makebox(0,0){\strut{}$t'/t$}}%
      \put(3000,1641){\makebox(0,0){\strut{}\scriptsize{(i)}}}%
      \put(2897,2467){\makebox(0,0){\strut{}\scriptsize{(iii)}}}%
      \put(2485,3135){\makebox(0,0){\strut{}\scriptsize{(ii)}}}%
    }%
    \gplgaddtomacro\gplfronttext{%
      \csname LTb\endcsname%
      \put(1971,1199){\makebox(0,0)[l]{\strut{}\fcolorbox{white}{white}{$U = 3t$}}}%
    }%
    \gplgaddtomacro\gplbacktext{%
      \csname LTb\endcsname%
      \put(1582,704){\makebox(0,0)[r]{\strut{}$10^{-3}$}}%
      \csname LTb\endcsname%
      \put(1582,1742){\makebox(0,0)[r]{\strut{}$10^{-2}$}}%
      \csname LTb\endcsname%
      \put(1582,2780){\makebox(0,0)[r]{\strut{}$10^{-1}$}}%
      \csname LTb\endcsname%
      \put(1714,484){\makebox(0,0){\strut{} 0}}%
      \csname LTb\endcsname%
      \put(2228,484){\makebox(0,0){\strut{} 0.1}}%
      \csname LTb\endcsname%
      \put(2742,484){\makebox(0,0){\strut{} 0.2}}%
      \csname LTb\endcsname%
      \put(3257,484){\makebox(0,0){\strut{} 0.3}}%
      \csname LTb\endcsname%
      \put(3771,484){\makebox(0,0){\strut{} 0.4}}%
      \csname LTb\endcsname%
      \put(4285,484){\makebox(0,0){\strut{} 0.5}}%
      \put(944,1989){\rotatebox{-270}{\makebox(0,0){\strut{}$\Omega_*/t$}}}%
      \put(2999,154){\makebox(0,0){\strut{}$t'/t$}}%
      \put(3000,1641){\makebox(0,0){\strut{}\scriptsize{(i)}}}%
      \put(2897,2467){\makebox(0,0){\strut{}\scriptsize{(iii)}}}%
      \put(2485,3135){\makebox(0,0){\strut{}\scriptsize{(ii)}}}%
    }%
    \gplgaddtomacro\gplfronttext{%
      \csname LTb\endcsname%
      \put(1971,1199){\makebox(0,0)[l]{\strut{}\fcolorbox{white}{white}{$U = 3t$}}}%
    }%
    \gplgaddtomacro\gplbacktext{%
      \csname LTb\endcsname%
      \put(1582,704){\makebox(0,0)[r]{\strut{}$10^{-3}$}}%
      \csname LTb\endcsname%
      \put(1582,1742){\makebox(0,0)[r]{\strut{}$10^{-2}$}}%
      \csname LTb\endcsname%
      \put(1582,2780){\makebox(0,0)[r]{\strut{}$10^{-1}$}}%
      \csname LTb\endcsname%
      \put(1714,484){\makebox(0,0){\strut{} 0}}%
      \csname LTb\endcsname%
      \put(2228,484){\makebox(0,0){\strut{} 0.1}}%
      \csname LTb\endcsname%
      \put(2742,484){\makebox(0,0){\strut{} 0.2}}%
      \csname LTb\endcsname%
      \put(3257,484){\makebox(0,0){\strut{} 0.3}}%
      \csname LTb\endcsname%
      \put(3771,484){\makebox(0,0){\strut{} 0.4}}%
      \csname LTb\endcsname%
      \put(4285,484){\makebox(0,0){\strut{} 0.5}}%
      \put(944,1989){\rotatebox{-270}{\makebox(0,0){\strut{}$\Omega_*/t$}}}%
      \put(2999,154){\makebox(0,0){\strut{}$t'/t$}}%
      \put(3000,1641){\makebox(0,0){\strut{}\scriptsize{(i)}}}%
      \put(2897,2467){\makebox(0,0){\strut{}\scriptsize{(iii)}}}%
      \put(2485,3135){\makebox(0,0){\strut{}\scriptsize{(ii)}}}%
    }%
    \gplgaddtomacro\gplfronttext{%
      \csname LTb\endcsname%
      \put(1971,1199){\makebox(0,0)[l]{\strut{}\fcolorbox{white}{white}{$U = 3t$}}}%
    }%
    \gplgaddtomacro\gplbacktext{%
      \csname LTb\endcsname%
      \put(1582,704){\makebox(0,0)[r]{\strut{}$10^{-3}$}}%
      \csname LTb\endcsname%
      \put(1582,1742){\makebox(0,0)[r]{\strut{}$10^{-2}$}}%
      \csname LTb\endcsname%
      \put(1582,2780){\makebox(0,0)[r]{\strut{}$10^{-1}$}}%
      \csname LTb\endcsname%
      \put(1714,484){\makebox(0,0){\strut{} 0}}%
      \csname LTb\endcsname%
      \put(2228,484){\makebox(0,0){\strut{} 0.1}}%
      \csname LTb\endcsname%
      \put(2742,484){\makebox(0,0){\strut{} 0.2}}%
      \csname LTb\endcsname%
      \put(3257,484){\makebox(0,0){\strut{} 0.3}}%
      \csname LTb\endcsname%
      \put(3771,484){\makebox(0,0){\strut{} 0.4}}%
      \csname LTb\endcsname%
      \put(4285,484){\makebox(0,0){\strut{} 0.5}}%
      \put(944,1989){\rotatebox{-270}{\makebox(0,0){\strut{}$\Omega_*/t$}}}%
      \put(2999,154){\makebox(0,0){\strut{}$t'/t$}}%
      \put(3000,1641){\makebox(0,0){\strut{}\scriptsize{(i)}}}%
      \put(2897,2467){\makebox(0,0){\strut{}\scriptsize{(iii)}}}%
      \put(2485,3135){\makebox(0,0){\strut{}\scriptsize{(ii)}}}%
    }%
    \gplgaddtomacro\gplfronttext{%
      \csname LTb\endcsname%
      \put(1971,1199){\makebox(0,0)[l]{\strut{}\fcolorbox{white}{white}{$U = 3t$}}}%
    }%
    \gplbacktext
    \put(0,0){\includegraphics{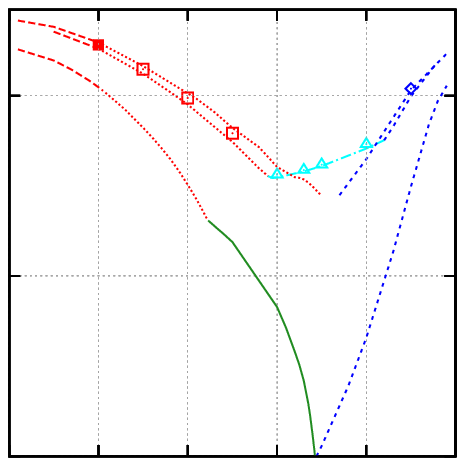}}%
    \gplfronttext
  \end{picture}%
\endgroup

%% file: FreqPap_withfreqNoSigma_Omcrit_compare.tex
% GNUPLOT: LaTeX picture with Postscript
\begingroup
  \makeatletter
  \providecommand\color[2][]{%
    \GenericError{(gnuplot) \space\space\space\@spaces}{%
      Package color not loaded in conjunction with
      terminal option `colourtext'%
    }{See the gnuplot documentation for explanation.%
    }{Either use 'blacktext' in gnuplot or load the package
      color.sty in LaTeX.}%
    \renewcommand\color[2][]{}%
  }%
  \providecommand\includegraphics[2][]{%
    \GenericError{(gnuplot) \space\space\space\@spaces}{%
      Package graphicx or graphics not loaded%
    }{See the gnuplot documentation for explanation.%
    }{The gnuplot epslatex terminal needs graphicx.sty or graphics.sty.}%
    \renewcommand\includegraphics[2][]{}%
  }%
  \providecommand\rotatebox[2]{#2}%
  \@ifundefined{ifGPcolor}{%
    \newif\ifGPcolor
    \GPcolortrue
  }{}%
  \@ifundefined{ifGPblacktext}{%
    \newif\ifGPblacktext
    \GPblacktexttrue
  }{}%
  % define a \g@addto@macro without @ in the name:
  \let\gplgaddtomacro\g@addto@macro
  % define empty templates for all commands taking text:
  \gdef\gplbacktext{}%
  \gdef\gplfronttext{}%
  \makeatother
  \ifGPblacktext
    % no textcolor at all
    \def\colorrgb#1{}%
    \def\colorgray#1{}%
  \else
    % gray or color?
    \ifGPcolor
      \def\colorrgb#1{\color[rgb]{#1}}%
      \def\colorgray#1{\color[gray]{#1}}%
      \expandafter\def\csname LTw\endcsname{\color{white}}%
      \expandafter\def\csname LTb\endcsname{\color{black}}%
      \expandafter\def\csname LTa\endcsname{\color{black}}%
      \expandafter\def\csname LT0\endcsname{\color[rgb]{1,0,0}}%
      \expandafter\def\csname LT1\endcsname{\color[rgb]{0,1,0}}%
      \expandafter\def\csname LT2\endcsname{\color[rgb]{0,0,1}}%
      \expandafter\def\csname LT3\endcsname{\color[rgb]{1,0,1}}%
      \expandafter\def\csname LT4\endcsname{\color[rgb]{0,1,1}}%
      \expandafter\def\csname LT5\endcsname{\color[rgb]{1,1,0}}%
      \expandafter\def\csname LT6\endcsname{\color[rgb]{0,0,0}}%
      \expandafter\def\csname LT7\endcsname{\color[rgb]{1,0.3,0}}%
      \expandafter\def\csname LT8\endcsname{\color[rgb]{0.5,0.5,0.5}}%
    \else
      % gray
      \def\colorrgb#1{\color{black}}%
      \def\colorgray#1{\color[gray]{#1}}%
      \expandafter\def\csname LTw\endcsname{\color{white}}%
      \expandafter\def\csname LTb\endcsname{\color{black}}%
      \expandafter\def\csname LTa\endcsname{\color{black}}%
      \expandafter\def\csname LT0\endcsname{\color{black}}%
      \expandafter\def\csname LT1\endcsname{\color{black}}%
      \expandafter\def\csname LT2\endcsname{\color{black}}%
      \expandafter\def\csname LT3\endcsname{\color{black}}%
      \expandafter\def\csname LT4\endcsname{\color{black}}%
      \expandafter\def\csname LT5\endcsname{\color{black}}%
      \expandafter\def\csname LT6\endcsname{\color{black}}%
      \expandafter\def\csname LT7\endcsname{\color{black}}%
      \expandafter\def\csname LT8\endcsname{\color{black}}%
    \fi
  \fi
  \setlength{\unitlength}{0.0500bp}%
  \begin{picture}(4680.00,3276.00)%
    \gplgaddtomacro\gplbacktext{%
      \csname LTb\endcsname%
      \put(1582,801){\makebox(0,0)[r]{\strut{}$10^{-3}$}}%
      \csname LTb\endcsname%
      \put(1582,1800){\makebox(0,0)[r]{\strut{}$10^{-2}$}}%
      \csname LTb\endcsname%
      \put(1582,2798){\makebox(0,0)[r]{\strut{}$10^{-1}$}}%
      \csname LTb\endcsname%
      \put(1714,484){\makebox(0,0){\strut{} 0}}%
      \csname LTb\endcsname%
      \put(2228,484){\makebox(0,0){\strut{} 0.1}}%
      \csname LTb\endcsname%
      \put(2742,484){\makebox(0,0){\strut{} 0.2}}%
      \csname LTb\endcsname%
      \put(3257,484){\makebox(0,0){\strut{} 0.3}}%
      \csname LTb\endcsname%
      \put(3771,484){\makebox(0,0){\strut{} 0.4}}%
      \csname LTb\endcsname%
      \put(4285,484){\makebox(0,0){\strut{} 0.5}}%
      \put(944,1989){\rotatebox{-270}{\makebox(0,0){\strut{}$\Omega_*/t$}}}%
      \put(2999,154){\makebox(0,0){\strut{}$t'/t$}}%
    }%
    \gplgaddtomacro\gplfronttext{%
    }%
    \gplgaddtomacro\gplbacktext{%
      \csname LTb\endcsname%
      \put(1582,801){\makebox(0,0)[r]{\strut{}$10^{-3}$}}%
      \csname LTb\endcsname%
      \put(1582,1800){\makebox(0,0)[r]{\strut{}$10^{-2}$}}%
      \csname LTb\endcsname%
      \put(1582,2798){\makebox(0,0)[r]{\strut{}$10^{-1}$}}%
      \csname LTb\endcsname%
      \put(1714,484){\makebox(0,0){\strut{} 0}}%
      \csname LTb\endcsname%
      \put(2228,484){\makebox(0,0){\strut{} 0.1}}%
      \csname LTb\endcsname%
      \put(2742,484){\makebox(0,0){\strut{} 0.2}}%
      \csname LTb\endcsname%
      \put(3257,484){\makebox(0,0){\strut{} 0.3}}%
      \csname LTb\endcsname%
      \put(3771,484){\makebox(0,0){\strut{} 0.4}}%
      \csname LTb\endcsname%
      \put(4285,484){\makebox(0,0){\strut{} 0.5}}%
      \put(944,1989){\rotatebox{-270}{\makebox(0,0){\strut{}$\Omega_*/t$}}}%
      \put(2999,154){\makebox(0,0){\strut{}$t'/t$}}%
    }%
    \gplgaddtomacro\gplfronttext{%
    }%
    \gplgaddtomacro\gplbacktext{%
      \csname LTb\endcsname%
      \put(1582,801){\makebox(0,0)[r]{\strut{}$10^{-3}$}}%
      \csname LTb\endcsname%
      \put(1582,1800){\makebox(0,0)[r]{\strut{}$10^{-2}$}}%
      \csname LTb\endcsname%
      \put(1582,2798){\makebox(0,0)[r]{\strut{}$10^{-1}$}}%
      \csname LTb\endcsname%
      \put(1714,484){\makebox(0,0){\strut{} 0}}%
      \csname LTb\endcsname%
      \put(2228,484){\makebox(0,0){\strut{} 0.1}}%
      \csname LTb\endcsname%
      \put(2742,484){\makebox(0,0){\strut{} 0.2}}%
      \csname LTb\endcsname%
      \put(3257,484){\makebox(0,0){\strut{} 0.3}}%
      \csname LTb\endcsname%
      \put(3771,484){\makebox(0,0){\strut{} 0.4}}%
      \csname LTb\endcsname%
      \put(4285,484){\makebox(0,0){\strut{} 0.5}}%
      \put(944,1989){\rotatebox{-270}{\makebox(0,0){\strut{}$\Omega_*/t$}}}%
      \put(2999,154){\makebox(0,0){\strut{}$t'/t$}}%
    }%
    \gplgaddtomacro\gplfronttext{%
    }%
    \gplbacktext
    \put(0,0){\includegraphics{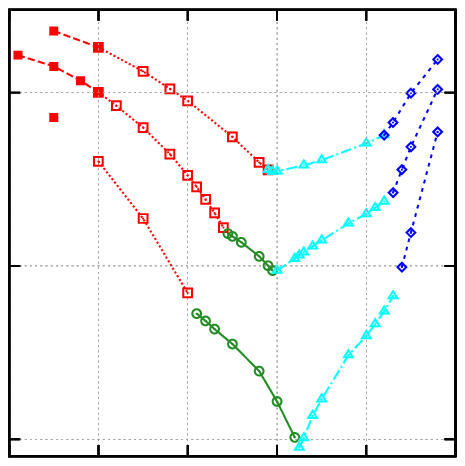}}%
    \gplfronttext
  \end{picture}%
\endgroup

%% file: FreqPap_K11strange_freqdep1.tex
% GNUPLOT: LaTeX picture with Postscript
\begingroup
  \makeatletter
  \providecommand\color[2][]{%
    \GenericError{(gnuplot) \space\space\space\@spaces}{%
      Package color not loaded in conjunction with
      terminal option `colourtext'%
    }{See the gnuplot documentation for explanation.%
    }{Either use 'blacktext' in gnuplot or load the package
      color.sty in LaTeX.}%
    \renewcommand\color[2][]{}%
  }%
  \providecommand\includegraphics[2][]{%
    \GenericError{(gnuplot) \space\space\space\@spaces}{%
      Package graphicx or graphics not loaded%
    }{See the gnuplot documentation for explanation.%
    }{The gnuplot epslatex terminal needs graphicx.sty or graphics.sty.}%
    \renewcommand\includegraphics[2][]{}%
  }%
  \providecommand\rotatebox[2]{#2}%
  \@ifundefined{ifGPcolor}{%
    \newif\ifGPcolor
    \GPcolortrue
  }{}%
  \@ifundefined{ifGPblacktext}{%
    \newif\ifGPblacktext
    \GPblacktexttrue
  }{}%
  % define a \g@addto@macro without @ in the name:
  \let\gplgaddtomacro\g@addto@macro
  % define empty templates for all commands taking text:
  \gdef\gplbacktext{}%
  \gdef\gplfronttext{}%
  \makeatother
  \ifGPblacktext
    % no textcolor at all
    \def\colorrgb#1{}%
    \def\colorgray#1{}%
  \else
    % gray or color?
    \ifGPcolor
      \def\colorrgb#1{\color[rgb]{#1}}%
      \def\colorgray#1{\color[gray]{#1}}%
      \expandafter\def\csname LTw\endcsname{\color{white}}%
      \expandafter\def\csname LTb\endcsname{\color{black}}%
      \expandafter\def\csname LTa\endcsname{\color{black}}%
      \expandafter\def\csname LT0\endcsname{\color[rgb]{1,0,0}}%
      \expandafter\def\csname LT1\endcsname{\color[rgb]{0,1,0}}%
      \expandafter\def\csname LT2\endcsname{\color[rgb]{0,0,1}}%
      \expandafter\def\csname LT3\endcsname{\color[rgb]{1,0,1}}%
      \expandafter\def\csname LT4\endcsname{\color[rgb]{0,1,1}}%
      \expandafter\def\csname LT5\endcsname{\color[rgb]{1,1,0}}%
      \expandafter\def\csname LT6\endcsname{\color[rgb]{0,0,0}}%
      \expandafter\def\csname LT7\endcsname{\color[rgb]{1,0.3,0}}%
      \expandafter\def\csname LT8\endcsname{\color[rgb]{0.5,0.5,0.5}}%
    \else
      % gray
      \def\colorrgb#1{\color{black}}%
      \def\colorgray#1{\color[gray]{#1}}%
      \expandafter\def\csname LTw\endcsname{\color{white}}%
      \expandafter\def\csname LTb\endcsname{\color{black}}%
      \expandafter\def\csname LTa\endcsname{\color{black}}%
      \expandafter\def\csname LT0\endcsname{\color{black}}%
      \expandafter\def\csname LT1\endcsname{\color{black}}%
      \expandafter\def\csname LT2\endcsname{\color{black}}%
      \expandafter\def\csname LT3\endcsname{\color{black}}%
      \expandafter\def\csname LT4\endcsname{\color{black}}%
      \expandafter\def\csname LT5\endcsname{\color{black}}%
      \expandafter\def\csname LT6\endcsname{\color{black}}%
      \expandafter\def\csname LT7\endcsname{\color{black}}%
      \expandafter\def\csname LT8\endcsname{\color{black}}%
    \fi
  \fi
  \setlength{\unitlength}{0.0500bp}%
  \begin{picture}(4320.00,3024.00)%
    \gplgaddtomacro\gplbacktext{%
      \csname LTb\endcsname%
      \put(1242,1157){\makebox(0,0)[r]{\strut{} 0}}%
      \csname LTb\endcsname%
      \put(1242,1624){\makebox(0,0)[r]{\strut{} 0.4}}%
      \csname LTb\endcsname%
      \put(1242,2090){\makebox(0,0)[r]{\strut{} 0.8}}%
      \csname LTb\endcsname%
      \put(1242,2557){\makebox(0,0)[r]{\strut{} 1.2}}%
      \csname LTb\endcsname%
      \put(1242,3023){\makebox(0,0)[r]{\strut{} 1.6}}%
      \csname LTb\endcsname%
      \put(1374,484){\makebox(0,0){\strut{}$10^{-3}$}}%
      \csname LTb\endcsname%
      \put(1838,484){\makebox(0,0){\strut{}$10^{-2}$}}%
      \csname LTb\endcsname%
      \put(2302,484){\makebox(0,0){\strut{}$10^{-1}$}}%
      \csname LTb\endcsname%
      \put(2765,484){\makebox(0,0){\strut{}$10^{0}$}}%
      \csname LTb\endcsname%
      \put(3229,484){\makebox(0,0){\strut{}$10^{1}$}}%
      \csname LTb\endcsname%
      \put(3693,484){\makebox(0,0){\strut{}$10^{2}$}}%
      \put(604,1863){\rotatebox{-270}{\makebox(0,0){\strut{}$K_{11}(l_0, \bzero) /t$}}}%
      \put(2533,154){\makebox(0,0){\strut{}$l_0 / t$}}%
    }%
    \gplgaddtomacro\gplfronttext{%
    }%
    \gplgaddtomacro\gplbacktext{%
      \csname LTb\endcsname%
      \put(1242,1160){\makebox(0,0)[r]{\strut{} 0}}%
      \csname LTb\endcsname%
      \put(1242,1626){\makebox(0,0)[r]{\strut{} 0.4}}%
      \csname LTb\endcsname%
      \put(1242,2091){\makebox(0,0)[r]{\strut{} 0.8}}%
      \csname LTb\endcsname%
      \put(1242,2557){\makebox(0,0)[r]{\strut{} 1.2}}%
      \csname LTb\endcsname%
      \put(1242,3023){\makebox(0,0)[r]{\strut{} 1.6}}%
      \csname LTb\endcsname%
      \put(1374,484){\makebox(0,0){\strut{}$10^{-3}$}}%
      \csname LTb\endcsname%
      \put(1838,484){\makebox(0,0){\strut{}$10^{-2}$}}%
      \csname LTb\endcsname%
      \put(2302,484){\makebox(0,0){\strut{}$10^{-1}$}}%
      \csname LTb\endcsname%
      \put(2765,484){\makebox(0,0){\strut{}$10^{0}$}}%
      \csname LTb\endcsname%
      \put(3229,484){\makebox(0,0){\strut{}$10^{1}$}}%
      \csname LTb\endcsname%
      \put(3693,484){\makebox(0,0){\strut{}$10^{2}$}}%
      \put(604,1863){\rotatebox{-270}{\makebox(0,0){\strut{}$K_{11}(l_0, \bzero) /t$}}}%
      \put(2533,154){\makebox(0,0){\strut{}$l_0 / t$}}%
    }%
    \gplgaddtomacro\gplfronttext{%
    }%
    \gplbacktext
    \put(0,0){\includegraphics{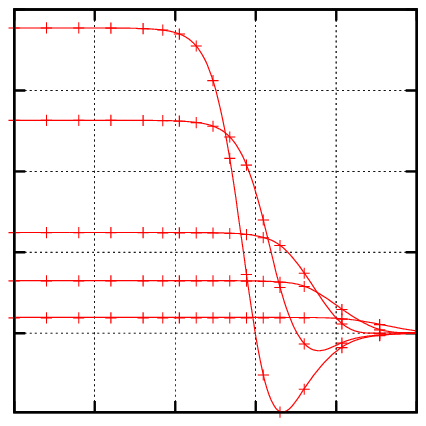}}%
    \gplfronttext
  \end{picture}%
\endgroup

%% file: FreqPap_K11strange_freqdep2.tex
% GNUPLOT: LaTeX picture with Postscript
\begingroup
  \makeatletter
  \providecommand\color[2][]{%
    \GenericError{(gnuplot) \space\space\space\@spaces}{%
      Package color not loaded in conjunction with
      terminal option `colourtext'%
    }{See the gnuplot documentation for explanation.%
    }{Either use 'blacktext' in gnuplot or load the package
      color.sty in LaTeX.}%
    \renewcommand\color[2][]{}%
  }%
  \providecommand\includegraphics[2][]{%
    \GenericError{(gnuplot) \space\space\space\@spaces}{%
      Package graphicx or graphics not loaded%
    }{See the gnuplot documentation for explanation.%
    }{The gnuplot epslatex terminal needs graphicx.sty or graphics.sty.}%
    \renewcommand\includegraphics[2][]{}%
  }%
  \providecommand\rotatebox[2]{#2}%
  \@ifundefined{ifGPcolor}{%
    \newif\ifGPcolor
    \GPcolortrue
  }{}%
  \@ifundefined{ifGPblacktext}{%
    \newif\ifGPblacktext
    \GPblacktexttrue
  }{}%
  % define a \g@addto@macro without @ in the name:
  \let\gplgaddtomacro\g@addto@macro
  % define empty templates for all commands taking text:
  \gdef\gplbacktext{}%
  \gdef\gplfronttext{}%
  \makeatother
  \ifGPblacktext
    % no textcolor at all
    \def\colorrgb#1{}%
    \def\colorgray#1{}%
  \else
    % gray or color?
    \ifGPcolor
      \def\colorrgb#1{\color[rgb]{#1}}%
      \def\colorgray#1{\color[gray]{#1}}%
      \expandafter\def\csname LTw\endcsname{\color{white}}%
      \expandafter\def\csname LTb\endcsname{\color{black}}%
      \expandafter\def\csname LTa\endcsname{\color{black}}%
      \expandafter\def\csname LT0\endcsname{\color[rgb]{1,0,0}}%
      \expandafter\def\csname LT1\endcsname{\color[rgb]{0,1,0}}%
      \expandafter\def\csname LT2\endcsname{\color[rgb]{0,0,1}}%
      \expandafter\def\csname LT3\endcsname{\color[rgb]{1,0,1}}%
      \expandafter\def\csname LT4\endcsname{\color[rgb]{0,1,1}}%
      \expandafter\def\csname LT5\endcsname{\color[rgb]{1,1,0}}%
      \expandafter\def\csname LT6\endcsname{\color[rgb]{0,0,0}}%
      \expandafter\def\csname LT7\endcsname{\color[rgb]{1,0.3,0}}%
      \expandafter\def\csname LT8\endcsname{\color[rgb]{0.5,0.5,0.5}}%
    \else
      % gray
      \def\colorrgb#1{\color{black}}%
      \def\colorgray#1{\color[gray]{#1}}%
      \expandafter\def\csname LTw\endcsname{\color{white}}%
      \expandafter\def\csname LTb\endcsname{\color{black}}%
      \expandafter\def\csname LTa\endcsname{\color{black}}%
      \expandafter\def\csname LT0\endcsname{\color{black}}%
      \expandafter\def\csname LT1\endcsname{\color{black}}%
      \expandafter\def\csname LT2\endcsname{\color{black}}%
      \expandafter\def\csname LT3\endcsname{\color{black}}%
      \expandafter\def\csname LT4\endcsname{\color{black}}%
      \expandafter\def\csname LT5\endcsname{\color{black}}%
      \expandafter\def\csname LT6\endcsname{\color{black}}%
      \expandafter\def\csname LT7\endcsname{\color{black}}%
      \expandafter\def\csname LT8\endcsname{\color{black}}%
    \fi
  \fi
  \setlength{\unitlength}{0.0500bp}%
  \begin{picture}(4320.00,3024.00)%
    \gplgaddtomacro\gplbacktext{%
      \csname LTb\endcsname%
      \put(1176,876){\makebox(0,0)[r]{\strut{}-20}}%
      \csname LTb\endcsname%
      \put(1176,1305){\makebox(0,0)[r]{\strut{}-15}}%
      \csname LTb\endcsname%
      \put(1176,1735){\makebox(0,0)[r]{\strut{}-10}}%
      \csname LTb\endcsname%
      \put(1176,2164){\makebox(0,0)[r]{\strut{}-5}}%
      \csname LTb\endcsname%
      \put(1176,2594){\makebox(0,0)[r]{\strut{} 0}}%
      \csname LTb\endcsname%
      \put(1176,3023){\makebox(0,0)[r]{\strut{} 5}}%
      \csname LTb\endcsname%
      \put(1308,484){\makebox(0,0){\strut{}$10^{-3}$}}%
      \csname LTb\endcsname%
      \put(1772,484){\makebox(0,0){\strut{}$10^{-2}$}}%
      \csname LTb\endcsname%
      \put(2236,484){\makebox(0,0){\strut{}$10^{-1}$}}%
      \csname LTb\endcsname%
      \put(2699,484){\makebox(0,0){\strut{}$10^{0}$}}%
      \csname LTb\endcsname%
      \put(3163,484){\makebox(0,0){\strut{}$10^{1}$}}%
      \csname LTb\endcsname%
      \put(3627,484){\makebox(0,0){\strut{}$10^{2}$}}%
      \put(670,1863){\rotatebox{-270}{\makebox(0,0){\strut{}$K_{11}(l_0, \bzero) /t$}}}%
      \put(2467,154){\makebox(0,0){\strut{}$l_0 / t$}}%
    }%
    \gplgaddtomacro\gplfronttext{%
    }%
    \gplgaddtomacro\gplbacktext{%
      \csname LTb\endcsname%
      \put(1176,876){\makebox(0,0)[r]{\strut{}-20}}%
      \csname LTb\endcsname%
      \put(1176,1305){\makebox(0,0)[r]{\strut{}-15}}%
      \csname LTb\endcsname%
      \put(1176,1735){\makebox(0,0)[r]{\strut{}-10}}%
      \csname LTb\endcsname%
      \put(1176,2164){\makebox(0,0)[r]{\strut{}-5}}%
      \csname LTb\endcsname%
      \put(1176,2594){\makebox(0,0)[r]{\strut{} 0}}%
      \csname LTb\endcsname%
      \put(1176,3023){\makebox(0,0)[r]{\strut{} 5}}%
      \csname LTb\endcsname%
      \put(1308,484){\makebox(0,0){\strut{}$10^{-3}$}}%
      \csname LTb\endcsname%
      \put(1772,484){\makebox(0,0){\strut{}$10^{-2}$}}%
      \csname LTb\endcsname%
      \put(2236,484){\makebox(0,0){\strut{}$10^{-1}$}}%
      \csname LTb\endcsname%
      \put(2699,484){\makebox(0,0){\strut{}$10^{0}$}}%
      \csname LTb\endcsname%
      \put(3163,484){\makebox(0,0){\strut{}$10^{1}$}}%
      \csname LTb\endcsname%
      \put(3627,484){\makebox(0,0){\strut{}$10^{2}$}}%
      \put(670,1863){\rotatebox{-270}{\makebox(0,0){\strut{}$K_{11}(l_0, \bzero) /t$}}}%
      \put(2467,154){\makebox(0,0){\strut{}$l_0 / t$}}%
    }%
    \gplgaddtomacro\gplfronttext{%
    }%
    \gplbacktext
    \put(0,0){\includegraphics{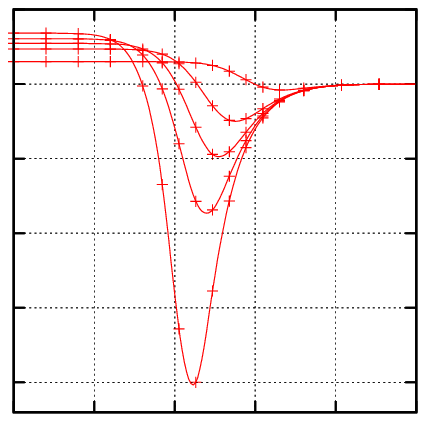}}%
    \gplfronttext
  \end{picture}%
\endgroup

%% file: FreqPap_LeadingFlow_with_ImDmm.tex
% GNUPLOT: LaTeX picture with Postscript
\begingroup
  \makeatletter
  \providecommand\color[2][]{%
    \GenericError{(gnuplot) \space\space\space\@spaces}{%
      Package color not loaded in conjunction with
      terminal option `colourtext'%
    }{See the gnuplot documentation for explanation.%
    }{Either use 'blacktext' in gnuplot or load the package
      color.sty in LaTeX.}%
    \renewcommand\color[2][]{}%
  }%
  \providecommand\includegraphics[2][]{%
    \GenericError{(gnuplot) \space\space\space\@spaces}{%
      Package graphicx or graphics not loaded%
    }{See the gnuplot documentation for explanation.%
    }{The gnuplot epslatex terminal needs graphicx.sty or graphics.sty.}%
    \renewcommand\includegraphics[2][]{}%
  }%
  \providecommand\rotatebox[2]{#2}%
  \@ifundefined{ifGPcolor}{%
    \newif\ifGPcolor
    \GPcolortrue
  }{}%
  \@ifundefined{ifGPblacktext}{%
    \newif\ifGPblacktext
    \GPblacktexttrue
  }{}%
  % define a \g@addto@macro without @ in the name:
  \let\gplgaddtomacro\g@addto@macro
  % define empty templates for all commands taking text:
  \gdef\gplbacktext{}%
  \gdef\gplfronttext{}%
  \makeatother
  \ifGPblacktext
    % no textcolor at all
    \def\colorrgb#1{}%
    \def\colorgray#1{}%
  \else
    % gray or color?
    \ifGPcolor
      \def\colorrgb#1{\color[rgb]{#1}}%
      \def\colorgray#1{\color[gray]{#1}}%
      \expandafter\def\csname LTw\endcsname{\color{white}}%
      \expandafter\def\csname LTb\endcsname{\color{black}}%
      \expandafter\def\csname LTa\endcsname{\color{black}}%
      \expandafter\def\csname LT0\endcsname{\color[rgb]{1,0,0}}%
      \expandafter\def\csname LT1\endcsname{\color[rgb]{0,1,0}}%
      \expandafter\def\csname LT2\endcsname{\color[rgb]{0,0,1}}%
      \expandafter\def\csname LT3\endcsname{\color[rgb]{1,0,1}}%
      \expandafter\def\csname LT4\endcsname{\color[rgb]{0,1,1}}%
      \expandafter\def\csname LT5\endcsname{\color[rgb]{1,1,0}}%
      \expandafter\def\csname LT6\endcsname{\color[rgb]{0,0,0}}%
      \expandafter\def\csname LT7\endcsname{\color[rgb]{1,0.3,0}}%
      \expandafter\def\csname LT8\endcsname{\color[rgb]{0.5,0.5,0.5}}%
    \else
      % gray
      \def\colorrgb#1{\color{black}}%
      \def\colorgray#1{\color[gray]{#1}}%
      \expandafter\def\csname LTw\endcsname{\color{white}}%
      \expandafter\def\csname LTb\endcsname{\color{black}}%
      \expandafter\def\csname LTa\endcsname{\color{black}}%
      \expandafter\def\csname LT0\endcsname{\color{black}}%
      \expandafter\def\csname LT1\endcsname{\color{black}}%
      \expandafter\def\csname LT2\endcsname{\color{black}}%
      \expandafter\def\csname LT3\endcsname{\color{black}}%
      \expandafter\def\csname LT4\endcsname{\color{black}}%
      \expandafter\def\csname LT5\endcsname{\color{black}}%
      \expandafter\def\csname LT6\endcsname{\color{black}}%
      \expandafter\def\csname LT7\endcsname{\color{black}}%
      \expandafter\def\csname LT8\endcsname{\color{black}}%
    \fi
  \fi
  \setlength{\unitlength}{0.0500bp}%
  \begin{picture}(4320.00,3024.00)%
    \gplgaddtomacro\gplbacktext{%
      \csname LTb\endcsname%
      \put(1132,704){\makebox(0,0)[r]{\strut{} 0}}%
      \csname LTb\endcsname%
      \put(1132,1284){\makebox(0,0)[r]{\strut{} 5}}%
      \csname LTb\endcsname%
      \put(1132,1864){\makebox(0,0)[r]{\strut{} 10}}%
      \csname LTb\endcsname%
      \put(1132,2443){\makebox(0,0)[r]{\strut{} 15}}%
      \csname LTb\endcsname%
      \put(1132,3023){\makebox(0,0)[r]{\strut{} 20}}%
      \csname LTb\endcsname%
      \put(1745,484){\makebox(0,0){\strut{}$10^{-1}$}}%
      \csname LTb\endcsname%
      \put(2664,484){\makebox(0,0){\strut{}$10^{0}$}}%
      \csname LTb\endcsname%
      \put(3583,484){\makebox(0,0){\strut{}$10^{1}$}}%
      \put(758,1863){\rotatebox{-270}{\makebox(0,0){\strut{}$B_{mm} /t$}}}%
      \put(2423,154){\makebox(0,0){\strut{}$\Om/t$}}%
    }%
    \gplgaddtomacro\gplfronttext{%
      \csname LTb\endcsname%
      \put(3541,2675){\makebox(0,0)[r]{\strut{}\wbox{$t' = 0.3t$, $U = 3t$}}}%
    }%
    \gplgaddtomacro\gplbacktext{%
      \csname LTb\endcsname%
      \put(1132,704){\makebox(0,0)[r]{\strut{} 0}}%
      \csname LTb\endcsname%
      \put(1132,1284){\makebox(0,0)[r]{\strut{} 5}}%
      \csname LTb\endcsname%
      \put(1132,1864){\makebox(0,0)[r]{\strut{} 10}}%
      \csname LTb\endcsname%
      \put(1132,2443){\makebox(0,0)[r]{\strut{} 15}}%
      \csname LTb\endcsname%
      \put(1132,3023){\makebox(0,0)[r]{\strut{} 20}}%
      \csname LTb\endcsname%
      \put(1745,484){\makebox(0,0){\strut{}$10^{-1}$}}%
      \csname LTb\endcsname%
      \put(2664,484){\makebox(0,0){\strut{}$10^{0}$}}%
      \csname LTb\endcsname%
      \put(3583,484){\makebox(0,0){\strut{}$10^{1}$}}%
      \put(758,1863){\rotatebox{-270}{\makebox(0,0){\strut{}$B_{mm} /t$}}}%
      \put(2423,154){\makebox(0,0){\strut{}$\Om/t$}}%
    }%
    \gplgaddtomacro\gplfronttext{%
      \csname LTb\endcsname%
      \put(3541,2675){\makebox(0,0)[r]{\strut{}\wbox{$t' = 0.3t$, $U = 3t$}}}%
    }%
    \gplbacktext
    \put(0,0){\includegraphics{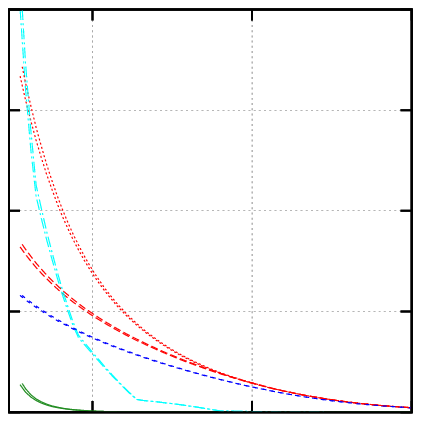}}%
    \gplfronttext
  \end{picture}%
\endgroup

%% file: FreqPap_precompT_2htol_light_t20_35_leading_flow.tex
% GNUPLOT: LaTeX picture with Postscript
\begingroup
  \makeatletter
  \providecommand\color[2][]{%
    \GenericError{(gnuplot) \space\space\space\@spaces}{%
      Package color not loaded in conjunction with
      terminal option `colourtext'%
    }{See the gnuplot documentation for explanation.%
    }{Either use 'blacktext' in gnuplot or load the package
      color.sty in LaTeX.}%
    \renewcommand\color[2][]{}%
  }%
  \providecommand\includegraphics[2][]{%
    \GenericError{(gnuplot) \space\space\space\@spaces}{%
      Package graphicx or graphics not loaded%
    }{See the gnuplot documentation for explanation.%
    }{The gnuplot epslatex terminal needs graphicx.sty or graphics.sty.}%
    \renewcommand\includegraphics[2][]{}%
  }%
  \providecommand\rotatebox[2]{#2}%
  \@ifundefined{ifGPcolor}{%
    \newif\ifGPcolor
    \GPcolortrue
  }{}%
  \@ifundefined{ifGPblacktext}{%
    \newif\ifGPblacktext
    \GPblacktexttrue
  }{}%
  % define a \g@addto@macro without @ in the name:
  \let\gplgaddtomacro\g@addto@macro
  % define empty templates for all commands taking text:
  \gdef\gplbacktext{}%
  \gdef\gplfronttext{}%
  \makeatother
  \ifGPblacktext
    % no textcolor at all
    \def\colorrgb#1{}%
    \def\colorgray#1{}%
  \else
    % gray or color?
    \ifGPcolor
      \def\colorrgb#1{\color[rgb]{#1}}%
      \def\colorgray#1{\color[gray]{#1}}%
      \expandafter\def\csname LTw\endcsname{\color{white}}%
      \expandafter\def\csname LTb\endcsname{\color{black}}%
      \expandafter\def\csname LTa\endcsname{\color{black}}%
      \expandafter\def\csname LT0\endcsname{\color[rgb]{1,0,0}}%
      \expandafter\def\csname LT1\endcsname{\color[rgb]{0,1,0}}%
      \expandafter\def\csname LT2\endcsname{\color[rgb]{0,0,1}}%
      \expandafter\def\csname LT3\endcsname{\color[rgb]{1,0,1}}%
      \expandafter\def\csname LT4\endcsname{\color[rgb]{0,1,1}}%
      \expandafter\def\csname LT5\endcsname{\color[rgb]{1,1,0}}%
      \expandafter\def\csname LT6\endcsname{\color[rgb]{0,0,0}}%
      \expandafter\def\csname LT7\endcsname{\color[rgb]{1,0.3,0}}%
      \expandafter\def\csname LT8\endcsname{\color[rgb]{0.5,0.5,0.5}}%
    \else
      % gray
      \def\colorrgb#1{\color{black}}%
      \def\colorgray#1{\color[gray]{#1}}%
      \expandafter\def\csname LTw\endcsname{\color{white}}%
      \expandafter\def\csname LTb\endcsname{\color{black}}%
      \expandafter\def\csname LTa\endcsname{\color{black}}%
      \expandafter\def\csname LT0\endcsname{\color{black}}%
      \expandafter\def\csname LT1\endcsname{\color{black}}%
      \expandafter\def\csname LT2\endcsname{\color{black}}%
      \expandafter\def\csname LT3\endcsname{\color{black}}%
      \expandafter\def\csname LT4\endcsname{\color{black}}%
      \expandafter\def\csname LT5\endcsname{\color{black}}%
      \expandafter\def\csname LT6\endcsname{\color{black}}%
      \expandafter\def\csname LT7\endcsname{\color{black}}%
      \expandafter\def\csname LT8\endcsname{\color{black}}%
    \fi
  \fi
  \setlength{\unitlength}{0.0500bp}%
  \begin{picture}(4320.00,3024.00)%
    \gplgaddtomacro\gplbacktext{%
      \csname LTb\endcsname%
      \put(1132,704){\makebox(0,0)[r]{\strut{} 0}}%
      \csname LTb\endcsname%
      \put(1132,1284){\makebox(0,0)[r]{\strut{} 5}}%
      \csname LTb\endcsname%
      \put(1132,1864){\makebox(0,0)[r]{\strut{} 10}}%
      \csname LTb\endcsname%
      \put(1132,2443){\makebox(0,0)[r]{\strut{} 15}}%
      \csname LTb\endcsname%
      \put(1132,3023){\makebox(0,0)[r]{\strut{} 20}}%
      \csname LTb\endcsname%
      \put(1264,484){\makebox(0,0){\strut{}$10^{-1}$}}%
      \csname LTb\endcsname%
      \put(2424,484){\makebox(0,0){\strut{}$10^{0}$}}%
      \csname LTb\endcsname%
      \put(3583,484){\makebox(0,0){\strut{}$10^{1}$}}%
      \put(758,1863){\rotatebox{-270}{\makebox(0,0){\strut{}$B_{mm}/t$}}}%
      \put(2423,154){\makebox(0,0){\strut{}$T/t$}}%
    }%
    \gplgaddtomacro\gplfronttext{%
      \csname LTb\endcsname%
      \put(3530,2675){\makebox(0,0)[r]{\strut{}\wbox{$t' = 0.35t$}}}%
      \put(3530,2385){\makebox(0,0)[r]{\strut{}\wbox{$U = 3t$}}}%
    }%
    \gplbacktext
    \put(0,0){\includegraphics{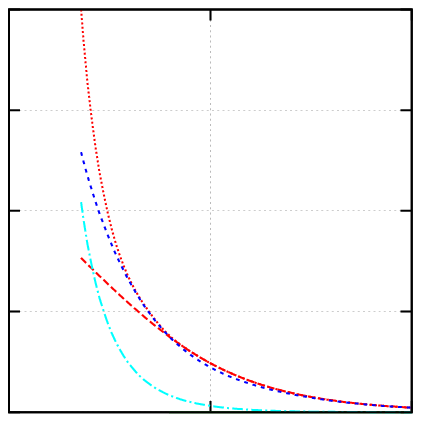}}%
    \gplfronttext
  \end{picture}%
\endgroup

%% file: FreqPap_precompT_2htol_light_t20_35_K11_flow.tex
% GNUPLOT: LaTeX picture with Postscript
\begingroup
  \makeatletter
  \providecommand\color[2][]{%
    \GenericError{(gnuplot) \space\space\space\@spaces}{%
      Package color not loaded in conjunction with
      terminal option `colourtext'%
    }{See the gnuplot documentation for explanation.%
    }{Either use 'blacktext' in gnuplot or load the package
      color.sty in LaTeX.}%
    \renewcommand\color[2][]{}%
  }%
  \providecommand\includegraphics[2][]{%
    \GenericError{(gnuplot) \space\space\space\@spaces}{%
      Package graphicx or graphics not loaded%
    }{See the gnuplot documentation for explanation.%
    }{The gnuplot epslatex terminal needs graphicx.sty or graphics.sty.}%
    \renewcommand\includegraphics[2][]{}%
  }%
  \providecommand\rotatebox[2]{#2}%
  \@ifundefined{ifGPcolor}{%
    \newif\ifGPcolor
    \GPcolortrue
  }{}%
  \@ifundefined{ifGPblacktext}{%
    \newif\ifGPblacktext
    \GPblacktexttrue
  }{}%
  % define a \g@addto@macro without @ in the name:
  \let\gplgaddtomacro\g@addto@macro
  % define empty templates for all commands taking text:
  \gdef\gplbacktext{}%
  \gdef\gplfronttext{}%
  \makeatother
  \ifGPblacktext
    % no textcolor at all
    \def\colorrgb#1{}%
    \def\colorgray#1{}%
  \else
    % gray or color?
    \ifGPcolor
      \def\colorrgb#1{\color[rgb]{#1}}%
      \def\colorgray#1{\color[gray]{#1}}%
      \expandafter\def\csname LTw\endcsname{\color{white}}%
      \expandafter\def\csname LTb\endcsname{\color{black}}%
      \expandafter\def\csname LTa\endcsname{\color{black}}%
      \expandafter\def\csname LT0\endcsname{\color[rgb]{1,0,0}}%
      \expandafter\def\csname LT1\endcsname{\color[rgb]{0,1,0}}%
      \expandafter\def\csname LT2\endcsname{\color[rgb]{0,0,1}}%
      \expandafter\def\csname LT3\endcsname{\color[rgb]{1,0,1}}%
      \expandafter\def\csname LT4\endcsname{\color[rgb]{0,1,1}}%
      \expandafter\def\csname LT5\endcsname{\color[rgb]{1,1,0}}%
      \expandafter\def\csname LT6\endcsname{\color[rgb]{0,0,0}}%
      \expandafter\def\csname LT7\endcsname{\color[rgb]{1,0.3,0}}%
      \expandafter\def\csname LT8\endcsname{\color[rgb]{0.5,0.5,0.5}}%
    \else
      % gray
      \def\colorrgb#1{\color{black}}%
      \def\colorgray#1{\color[gray]{#1}}%
      \expandafter\def\csname LTw\endcsname{\color{white}}%
      \expandafter\def\csname LTb\endcsname{\color{black}}%
      \expandafter\def\csname LTa\endcsname{\color{black}}%
      \expandafter\def\csname LT0\endcsname{\color{black}}%
      \expandafter\def\csname LT1\endcsname{\color{black}}%
      \expandafter\def\csname LT2\endcsname{\color{black}}%
      \expandafter\def\csname LT3\endcsname{\color{black}}%
      \expandafter\def\csname LT4\endcsname{\color{black}}%
      \expandafter\def\csname LT5\endcsname{\color{black}}%
      \expandafter\def\csname LT6\endcsname{\color{black}}%
      \expandafter\def\csname LT7\endcsname{\color{black}}%
      \expandafter\def\csname LT8\endcsname{\color{black}}%
    \fi
  \fi
  \setlength{\unitlength}{0.0500bp}%
  \begin{picture}(4320.00,3024.00)%
    \gplgaddtomacro\gplbacktext{%
      \csname LTb\endcsname%
      \put(1132,704){\makebox(0,0)[r]{\strut{}-12}}%
      \csname LTb\endcsname%
      \put(1132,1035){\makebox(0,0)[r]{\strut{}-10}}%
      \csname LTb\endcsname%
      \put(1132,1367){\makebox(0,0)[r]{\strut{}-8}}%
      \csname LTb\endcsname%
      \put(1132,1698){\makebox(0,0)[r]{\strut{}-6}}%
      \csname LTb\endcsname%
      \put(1132,2029){\makebox(0,0)[r]{\strut{}-4}}%
      \csname LTb\endcsname%
      \put(1132,2360){\makebox(0,0)[r]{\strut{}-2}}%
      \csname LTb\endcsname%
      \put(1132,2692){\makebox(0,0)[r]{\strut{} 0}}%
      \csname LTb\endcsname%
      \put(1132,3023){\makebox(0,0)[r]{\strut{} 2}}%
      \csname LTb\endcsname%
      \put(1264,484){\makebox(0,0){\strut{}$10^{-1}$}}%
      \csname LTb\endcsname%
      \put(2424,484){\makebox(0,0){\strut{}$10^{0}$}}%
      \csname LTb\endcsname%
      \put(3583,484){\makebox(0,0){\strut{}$10^{1}$}}%
      \put(758,1863){\rotatebox{-270}{\makebox(0,0){\strut{}$K_{11}( \om, \bzero) /t$}}}%
      \put(2423,154){\makebox(0,0){\strut{}$T/t$}}%
    }%
    \gplgaddtomacro\gplfronttext{%
      \csname LTb\endcsname%
      \put(3530,1201){\makebox(0,0)[r]{\strut{}\wbox{$t' = 0.35t$}}}%
      \put(3530,952){\makebox(0,0)[r]{\strut{}\wbox{$U = 3t$}}}%
    }%
    \gplbacktext
    \put(0,0){\includegraphics{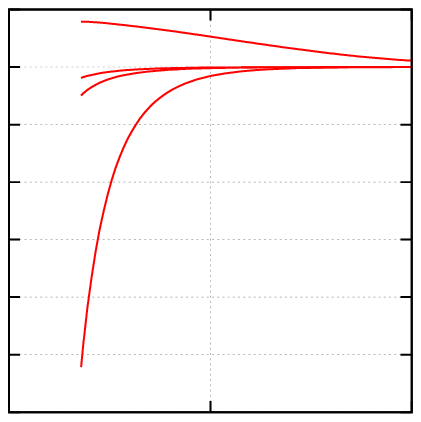}}%
    \gplfronttext
  \end{picture}%
\endgroup

%% file: FreqPap_withfreqwithSigma_Omcrit_compare.tex
% GNUPLOT: LaTeX picture with Postscript
\begingroup
  \makeatletter
  \providecommand\color[2][]{%
    \GenericError{(gnuplot) \space\space\space\@spaces}{%
      Package color not loaded in conjunction with
      terminal option `colourtext'%
    }{See the gnuplot documentation for explanation.%
    }{Either use 'blacktext' in gnuplot or load the package
      color.sty in LaTeX.}%
    \renewcommand\color[2][]{}%
  }%
  \providecommand\includegraphics[2][]{%
    \GenericError{(gnuplot) \space\space\space\@spaces}{%
      Package graphicx or graphics not loaded%
    }{See the gnuplot documentation for explanation.%
    }{The gnuplot epslatex terminal needs graphicx.sty or graphics.sty.}%
    \renewcommand\includegraphics[2][]{}%
  }%
  \providecommand\rotatebox[2]{#2}%
  \@ifundefined{ifGPcolor}{%
    \newif\ifGPcolor
    \GPcolortrue
  }{}%
  \@ifundefined{ifGPblacktext}{%
    \newif\ifGPblacktext
    \GPblacktexttrue
  }{}%
  % define a \g@addto@macro without @ in the name:
  \let\gplgaddtomacro\g@addto@macro
  % define empty templates for all commands taking text:
  \gdef\gplbacktext{}%
  \gdef\gplfronttext{}%
  \makeatother
  \ifGPblacktext
    % no textcolor at all
    \def\colorrgb#1{}%
    \def\colorgray#1{}%
  \else
    % gray or color?
    \ifGPcolor
      \def\colorrgb#1{\color[rgb]{#1}}%
      \def\colorgray#1{\color[gray]{#1}}%
      \expandafter\def\csname LTw\endcsname{\color{white}}%
      \expandafter\def\csname LTb\endcsname{\color{black}}%
      \expandafter\def\csname LTa\endcsname{\color{black}}%
      \expandafter\def\csname LT0\endcsname{\color[rgb]{1,0,0}}%
      \expandafter\def\csname LT1\endcsname{\color[rgb]{0,1,0}}%
      \expandafter\def\csname LT2\endcsname{\color[rgb]{0,0,1}}%
      \expandafter\def\csname LT3\endcsname{\color[rgb]{1,0,1}}%
      \expandafter\def\csname LT4\endcsname{\color[rgb]{0,1,1}}%
      \expandafter\def\csname LT5\endcsname{\color[rgb]{1,1,0}}%
      \expandafter\def\csname LT6\endcsname{\color[rgb]{0,0,0}}%
      \expandafter\def\csname LT7\endcsname{\color[rgb]{1,0.3,0}}%
      \expandafter\def\csname LT8\endcsname{\color[rgb]{0.5,0.5,0.5}}%
    \else
      % gray
      \def\colorrgb#1{\color{black}}%
      \def\colorgray#1{\color[gray]{#1}}%
      \expandafter\def\csname LTw\endcsname{\color{white}}%
      \expandafter\def\csname LTb\endcsname{\color{black}}%
      \expandafter\def\csname LTa\endcsname{\color{black}}%
      \expandafter\def\csname LT0\endcsname{\color{black}}%
      \expandafter\def\csname LT1\endcsname{\color{black}}%
      \expandafter\def\csname LT2\endcsname{\color{black}}%
      \expandafter\def\csname LT3\endcsname{\color{black}}%
      \expandafter\def\csname LT4\endcsname{\color{black}}%
      \expandafter\def\csname LT5\endcsname{\color{black}}%
      \expandafter\def\csname LT6\endcsname{\color{black}}%
      \expandafter\def\csname LT7\endcsname{\color{black}}%
      \expandafter\def\csname LT8\endcsname{\color{black}}%
    \fi
  \fi
  \setlength{\unitlength}{0.0500bp}%
  \begin{picture}(4680.00,3276.00)%
    \gplgaddtomacro\gplbacktext{%
      \csname LTb\endcsname%
      \put(1582,704){\makebox(0,0)[r]{\strut{}$10^{-5}$}}%
      \csname LTb\endcsname%
      \put(1582,1218){\makebox(0,0)[r]{\strut{}$10^{-4}$}}%
      \csname LTb\endcsname%
      \put(1582,1732){\makebox(0,0)[r]{\strut{}$10^{-3}$}}%
      \csname LTb\endcsname%
      \put(1582,2247){\makebox(0,0)[r]{\strut{}$10^{-2}$}}%
      \csname LTb\endcsname%
      \put(1582,2761){\makebox(0,0)[r]{\strut{}$10^{-1}$}}%
      \csname LTb\endcsname%
      \put(1582,3275){\makebox(0,0)[r]{\strut{}$10^{0}$}}%
      \csname LTb\endcsname%
      \put(1714,484){\makebox(0,0){\strut{} 0}}%
      \csname LTb\endcsname%
      \put(2228,484){\makebox(0,0){\strut{} 0.1}}%
      \csname LTb\endcsname%
      \put(2742,484){\makebox(0,0){\strut{} 0.2}}%
      \csname LTb\endcsname%
      \put(3257,484){\makebox(0,0){\strut{} 0.3}}%
      \csname LTb\endcsname%
      \put(3771,484){\makebox(0,0){\strut{} 0.4}}%
      \csname LTb\endcsname%
      \put(4285,484){\makebox(0,0){\strut{} 0.5}}%
      \put(944,1989){\rotatebox{-270}{\makebox(0,0){\strut{}$\Omega_*/t$}}}%
      \put(2999,154){\makebox(0,0){\strut{}$t'/t$}}%
    }%
    \gplgaddtomacro\gplfronttext{%
      \csname LTb\endcsname%
      \put(1971,1464){\makebox(0,0)[l]{\strut{}\fcolorbox{white}{white}{$U = 3t$}}}%
    }%
    \gplgaddtomacro\gplbacktext{%
      \csname LTb\endcsname%
      \put(1582,704){\makebox(0,0)[r]{\strut{}$10^{-5}$}}%
      \csname LTb\endcsname%
      \put(1582,1218){\makebox(0,0)[r]{\strut{}$10^{-4}$}}%
      \csname LTb\endcsname%
      \put(1582,1732){\makebox(0,0)[r]{\strut{}$10^{-3}$}}%
      \csname LTb\endcsname%
      \put(1582,2247){\makebox(0,0)[r]{\strut{}$10^{-2}$}}%
      \csname LTb\endcsname%
      \put(1582,2761){\makebox(0,0)[r]{\strut{}$10^{-1}$}}%
      \csname LTb\endcsname%
      \put(1582,3275){\makebox(0,0)[r]{\strut{}$10^{0}$}}%
      \csname LTb\endcsname%
      \put(1714,484){\makebox(0,0){\strut{} 0}}%
      \csname LTb\endcsname%
      \put(2228,484){\makebox(0,0){\strut{} 0.1}}%
      \csname LTb\endcsname%
      \put(2742,484){\makebox(0,0){\strut{} 0.2}}%
      \csname LTb\endcsname%
      \put(3257,484){\makebox(0,0){\strut{} 0.3}}%
      \csname LTb\endcsname%
      \put(3771,484){\makebox(0,0){\strut{} 0.4}}%
      \csname LTb\endcsname%
      \put(4285,484){\makebox(0,0){\strut{} 0.5}}%
      \put(944,1989){\rotatebox{-270}{\makebox(0,0){\strut{}$\Omega_*/t$}}}%
      \put(2999,154){\makebox(0,0){\strut{}$t'/t$}}%
    }%
    \gplgaddtomacro\gplfronttext{%
      \csname LTb\endcsname%
      \put(1971,1464){\makebox(0,0)[l]{\strut{}\fcolorbox{white}{white}{$U = 3t$}}}%
    }%
    \gplgaddtomacro\gplbacktext{%
      \csname LTb\endcsname%
      \put(1582,704){\makebox(0,0)[r]{\strut{}$10^{-5}$}}%
      \csname LTb\endcsname%
      \put(1582,1218){\makebox(0,0)[r]{\strut{}$10^{-4}$}}%
      \csname LTb\endcsname%
      \put(1582,1732){\makebox(0,0)[r]{\strut{}$10^{-3}$}}%
      \csname LTb\endcsname%
      \put(1582,2247){\makebox(0,0)[r]{\strut{}$10^{-2}$}}%
      \csname LTb\endcsname%
      \put(1582,2761){\makebox(0,0)[r]{\strut{}$10^{-1}$}}%
      \csname LTb\endcsname%
      \put(1582,3275){\makebox(0,0)[r]{\strut{}$10^{0}$}}%
      \csname LTb\endcsname%
      \put(1714,484){\makebox(0,0){\strut{} 0}}%
      \csname LTb\endcsname%
      \put(2228,484){\makebox(0,0){\strut{} 0.1}}%
      \csname LTb\endcsname%
      \put(2742,484){\makebox(0,0){\strut{} 0.2}}%
      \csname LTb\endcsname%
      \put(3257,484){\makebox(0,0){\strut{} 0.3}}%
      \csname LTb\endcsname%
      \put(3771,484){\makebox(0,0){\strut{} 0.4}}%
      \csname LTb\endcsname%
      \put(4285,484){\makebox(0,0){\strut{} 0.5}}%
      \put(944,1989){\rotatebox{-270}{\makebox(0,0){\strut{}$\Omega_*/t$}}}%
      \put(2999,154){\makebox(0,0){\strut{}$t'/t$}}%
    }%
    \gplgaddtomacro\gplfronttext{%
      \csname LTb\endcsname%
      \put(1971,1464){\makebox(0,0)[l]{\strut{}\fcolorbox{white}{white}{$U = 3t$}}}%
    }%
    \gplbacktext
    \put(0,0){\includegraphics{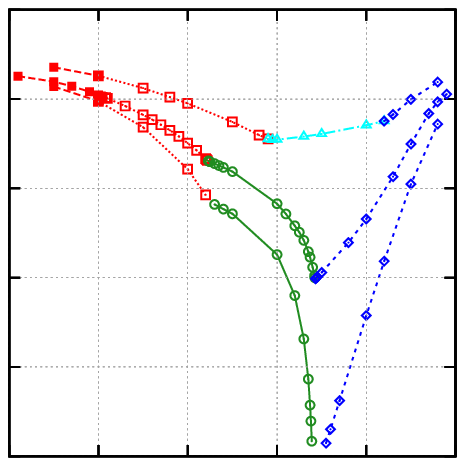}}%
    \gplfronttext
  \end{picture}%
\endgroup

%% file: FreqPap_Zdot_LorentzTest.tex
% GNUPLOT: LaTeX picture with Postscript
\begingroup
  \makeatletter
  \providecommand\color[2][]{%
    \GenericError{(gnuplot) \space\space\space\@spaces}{%
      Package color not loaded in conjunction with
      terminal option `colourtext'%
    }{See the gnuplot documentation for explanation.%
    }{Either use 'blacktext' in gnuplot or load the package
      color.sty in LaTeX.}%
    \renewcommand\color[2][]{}%
  }%
  \providecommand\includegraphics[2][]{%
    \GenericError{(gnuplot) \space\space\space\@spaces}{%
      Package graphicx or graphics not loaded%
    }{See the gnuplot documentation for explanation.%
    }{The gnuplot epslatex terminal needs graphicx.sty or graphics.sty.}%
    \renewcommand\includegraphics[2][]{}%
  }%
  \providecommand\rotatebox[2]{#2}%
  \@ifundefined{ifGPcolor}{%
    \newif\ifGPcolor
    \GPcolortrue
  }{}%
  \@ifundefined{ifGPblacktext}{%
    \newif\ifGPblacktext
    \GPblacktexttrue
  }{}%
  % define a \g@addto@macro without @ in the name:
  \let\gplgaddtomacro\g@addto@macro
  % define empty templates for all commands taking text:
  \gdef\gplbacktext{}%
  \gdef\gplfronttext{}%
  \makeatother
  \ifGPblacktext
    % no textcolor at all
    \def\colorrgb#1{}%
    \def\colorgray#1{}%
  \else
    % gray or color?
    \ifGPcolor
      \def\colorrgb#1{\color[rgb]{#1}}%
      \def\colorgray#1{\color[gray]{#1}}%
      \expandafter\def\csname LTw\endcsname{\color{white}}%
      \expandafter\def\csname LTb\endcsname{\color{black}}%
      \expandafter\def\csname LTa\endcsname{\color{black}}%
      \expandafter\def\csname LT0\endcsname{\color[rgb]{1,0,0}}%
      \expandafter\def\csname LT1\endcsname{\color[rgb]{0,1,0}}%
      \expandafter\def\csname LT2\endcsname{\color[rgb]{0,0,1}}%
      \expandafter\def\csname LT3\endcsname{\color[rgb]{1,0,1}}%
      \expandafter\def\csname LT4\endcsname{\color[rgb]{0,1,1}}%
      \expandafter\def\csname LT5\endcsname{\color[rgb]{1,1,0}}%
      \expandafter\def\csname LT6\endcsname{\color[rgb]{0,0,0}}%
      \expandafter\def\csname LT7\endcsname{\color[rgb]{1,0.3,0}}%
      \expandafter\def\csname LT8\endcsname{\color[rgb]{0.5,0.5,0.5}}%
    \else
      % gray
      \def\colorrgb#1{\color{black}}%
      \def\colorgray#1{\color[gray]{#1}}%
      \expandafter\def\csname LTw\endcsname{\color{white}}%
      \expandafter\def\csname LTb\endcsname{\color{black}}%
      \expandafter\def\csname LTa\endcsname{\color{black}}%
      \expandafter\def\csname LT0\endcsname{\color{black}}%
      \expandafter\def\csname LT1\endcsname{\color{black}}%
      \expandafter\def\csname LT2\endcsname{\color{black}}%
      \expandafter\def\csname LT3\endcsname{\color{black}}%
      \expandafter\def\csname LT4\endcsname{\color{black}}%
      \expandafter\def\csname LT5\endcsname{\color{black}}%
      \expandafter\def\csname LT6\endcsname{\color{black}}%
      \expandafter\def\csname LT7\endcsname{\color{black}}%
      \expandafter\def\csname LT8\endcsname{\color{black}}%
    \fi
  \fi
  \setlength{\unitlength}{0.0500bp}%
  \begin{picture}(3600.00,2520.00)%
    \gplgaddtomacro\gplbacktext{%
      \csname LTb\endcsname%
      \put(1265,704){\makebox(0,0)[r]{\strut{}-60}}%
      \csname LTb\endcsname%
      \put(1265,1221){\makebox(0,0)[r]{\strut{}-40}}%
      \csname LTb\endcsname%
      \put(1265,1739){\makebox(0,0)[r]{\strut{}-20}}%
      \csname LTb\endcsname%
      \put(1265,2256){\makebox(0,0)[r]{\strut{} 0}}%
      \csname LTb\endcsname%
      \put(1397,484){\makebox(0,0){\strut{}$10^{-2}$}}%
      \csname LTb\endcsname%
      \put(2174,484){\makebox(0,0){\strut{}$10^{-1}$}}%
      \csname LTb\endcsname%
      \put(2950,484){\makebox(0,0){\strut{}$10^{0}$}}%
      \put(627,1480){\rotatebox{-270}{\makebox(0,0){\strut{}$\dot Z_{(0, \pi)} /t$}}}%
      \put(2173,154){\makebox(0,0){\strut{}$\Om/t$}}%
    }%
    \gplgaddtomacro\gplfronttext{%
      \csname LTb\endcsname%
      \put(2914,1221){\makebox(0,0)[r]{\strut{}\wbox{$t' = 0.2t$}}}%
      \put(2914,963){\makebox(0,0)[r]{\strut{}\wbox{$U = 3t$}}}%
    }%
    \gplbacktext
    \put(0,0){\includegraphics{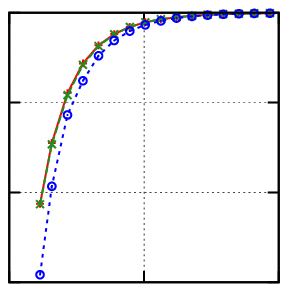}}%
    \gplfronttext
  \end{picture}%
\endgroup

%% file: FreqPap.bbl
\begin{thebibliography}{38}
\expandafter\ifx\csname natexlab\endcsname\relax\def\natexlab#1{#1}\fi
\expandafter\ifx\csname bibnamefont\endcsname\relax
  \def\bibnamefont#1{#1}\fi
\expandafter\ifx\csname bibfnamefont\endcsname\relax
  \def\bibfnamefont#1{#1}\fi
\expandafter\ifx\csname citenamefont\endcsname\relax
  \def\citenamefont#1{#1}\fi
\expandafter\ifx\csname url\endcsname\relax
  \def\url#1{\texttt{#1}}\fi
\expandafter\ifx\csname urlprefix\endcsname\relax\def\urlprefix{URL }\fi
\providecommand{\bibinfo}[2]{#2}
\providecommand{\eprint}[2][]{\url{#2}}

\bibitem[{\citenamefont{Lee et~al.}(2006)\citenamefont{Lee, Nagaosa, and
  Wen}}]{LeeNagaosaWen2006}
\bibinfo{author}{\bibfnamefont{P.~A.} \bibnamefont{Lee}},
  \bibinfo{author}{\bibfnamefont{N.}~\bibnamefont{Nagaosa}}, \bibnamefont{and}
  \bibinfo{author}{\bibfnamefont{X.-G.} \bibnamefont{Wen}},
  \bibinfo{journal}{Rev. Mod. Phys.} \textbf{\bibinfo{volume}{78}},
  \bibinfo{pages}{17} (\bibinfo{year}{2006}).

\bibitem[{\citenamefont{Mackenzie and Maeno}(2003)}]{MackenzieMaeno}
\bibinfo{author}{\bibfnamefont{A.~P.} \bibnamefont{Mackenzie}}
  \bibnamefont{and} \bibinfo{author}{\bibfnamefont{Y.}~\bibnamefont{Maeno}},
  \bibinfo{journal}{Rev. Mod. Phys.} \textbf{\bibinfo{volume}{75}},
  \bibinfo{pages}{657} (\bibinfo{year}{2003}).

\bibitem[{\citenamefont{Kamihara et~al.}(2008)\citenamefont{Kamihara, Watanabe,
  Hirano, and Hosono}}]{pnictides}
\bibinfo{author}{\bibfnamefont{Y.}~\bibnamefont{Kamihara}},
  \bibinfo{author}{\bibfnamefont{T.}~\bibnamefont{Watanabe}},
  \bibinfo{author}{\bibfnamefont{M.}~\bibnamefont{Hirano}}, \bibnamefont{and}
  \bibinfo{author}{\bibfnamefont{H.}~\bibnamefont{Hosono}},
  \bibinfo{journal}{Journal of the American Chemical Society}
  \textbf{\bibinfo{volume}{130}}, \bibinfo{pages}{3296} (\bibinfo{year}{2008}).

\bibitem[{\citenamefont{Paglione and Greene}(2010)}]{PaglioneGreene2010}
\bibinfo{author}{\bibfnamefont{J.}~\bibnamefont{Paglione}} \bibnamefont{and}
  \bibinfo{author}{\bibfnamefont{R.~L.} \bibnamefont{Greene}},
  \bibinfo{journal}{Nature Physics} \textbf{\bibinfo{volume}{6}},
  \bibinfo{pages}{645} (\bibinfo{year}{2010}).

\bibitem[{\citenamefont{Metzner et~al.}(2011)\citenamefont{Metzner, Salmhofer,
  Honerkamp, Meden, and Sch\"onhammer}}]{RGreview2011}
\bibinfo{author}{\bibfnamefont{W.}~\bibnamefont{Metzner}},
  \bibinfo{author}{\bibfnamefont{M.}~\bibnamefont{Salmhofer}},
  \bibinfo{author}{\bibfnamefont{C.}~\bibnamefont{Honerkamp}},
  \bibinfo{author}{\bibfnamefont{V.}~\bibnamefont{Meden}}, \bibnamefont{and}
  \bibinfo{author}{\bibfnamefont{K.}~\bibnamefont{Sch\"onhammer}}
  (\bibinfo{year}{2011}), \bibinfo{note}{to appear in Rev. Mod. Phys.},
  \eprint{arXiv:1105.5289}.

\bibitem[{\citenamefont{Feldman and Trubowitz}(1990)}]{FT1990}
\bibinfo{author}{\bibfnamefont{J.}~\bibnamefont{Feldman}} \bibnamefont{and}
  \bibinfo{author}{\bibfnamefont{E.}~\bibnamefont{Trubowitz}},
  \bibinfo{journal}{Helv. Phys. Acta} \textbf{\bibinfo{volume}{63}},
  \bibinfo{pages}{156} (\bibinfo{year}{1990}).

\bibitem[{\citenamefont{Feldman et~al.}(1992)\citenamefont{Feldman, Magnen,
  Rivasseau, and Trubowitz}}]{FMRT}
\bibinfo{author}{\bibfnamefont{J.}~\bibnamefont{Feldman}},
  \bibinfo{author}{\bibfnamefont{J.}~\bibnamefont{Magnen}},
  \bibinfo{author}{\bibfnamefont{V.}~\bibnamefont{Rivasseau}},
  \bibnamefont{and}
  \bibinfo{author}{\bibfnamefont{E.}~\bibnamefont{Trubowitz}},
  \bibinfo{journal}{Helv. Phys. Acta} \textbf{\bibinfo{volume}{65}},
  \bibinfo{pages}{679} (\bibinfo{year}{1992}).

\bibitem[{\citenamefont{Feldman et~al.}(2004)\citenamefont{Feldman, Kn\"orrer,
  and Trubowitz}}]{FKTperturbationbound2004}
\bibinfo{author}{\bibfnamefont{J.}~\bibnamefont{Feldman}},
  \bibinfo{author}{\bibfnamefont{H.}~\bibnamefont{Kn\"orrer}},
  \bibnamefont{and}
  \bibinfo{author}{\bibfnamefont{E.}~\bibnamefont{Trubowitz}},
  \bibinfo{journal}{Comm. Math. Phys.} \textbf{\bibinfo{volume}{247}},
  \bibinfo{pages}{195} (\bibinfo{year}{2004}).

\bibitem[{\citenamefont{Salmhofer and
  Wieczerkowski}(2000)}]{SalmhoferWieczerkowski2000}
\bibinfo{author}{\bibfnamefont{M.}~\bibnamefont{Salmhofer}} \bibnamefont{and}
  \bibinfo{author}{\bibfnamefont{C.}~\bibnamefont{Wieczerkowski}},
  \bibinfo{journal}{J. Stat. Phys.} \textbf{\bibinfo{volume}{99}},
  \bibinfo{pages}{557} (\bibinfo{year}{2000}).

\bibitem[{\citenamefont{Disertori and Rivasseau}(2000)}]{DR2000}
\bibinfo{author}{\bibfnamefont{M.}~\bibnamefont{Disertori}} \bibnamefont{and}
  \bibinfo{author}{\bibfnamefont{V.}~\bibnamefont{Rivasseau}},
  \bibinfo{journal}{Commun. Math. Phys.} \textbf{\bibinfo{volume}{215}},
  \bibinfo{pages}{251} (\bibinfo{year}{2000}).

\bibitem[{\citenamefont{Benfatto et~al.}(2006)\citenamefont{Benfatto, Giuliani,
  and Mastropietro}}]{BenfattoGiulianiMastropietro2006}
\bibinfo{author}{\bibfnamefont{G.}~\bibnamefont{Benfatto}},
  \bibinfo{author}{\bibfnamefont{A.}~\bibnamefont{Giuliani}}, \bibnamefont{and}
  \bibinfo{author}{\bibfnamefont{V.}~\bibnamefont{Mastropietro}},
  \bibinfo{journal}{Ann. H. Poincar\'e} \textbf{\bibinfo{volume}{5}},
  \bibinfo{pages}{809} (\bibinfo{year}{2006}).

\bibitem[{\citenamefont{Pedra and Salmhofer}(2008)}]{PedraSalmhofer2008}
\bibinfo{author}{\bibfnamefont{W.}~\bibnamefont{Pedra}} \bibnamefont{and}
  \bibinfo{author}{\bibfnamefont{M.}~\bibnamefont{Salmhofer}},
  \bibinfo{journal}{Comm. Math. Phys.} \textbf{\bibinfo{volume}{282}},
  \bibinfo{pages}{797} (\bibinfo{year}{2008}).

\bibitem[{\citenamefont{Husemann and Salmhofer}(2009)}]{decomposition}
\bibinfo{author}{\bibfnamefont{C.}~\bibnamefont{Husemann}} \bibnamefont{and}
  \bibinfo{author}{\bibfnamefont{M.}~\bibnamefont{Salmhofer}},
  \bibinfo{journal}{Phys. Rev. B} \textbf{\bibinfo{volume}{79}},
  \bibinfo{pages}{195125} (\bibinfo{year}{2009}).

\bibitem[{\citenamefont{Zanchi and Schulz}(1998)}]{ZanchiSchulz1}
\bibinfo{author}{\bibfnamefont{D.}~\bibnamefont{Zanchi}} \bibnamefont{and}
  \bibinfo{author}{\bibfnamefont{H.~J.} \bibnamefont{Schulz}},
  \bibinfo{journal}{Europhys. Lett.} \textbf{\bibinfo{volume}{44}},
  \bibinfo{pages}{235} (\bibinfo{year}{1998}).

\bibitem[{\citenamefont{Zanchi and Schulz}(2000)}]{ZanchiSchulz2000}
\bibinfo{author}{\bibfnamefont{D.}~\bibnamefont{Zanchi}} \bibnamefont{and}
  \bibinfo{author}{\bibfnamefont{H.~J.} \bibnamefont{Schulz}},
  \bibinfo{journal}{Phys. Rev. B} \textbf{\bibinfo{volume}{61}},
  \bibinfo{pages}{13609} (\bibinfo{year}{2000}).

\bibitem[{\citenamefont{Halboth and Metzner}(2000)}]{HalbothMetzner}
\bibinfo{author}{\bibfnamefont{C.~J.} \bibnamefont{Halboth}} \bibnamefont{and}
  \bibinfo{author}{\bibfnamefont{W.}~\bibnamefont{Metzner}},
  \bibinfo{journal}{Phys. Rev. B} \textbf{\bibinfo{volume}{61}},
  \bibinfo{pages}{7364} (\bibinfo{year}{2000}).

\bibitem[{\citenamefont{Honerkamp et~al.}(2001)\citenamefont{Honerkamp,
  Salmhofer, Furukawa, and Rice}}]{Umklapp}
\bibinfo{author}{\bibfnamefont{C.}~\bibnamefont{Honerkamp}},
  \bibinfo{author}{\bibfnamefont{M.}~\bibnamefont{Salmhofer}},
  \bibinfo{author}{\bibfnamefont{N.}~\bibnamefont{Furukawa}}, \bibnamefont{and}
  \bibinfo{author}{\bibfnamefont{T.~M.} \bibnamefont{Rice}},
  \bibinfo{journal}{Phys. Rev. B} \textbf{\bibinfo{volume}{63}},
  \bibinfo{pages}{035109} (\bibinfo{year}{2001}).

\bibitem[{\citenamefont{Honerkamp and
  Salmhofer}(2003)}]{HonerkampSalmhofer_QuasiparticleWeight}
\bibinfo{author}{\bibfnamefont{C.}~\bibnamefont{Honerkamp}} \bibnamefont{and}
  \bibinfo{author}{\bibfnamefont{M.}~\bibnamefont{Salmhofer}},
  \bibinfo{journal}{Phys. Rev. B} \textbf{\bibinfo{volume}{67}},
  \bibinfo{pages}{174504} (\bibinfo{year}{2003}).

\bibitem[{\citenamefont{{Honerkamp} and
  {Salmhofer}}(2001)}]{HonerkampSalmhofer_Tflow}
\bibinfo{author}{\bibfnamefont{C.}~\bibnamefont{{Honerkamp}}} \bibnamefont{and}
  \bibinfo{author}{\bibfnamefont{M.}~\bibnamefont{{Salmhofer}}},
  \bibinfo{journal}{\prb} \textbf{\bibinfo{volume}{64}},
  \bibinfo{pages}{184516} (\bibinfo{year}{2001}).

\bibitem[{\citenamefont{Honerkamp et~al.}(2004)\citenamefont{Honerkamp, Rohe,
  Andergassen, and Enss}}]{HonerkampInteractionFlow2004}
\bibinfo{author}{\bibfnamefont{C.}~\bibnamefont{Honerkamp}},
  \bibinfo{author}{\bibfnamefont{D.}~\bibnamefont{Rohe}},
  \bibinfo{author}{\bibfnamefont{S.}~\bibnamefont{Andergassen}},
  \bibnamefont{and} \bibinfo{author}{\bibfnamefont{T.}~\bibnamefont{Enss}},
  \bibinfo{journal}{Phys. Rev. B} \textbf{\bibinfo{volume}{70}},
  \bibinfo{pages}{235115} (\bibinfo{year}{2004}).

\bibitem[{\citenamefont{Ortloff et~al.}(2011)\citenamefont{Ortloff, Husemann,
  Honerkamp, and Salmhofer}}]{remainder}
\bibinfo{author}{\bibfnamefont{J.}~\bibnamefont{Ortloff}},
  \bibinfo{author}{\bibfnamefont{C.}~\bibnamefont{Husemann}},
  \bibinfo{author}{\bibfnamefont{C.}~\bibnamefont{Honerkamp}},
  \bibnamefont{and} \bibinfo{author}{\bibfnamefont{M.}~\bibnamefont{Salmhofer}}
  (\bibinfo{year}{2011}), \eprint{in preparation.}

\bibitem[{\citenamefont{Salmhofer and Honerkamp}(2001)}]{salmfrgtt}
\bibinfo{author}{\bibfnamefont{M.}~\bibnamefont{Salmhofer}} \bibnamefont{and}
  \bibinfo{author}{\bibfnamefont{C.}~\bibnamefont{Honerkamp}},
  \bibinfo{journal}{Prog. Theor. Phys.} \textbf{\bibinfo{volume}{105}},
  \bibinfo{pages}{1} (\bibinfo{year}{2001}).

\bibitem[{\citenamefont{Honerkamp}(2001)}]{Honerkamp_ScatRate}
\bibinfo{author}{\bibfnamefont{C.}~\bibnamefont{Honerkamp}},
  \bibinfo{journal}{Eur. Phys. J. B} \textbf{\bibinfo{volume}{21}},
  \bibinfo{pages}{81} (\bibinfo{year}{2001}).

\bibitem[{\citenamefont{Rohe and Metzner}(2005)}]{RoheMetzner}
\bibinfo{author}{\bibfnamefont{D.}~\bibnamefont{Rohe}} \bibnamefont{and}
  \bibinfo{author}{\bibfnamefont{W.}~\bibnamefont{Metzner}},
  \bibinfo{journal}{Phys. Rev. B} \textbf{\bibinfo{volume}{71}},
  \bibinfo{pages}{115116} (\bibinfo{year}{2005}).

\bibitem[{\citenamefont{Katanin and Kampf}(2004)}]{KataninKampf2004}
\bibinfo{author}{\bibfnamefont{A.~A.} \bibnamefont{Katanin}} \bibnamefont{and}
  \bibinfo{author}{\bibfnamefont{A.~P.} \bibnamefont{Kampf}},
  \bibinfo{journal}{Phys. Rev. Lett.} \textbf{\bibinfo{volume}{93}},
  \bibinfo{pages}{106406} (\bibinfo{year}{2004}).

\bibitem[{\citenamefont{Baier et~al.}(2004)\citenamefont{Baier, Bick, and
  Wetterich}}]{BaierBickWetterich}
\bibinfo{author}{\bibfnamefont{T.}~\bibnamefont{Baier}},
  \bibinfo{author}{\bibfnamefont{E.}~\bibnamefont{Bick}}, \bibnamefont{and}
  \bibinfo{author}{\bibfnamefont{C.}~\bibnamefont{Wetterich}},
  \bibinfo{journal}{Phys. Rev. B} \textbf{\bibinfo{volume}{70}},
  \bibinfo{pages}{125111} (\bibinfo{year}{2004}).

\bibitem[{\citenamefont{Friederich et~al.}(2011)\citenamefont{Friederich,
  Krahl, and Wetterich}}]{Friederich2011}
\bibinfo{author}{\bibfnamefont{S.}~\bibnamefont{Friederich}},
  \bibinfo{author}{\bibfnamefont{H.~C.} \bibnamefont{Krahl}}, \bibnamefont{and}
  \bibinfo{author}{\bibfnamefont{C.}~\bibnamefont{Wetterich}},
  \bibinfo{journal}{Phys. Rev. B} \textbf{\bibinfo{volume}{83}},
  \bibinfo{pages}{155125} (\bibinfo{year}{2011}).

\bibitem[{\citenamefont{Karrasch et~al.}(2008)\citenamefont{Karrasch, Hedden,
  Peters, Pruschke, Sch\"onhammer, and Meden}}]{KarraschHeddenMeden2008}
\bibinfo{author}{\bibfnamefont{C.}~\bibnamefont{Karrasch}},
  \bibinfo{author}{\bibfnamefont{R.}~\bibnamefont{Hedden}},
  \bibinfo{author}{\bibfnamefont{R.}~\bibnamefont{Peters}},
  \bibinfo{author}{\bibfnamefont{T.}~\bibnamefont{Pruschke}},
  \bibinfo{author}{\bibfnamefont{K.}~\bibnamefont{Sch\"onhammer}},
  \bibnamefont{and} \bibinfo{author}{\bibfnamefont{V.}~\bibnamefont{Meden}},
  \bibinfo{journal}{J. Phys.: Cond. Matt.} \textbf{\bibinfo{volume}{20}},
  \bibinfo{pages}{345205} (\bibinfo{year}{2008}).

\bibitem[{\citenamefont{Jakobs et~al.}(2010)\citenamefont{Jakobs, Pletyukhov,
  and Schoeller}}]{SeverinPletyukovSchoeller2010}
\bibinfo{author}{\bibfnamefont{S.~G.} \bibnamefont{Jakobs}},
  \bibinfo{author}{\bibfnamefont{M.}~\bibnamefont{Pletyukhov}},
  \bibnamefont{and}
  \bibinfo{author}{\bibfnamefont{H.}~\bibnamefont{Schoeller}},
  \bibinfo{journal}{Phys. Rev. B} \textbf{\bibinfo{volume}{81}},
  \bibinfo{pages}{195109} (\bibinfo{year}{2010}).

\bibitem[{\citenamefont{Bartosch et~al.}(2009)\citenamefont{Bartosch, Freire,
  Cardenas, and Kopietz}}]{BartoschKopietz2009}
\bibinfo{author}{\bibfnamefont{L.}~\bibnamefont{Bartosch}},
  \bibinfo{author}{\bibfnamefont{H.}~\bibnamefont{Freire}},
  \bibinfo{author}{\bibfnamefont{J.~J.~R.} \bibnamefont{Cardenas}},
  \bibnamefont{and} \bibinfo{author}{\bibfnamefont{P.}~\bibnamefont{Kopietz}},
  \bibinfo{journal}{Journal of Physics: Condensed Matter}
  \textbf{\bibinfo{volume}{21}}, \bibinfo{pages}{305602}
  (\bibinfo{year}{2009}).

\bibitem[{\citenamefont{Schmidt and Enss}(2011)}]{SchmidtEnss}
\bibinfo{author}{\bibfnamefont{R.}~\bibnamefont{Schmidt}} \bibnamefont{and}
  \bibinfo{author}{\bibfnamefont{T.}~\bibnamefont{Enss}},
  \bibinfo{journal}{Phys. Rev. A} \textbf{\bibinfo{volume}{83}},
  \bibinfo{pages}{063620} (\bibinfo{year}{2011}).

\bibitem[{\citenamefont{Zinn-Justin}(2002)}]{ZinnJustin}
\bibinfo{author}{\bibfnamefont{J.}~\bibnamefont{Zinn-Justin}},
  \emph{\bibinfo{title}{Quantum Field Theory and Critical Phenomena}}
  (\bibinfo{publisher}{Oxford Univ. Press}, \bibinfo{year}{2002}).

\bibitem[{\citenamefont{Katanin}(2004)}]{KataninWard}
\bibinfo{author}{\bibfnamefont{A.~A.} \bibnamefont{Katanin}},
  \bibinfo{journal}{Phys. Rev. B} \textbf{\bibinfo{volume}{70}},
  \bibinfo{pages}{115109} (\bibinfo{year}{2004}).

\bibitem[{\citenamefont{Salmhofer et~al.}(2004)\citenamefont{Salmhofer,
  Honerkamp, Metzner, and Lauscher}}]{symmetrybrokenBCS}
\bibinfo{author}{\bibfnamefont{M.}~\bibnamefont{Salmhofer}},
  \bibinfo{author}{\bibfnamefont{C.}~\bibnamefont{Honerkamp}},
  \bibinfo{author}{\bibfnamefont{W.}~\bibnamefont{Metzner}}, \bibnamefont{and}
  \bibinfo{author}{\bibfnamefont{O.}~\bibnamefont{Lauscher}},
  \bibinfo{journal}{Prog. Theor. Phys.} \textbf{\bibinfo{volume}{112}},
  \bibinfo{pages}{943} (\bibinfo{year}{2004}).

\bibitem[{\citenamefont{Igoshev et~al.}(2011)\citenamefont{Igoshev, Irkhin, and
  Katanin}}]{Katanin_SigmaFM}
\bibinfo{author}{\bibfnamefont{P.~A.} \bibnamefont{Igoshev}},
  \bibinfo{author}{\bibfnamefont{V.~Y.} \bibnamefont{Irkhin}},
  \bibnamefont{and} \bibinfo{author}{\bibfnamefont{A.~A.}
  \bibnamefont{Katanin}}, \bibinfo{journal}{Phys. Rev. B}
  \textbf{\bibinfo{volume}{83}}, \bibinfo{pages}{245118}
  (\bibinfo{year}{2011}).

\bibitem[{\citenamefont{Giering}(in preparation)}]{giering_phd}
\bibinfo{author}{\bibfnamefont{K.-U.} \bibnamefont{Giering}}, Ph.D. thesis
  (\bibinfo{year}{in preparation}).

\bibitem[{\citenamefont{{Feldman} and
  {Salmhofer}}(2008)}]{FeldmannSalmhofer2008b}
\bibinfo{author}{\bibfnamefont{J.}~\bibnamefont{{Feldman}}} \bibnamefont{and}
  \bibinfo{author}{\bibfnamefont{M.}~\bibnamefont{{Salmhofer}}},
  \bibinfo{journal}{Rev. Math. Phys.} \textbf{\bibinfo{volume}{20}},
  \bibinfo{pages}{275} (\bibinfo{year}{2008}).

\bibitem[{\citenamefont{Giering and Salmhofer}(2011)}]{GieringSalmhofer2011}
\bibinfo{author}{\bibfnamefont{K.-U.} \bibnamefont{Giering}} \bibnamefont{and}
  \bibinfo{author}{\bibfnamefont{M.}~\bibnamefont{Salmhofer}}
  (\bibinfo{year}{2011}), \eprint{in preparation.}

\end{thebibliography}
